\documentclass[aip,amsmath,amssymb,preprint]{revtex4-1}
\usepackage[utf8]{inputenc}

\usepackage{amsmath}%
\usepackage{graphicx}%
\usepackage{dcolumn}%
\usepackage{bm}%
\usepackage[caption=false]{subfig}%
\usepackage{multirow}%
\usepackage{verbatim}
\usepackage{mathrsfs}%
\usepackage{amssymb}
\usepackage{mathrsfs}
\usepackage{xcolor}

\begin{document}

\title{How to build coarse-grain transport models consistent from the kinetic to fluid regimes}

\date{\today}

\author{Erik \surname{Torres}}
\email[Corresponding author: ]{etorres@umn.edu}
\affiliation{Department of Aerospace Engineering, University of Minnesota, 110 Union St. SE, Minneapolis, MN, 55455, USA}
\author{Georgios \surname{Bellas-Chatzigeorgis}}
\email[Currently at NASA Ames Research Center, Moffett Field, CA, 94035, USA]{}
\affiliation{Aeronautics and Aerospace Department, von Karman Institute for Fluid Dynamics, Chauss\'ee de Waterloo 72, 1640 Rhode-Saint-Gen\`ese, Belgium}
\author{Thierry E. \surname{Magin}}
\affiliation{Aeronautics and Aerospace Department, von Karman Institute for Fluid Dynamics, Chauss\'ee de Waterloo 72, 1640 Rhode-Saint-Gen\`ese, Belgium}

\begin{abstract}
 In this paper, we examine how to build coarse-grain transport models consistently from the kinetic to fluid regimes. The internal energy of the gas particles is described through a state-to-state approach. A kinetic equation allows us to study transport phenomena in phase space for a non-homogeneous gas mixture. Internal energy excitation is modeled using a binary collision operator, whereas the gas chemical processes rely on a reactive collision operator. We obtain an asymptotic fluid model by means of a Chapman-Enskog perturbative solution to the Boltzmann equation in the Maxwellian reaction regime. The macroscopic conservation equations of species mass, mixture momentum, and energy are given, as well as expressions of the transport properties. Reversibility relations for elementary processes are formulated in the coarse-grain model at the kinetic level and are enforced in the collision algorithm of the direct simulation Monte Carlo method used to solve the kinetic equation. Furthermore, respecting these reversibility relations is key to deriving a fluid model that is well-posed and compatible with the second law of thermodynamics. Consistency between the kinetic and fluid simulations is assessed for the simulation of a shock wave in a nitrogen gas using the uniform rovibrational collisional coarse-grain model. The kinetic and fluid simulations show consistency for the macroscopic properties and transport fluxes between both regimes.
\end{abstract}


\maketitle



\section{Introduction} \label{sec:introduction}

Successful prediction of the heat loads on a spacecraft during atmospheric entry relies, among other things, on the completeness and accuracy of the model used to describe thermo-chemical nonequilibrium and transport phenomena in the flow~\cite{park90a}. 
Modeling of such effects in the continuum limit is usually done with hydrodynamic-scale Computational Fluid Dynamics~\cite{hirsch88a} (CFD) methods, which require chemical-kinetic databases for calculating the rate coefficients of internal energy excitation and molecular dissociation, as well as transport properties for modeling viscous and diffusion effects. On the other hand, kinetic-scale direct simulation Monte Carlo~\cite{bird94a} (DSMC) methods allow for accurate description of the flow encountered in regions with continuum breakdown and rely on cross section models to predict the outcome of elastic and inelastic collisions. 

With increasing computational power it is becoming commonplace to generate high-fidelity kinetic data free from empiricism through the methods of computational chemistry. This typically involves the generation of potential energy surfaces (PES) for the molecular systems in question and subsequent quasi-classical trajectory (QCT) calculations on these surfaces to obtain reaction cross sections and the related rate coefficients (e.g. for $\mathrm{N_2}$-$\mathrm{N}$~\cite{esposito99a, esposito06a, jaffe15a}, $\mathrm{N_2}$-$\mathrm{N_2}$~\cite{bender15a, macdonald18b}, $\mathrm{O_2}$-$\mathrm{O}$~\cite{esposito08a} and $\mathrm{N_2}$-$\mathrm{O_2}$~\cite{chaudhry18b}). Due to the vast number of internal energy transfer and elementary chemical processes that must be tracked for all mixture components in Earth atmospheric entry flows, detailed chemistry CFD simulations are still too computationally expensive for practical applications. Even for relative simple mixtures consisting only of nitrogen molecules and atoms, rovibrational-specific state-to-state calculations have been limited to master equation studies involving space-homogeneous heat baths~\cite{panesi13a, kim13a, macdonald20b} and, at most, one-dimensional flows behind inviscid normal shocks~\cite{panesi14a}. Electronic-specific state-to-state CFD models have been developed to simulate electronic excitation and partial ionization in gases such as argon~\cite{kapper11a}, although in this case the number of discrete internal energy levels was much smaller than for the aforementioned molecular systems. Equivalent DSMC studies are even less common. Bruno et al~\cite{bruno02a} were the first to incorporate QCT-derived vibrational-specific $\mathrm{N_2}$-$\mathrm{N}$ cross sections into a DSMC solver to study internal energy exchange and dissociation of nitrogen across a normal shock. To date, the only state-to-state DSMC simulations using a complete set of rovibrational-specific reaction cross sections for the $\mathrm{N_2}$-$\mathrm{N}$ system have been carried out by Kim and Boyd~\cite{kim14a}.

One appealing way to reduce the computational cost of state-to-state nonequilibrium reacting flow calculations has been to develop coarse-grain models. The details of the reduction vary, but can broadly be classified into vibrational-specific,~\cite{esposito00a, munafo12a} energy bin,~\cite{magin12a, munafo14c, munafo14d} hybrids of both,~\cite{liu15a, macdonald18b} or more recently adaptive grouping of rovibrational states.~\cite{sahai17a, sahai19a, sharma20a} The basic concept is always to approximate the behavior of the full kinetic database with a much smaller set of cross sections/rate coefficients by grouping together many individual processes. In addition to air chemistry, the approach has also been applied to electronic-specific simulations of argon plasma~\cite{le13a}. On the DSMC side, coarse-grain models have been investigated as well.~\cite{zhu16a, torres18b} In every case, the lumping-together of internal energy levels leads to a reduction in the number of associated state-to-state reaction rate coefficients / cross sections and greatly reduces the cost of simulations. 

But with the size reduction of the kinetic databases also comes a loss of fidelity in the thermodynamic and chemical-kinetic description, especially if the binning strategy chosen is inadequate. The problem was recognized early on with the uniform rovibrational collisional (URVC) bin model~\cite{magin12a}. Its main underlying assumption of ``freezing'' the populations of rovibrational levels within a given bin to constant values (short recap in Sec.~\ref{sec:coarse_grain_model}) was shown to produce a set of mass balance equations at the coarse-grain level, which would be incompatible with micro-reversibility relations linking forward and backward rates between rovibrational states. This meant that the fluid equations could not satisfy the second law of thermodynamics, and the formulation of an associated entropy equation with unambiguous non-negative entropy production terms was not possible. Thus, one would not be guaranteed to retrieve the correct thermodynamic equilibrium with the URVC model. Furthermore, it was shown~\cite{munafo14a} that for the original formulation of the URVC bin model a large number of bins (up to 50) was required to approach the chemical dynamics of the full state-to-state system.

For the inviscid fluid limit, this problem was first remedied by allowing the level populations within each bin to assume a Boltzmann population at the gas temperature $T$ (so-called Boltzmann binning~\cite{munafo14c, munafo14d}), while the bin populations themselves could still be relaxing toward equilibrium. With this change, the gas would be guaranteed to eventually assume the correct equilibrium populations regardless of the number of bins used. Building on this approach, so-called Maximum Entropy grouping~\cite{liu15a, sahai17a, sahai19a, sharma20a} adopted the use of bin-specific, or ``group-internal'' temperatures to allow the level populations within each bin even more degrees of freedom for local adjustment. This made it possible to almost exactly match the thermodynamic equilibrium and reaction rates of full state-to-state calculations with as few as two, or three overall bins. As a consequence, most research has so far concentrated on refining the coarse-grain models to best approximate the full chemical kinetics in the inviscid limit. 

In the few cases where viscous phenomena have been taken into account,~\cite{munafo14d, bellas20a} the transport properties were assumed to be independent of the molecules' internal energy states and were computed based on the current state-of-the-art collision integrals~\cite{stallcop01a}. It has however been theorized~\cite{mccourt90a, giovangigli99a, nagnibeda09a, brun09a} that transport properties should exhibit a dependence on the internal energy distributions. Indeed, in the state-to-state framework such a dependence appears naturally when deriving the Navier-Stokes equations as asymptotic solutions to the Boltzmann equation. Therefore, it must be taken into account for a well-posed fluid model. None of the coarse-grain models proposed so far have really addressed this issue. The aforementioned Boltzmann bin and Maximum entropy reductions also imply that the partition function of each energy bin must be temperature-dependent. This may work well within the context of a CFD fluid model, where detailed balance relations involve ratios of temperature-dependent forward and backward rate coefficients, but breaks down in DSMC~\cite{torres13a}, where these same relations have to be expressed in terms of collision energy-dependent cross sections. Such coarse-grain models effectively require individual molecules in the gas to ``be aware'' of the surrounding temperature, which is not compatible with the kinetic-scale description. 

In this work we formulate a coarse-grain model, which works within the context of DSMC and simultaneously allows for the derivation of viscous fluid equations with consistent transport terms. We propose a small, but important change~\cite{torres20a} to the early URVC bin model~\cite{magin12a}. Instead of attempting to enforce micro-reversibility relations for all rovibrational levels, we will postulate that such relations must hold only for the bin populations. We will therefore sacrifice some of the fine-grain detail of the original system in exchange for a simpler coarse-grain model. Our model still assumes constant populations for all rovibrational levels lumped into a given bin.
Its main usefulness lies not so much in the ability to reproduce the full chemical kinetics with the smallest number of bins, but in the rather simple manner with which reversibility relations can be expressed in terms of coarse-grain cross sections. Furthermore, it allows us to postulate a Boltzmann equation at the coarse-grain level, which forms the starting point for a straightforward application of the Chapman-Enskog method to derive the fluid equations. As part of this, we obtain expressions for the transport properties directly based on the same coarse-grain cross sections appearing in the kinetic equation. As a consequence, the resulting transport properties are fully consistent with the corresponding coarse-grain DSMC collision model~\cite{torres18b} and naturally account for the transfer of internal energy without the need for \emph{ad hoc} fixes, such as the Eucken correction~\cite{stephani12a, liechty19a}.

%

%
%
%
%
%
%
%
%



Our main objectives with this paper are: (1) Formulate the state-to-state kinetic equation for the coarse-grain model including fast (elastic) and slow (inelastic and reactive) collision terms, with reversibility expressed at the kinetic scale. (2) Derive the fluid equations for the coarse-grain model as an asymptotic solution to the kinetic equation by means of the Chapman-Enskog method. This includes expressions for the chemical source terms, the viscous fluxes and an entropy equation, with reversibility found at the macroscopic scale. (3) Verify the consistency of the hydrodynamic (Euler, or Navier-Stokes eqs.) and kinetic (Boltzmann eq.) coarse-grain models by simulation of normal shocks in nitrogen with CFD and DSMC methods. Assess the degree to which continuum breakdown across the shock causes the flow fields in the hydrodynamic and kinetic solutions to depart from one another.

This paper is organized as follows. In Sec.~\ref{sec:coarse_grain_model}, we introduce the coarse-grain model for inelastic processes in molecular gas mixtures and recall its main features. In Sec.~\ref{sec:boltzmann_equation}, we discuss the governing kinetic equation and detail its constituting terms, including reversibility relations. In Sec.~\ref{sec:hydrodynamic_description}, we apply the Chapman-Enskog method to derive the corresponding fluid equations, along with expressions for all necessary transport and chemical source terms. In addition, we show that the entropy production terms due to transport and chemistry are always non-negative and thus the coarse-grain fluid equations satisfy the second law of thermodynamics. In Sec.~\ref{sec:normal_shock_bins}, we apply the coarse-grain model to reveal the structure of normal shock waves in a reacting gas mixture using three distinct simulation techniques. We first obtain the flow field in the inviscid limit by solving the system of master equations coupled to total momentum and energy balances behind the shock front. Then, we solve the full fluid equations across the shock with added viscous terms (Navier-Stokes) by means of the Finite Volume method. Finally, we directly solve the Boltzmann kinetic equation for the coarse-grain model by means of direct simulation Monte Carlo. These high-fidelity calculations provide a check on the fluid model and reveal additional features of the flow field. Finally, in Sec.~\ref{sec:conclusions}, we state the conclusions and discuss possible future work. 


\section{Coarse-grain model for N3 system} \label{sec:coarse_grain_model}

Throughout the remainder of this paper, we will consider as an example a mixture of molecular and atomic nitrogen, with both species in their ground electronic states, and use a set of cross sections derived from QCT calculations on an \emph{ab initio} PES for the $\mathrm{N_2}(v,J) + \mathrm{N}$ system, originally compiled at NASA Ames Research Center~\cite{jaffe08a, jaffe15a}. The 9390 rovibrational levels of the $\mathrm{N_2}$ molecule in its ground electronic state have been grouped together into a much smaller number of discrete internal energy bins according to the uniform rovibrational collisional (URVC) energy bin model~\cite{magin12a, torres20a, torres18b}. As a result, our mixture is composed of energy bins (labeled with indices $k = 1, 2, \ldots, \mathcal{N}_\mathrm{bins}$) that encompass all bound- and pre-dissociated levels $i \in \mathcal{I}_\mathrm{N_2}$, plus atomic nitrogen in the ground electronic state. 

The number of molecules per unit volume populating bin $k$ is the sum over all level populations belonging to it, i.e.: $n_k = \sum_{i \in \mathcal{I}_k} \{ \mathsf{n}_i \}$. The core assumptions in the URVC reduction are that the level populations within a bin are fixed by the relation $\mathsf{n}_i = n_k \, \mathsf{a}_i / a_k$ and that each bin possesses an internal energy defined as the weighted average over the energies of its constituting rovibrational levels, i.e.: $E_k = \sum_{i \in \mathcal{I}_k} \{ \mathsf{a}_i \, \mathsf{E}_i \} / a_k$. Here, the overall degeneracy of each bin is the sum over degeneracies of all rovibrational levels belonging to it: $a_k = \sum_{i \in \mathcal{I}_k} \{ \mathsf{a}_i \}$. 

The set $\mathcal{K}_\mathrm{N_2}$ 
contains the indices pointing to every one of the bins. For simplicity, it is assumed that atomic nitrogen only occupies a single internal energy state. The full set of $\mathcal{N}_\mathrm{s} = 1 + \mathcal{N}_\mathrm{bins}$ (pseudo)-species in the mixture then becomes $S = \left\lbrace \mathrm{N}, \mathrm{N_2} \left( k \right) \, \forall \, (k \in \mathcal{K}_\mathrm{N_2}) \right\rbrace$. This reduction effectively replaces the highly resolved representation of molecular nitrogen's thermodynamic state provided by the full set of level populations $\mathsf{n}_i, (i \in \mathcal{I}_\mathrm{N_2})$ with a similar, but lower-resolution one that only relies on the bin populations $n_k, (k \in \mathcal{K}_\mathrm{N_2})$. By applying the URVC binning approach, the level-specific reaction rate/cross section data from the Ames database are condensed into bin-resolved rate coefficients/cross sections, first for inelastic collisions between molecular and atomic nitrogen:
\begin{equation}
 \mathrm{N_2} \left( k \right) + \mathrm{N} \underset{k_{k \rightarrow l}^{E \mathrm{b}}}{\overset{k_{k \rightarrow l}^{E \mathrm{f}}}{\rightleftharpoons}} \mathrm{N_2} \left( l \right) + \mathrm{N} \quad \begin{array}{l} k, l \in \mathcal{K}_\mathrm{N_2}. \\ (k < l) \end{array}                                                                                                                                                                                                     \label{eq:n3_excitation}
\end{equation}

Here we have labeled the \emph{forward} rate coefficient for the transition between molecules populating bins $\mathrm{N_2}(k)$ and $\mathrm{N_2}(l)$ as $k_{k \rightarrow l}^{E \mathrm{f}}$ (i.e. when Eq.~(\ref{eq:n3_excitation}) is read from left to right), whereas the \emph{backward} rate coefficient in the opposite sense is labeled as $k_{k \rightarrow l}^{E \mathrm{b}}$. Second, we have dissociation/recombination of an $\mathrm{N_2} \left( k \right)$-molecule by collision with an $\mathrm{N}$-atom:
\begin{equation}
 \mathrm{N_2} \left( k \right) + \mathrm{N} \underset{k_{k}^{D \mathrm{b}}}{\overset{k_{k}^{D \mathrm{f}}}{\rightleftharpoons}} 3 \, \mathrm{N}, \qquad k \in \mathcal{K}_\mathrm{N_2} \label{eq:n3_dissociation}
\end{equation}
where we have labeled the rate coefficients for dissociation $k_{k}^{D \mathrm{f}}$ and $k_{k}^{D \mathrm{b}}$ recombination respectively. Third, $\mathrm{N_2} \left( k \right) + \mathrm{N}$-collisions in which no transition to another bin occurs, are referred to as ``intra-bin scattering''. Such processes are the equivalent of elastic collisions in our framework, since no internal energy is exchanged.  We make the rather strong assumption that, after lumping together the set of rovibrational levels into bins, detailed information about the rovibrational population distributions within each bin is irretrievably lost. This means that only the coarse-grain thermodynamic state represented by the bin populations can be tracked by the governing equations and one should not expect to retrieve any microscopically-resolved information (i.e. rovibrational populations) from the solutions to these equations. This simplification is valuable nonetheless, because it allows us to derive the governing equations at the hydrodynamic scale (i.e. Navier-Stokes) from the corresponding kinetic-scale equations (i.e. Boltzmann) in a fully consistent manner through application the Chapman-Enskog method. 
Finally, note that no QCT data equivalent to the N3 database was available for $\mathrm{N_2}$-$\mathrm{N_2}$, or $\mathrm{N}$-$\mathrm{N}$ collisions, and the corresponding cross sections have been replaced with very simple ones only accounting for elastic scattering. However, this does not constitute a problem for the purposes of our comparisons, as long as the simplification is done in a consistent manner when evaluating the collision terms of the Boltzmann equation and when calculating the dissipative fluxes and chemical reaction rates in the Navier-Stokes equations.



\section{Kinetic description: Boltzmann equation for coarse-grain model} \label{sec:boltzmann_equation}

At the kinetic scale, the evolution of the gas mixture is governed by a system of Boltzmann equations:
\begin{equation}
 \mathscr{D}_i \left( f_i \right) = \mathcal{J}_i (f) + \mathcal{C}_i (f), \qquad i \in S. \label{eq:generalized_boltzmann_equations}
\end{equation}

Here, $f_i = f_i \left( x, \boldsymbol{c}_i, t \right)$ are the velocity distributions of the N-atoms and the $\mathrm{N_2}(k)$-molecules populating each one of the discrete internal states $k \in \mathcal{K}_\mathrm{N_2}$. The distributions depend on position $\boldsymbol{x}$ in physical space, particle velocity $\boldsymbol{c}_i$ and time $t$. The term $\mathscr{D}_i \left( f_i \right) = \partial f_i / \partial t + \boldsymbol{c}_i \cdot \nabla_{\boldsymbol{x}} \, f_i$ on the left hand side of Eq.~(\ref{eq:generalized_boltzmann_equations}) is the streaming operator. It accounts for local time evolution and advection of the $\mathrm{N_2}(k)$- and $\mathrm{N}$-velocity distributions in phase space. Any influence of external forces (e.g. gravitational potential) has been neglected in Eq.~(\ref{eq:generalized_boltzmann_equations}). 

The terms on the right hand side are the collision operators. Together they account for any changes in the velocity distributions due to collisions between the mixture species. The precise mathematical form of these operators depends on the collision types considered. In the present work we take into account the processes listed in Table~\ref{tab:collisional_processes}. There the collision types have been sub-divided into so-called \emph{fast} and \emph{slow} processes, based on their relative time scales. The \emph{fast} scattering processes are responsible for driving the mixture toward a Maxwell-Boltzmann distribution at a common kinetic temperature $T$ (i.e. thermalization) and for diffusive transport phenomena, whereas the \emph{slow} processes can be either excitation/deexcitation reactions (responsible for relaxation of internal energy) and molecular dissociation-recombination reactions. The \emph{slow} processes typically involve some energy threshold and any individual collision is far less likely to produce a significant change in the colliding particles' states than the \emph{fast} collision types. This is reflected in the relative sizes of the associated cross sections. The fast processes possess differential cross sections typically orders of magnitude greater than the slow ones, i.e. $\sigma^\mathrm{slow} \ll \sigma^\mathrm{fast}$. The sub-division into fast and slow processes is of little concern when Eq.~(\ref{eq:generalized_boltzmann_equations}) is solved directly, e.g. by means of the DSMC method. However, as discussed in Sec.~\ref{sec:macroscopic_balance_equation}, the associated difference in time scales is exploited to derive the corresponding governing equations at the hydrodynamic scale.


\begin{table}
 \centering
 \caption{Collision types being modeled, separated into \emph{fast} and \emph{slow} processes}
 \label{tab:collisional_processes}
 
 \begin{tabular}{l l l}
  \multicolumn{3}{l}{\textbf{Fast collision processes}} \\  \hline
  \\[-0.5em]
  N-N elastic & $\mathrm{N} \left( \boldsymbol{c}_1 \right) + \mathrm{N} \left( \boldsymbol{c}_2 \right) \rightleftharpoons$ & \\[0.1em]
  scattering & $\qquad \qquad \qquad \quad \mathrm{N} \left( \boldsymbol{c}_1^\prime \right) + \mathrm{N} \left( \boldsymbol{c}_2^\prime \right)$ & \\[0.5em]
  $\mathrm{N_2} (k)$-N & $\mathrm{N_2} \left( \boldsymbol{c}_1, E_k \right) + \mathrm{N} \left( \boldsymbol{c}_2 \right) \rightleftharpoons$ & \multirow{2}{*}{$k \in \mathcal{K}_\mathrm{N_2}$} \\[0.1em]
  intra-bin & $\qquad \qquad \quad \mathrm{N_2} \left( \boldsymbol{c}_1^\prime, E_k \right) + \mathrm{N} \left( \boldsymbol{c}_2^\prime \right)$ & \\
  scattering & & \\[0.5em]
  $\mathrm{N_2} (k)$-$\mathrm{N_2} (l)$ & $\mathrm{N_2} \left( \boldsymbol{c}_1, E_k \right) + \mathrm{N_2} \left( \boldsymbol{c}_2, E_l \right) \rightleftharpoons \qquad \quad$ & \multirow{2}{*}{$k,l \in \mathcal{K}_\mathrm{N_2}$} \\[0.1em] 
  intra-bin & $\qquad \quad \mathrm{N_2} \left( \boldsymbol{c}_1^\prime, E_k \right) +  \mathrm{N_2} \left( \boldsymbol{c}_2^\prime, E_l \right)$ & \\
  scattering & & \\
   & & \\
  \multicolumn{3}{l}{\textbf{Slow collision processes}} \\ \hline
   & & \\[-0.5em]
  $\mathrm{N_2} (k)$-$\mathrm{N}$ & $\mathrm{N_2} \left( \boldsymbol{c}_1, E_k \right) + \mathrm{N} \left( \boldsymbol{c}_2 \right) \rightleftharpoons$ &  $k,l \in \mathcal{K}_\mathrm{N_2}$ \\[0.1em]
  de/excitation & $\qquad \qquad \quad \mathrm{N_2} \left( \boldsymbol{c}_1^\prime, E_l \right) +  \mathrm{N} \left( \boldsymbol{c}_2^\prime \right)$ & $\left( k < l \right)$ \\[0.5em]
  $\mathrm{N_2} (k)$-$\mathrm{N}$ & $\mathrm{N_2} \left( \boldsymbol{c}_1, E_k \right) + \mathrm{N} \left( \boldsymbol{c}_2 \right) \rightleftharpoons$ & \multirow{2}{*}{$k \in \mathcal{K}_\mathrm{N_2}$} \\[0.1em]
   dissociation- & $\qquad \quad \mathrm{N} \left( \boldsymbol{c}_3 \right) +  \mathrm{N} \left( \boldsymbol{c}_4 \right) +  \mathrm{N} \left( \boldsymbol{c}_5 \right)$ & \\
   recombination & &
 \end{tabular}
\end{table}

\subsection{Fast collision operators} \label{sec:fast_collision_operator}


The fast collision operator in Eq.~(\ref{eq:generalized_boltzmann_equations}) corresponds to the sum $\mathcal{J}_i (f) = \sum_{j \in S} \{ \mathcal{J}_{ij} (f_i, f_j) \}$. The partial collision operators:
\begin{equation}
 \begin{split}
  \mathcal{J}_{ij} (f_i, f_j) = \int\limits_{\mathcal{R}^3} \int\limits_{\mathcal{S}^2} \Bigl( f_i^\prime \, f_j^\prime  - f_i \, f_j \Bigr) g \, \sigma_{ij} \, \mathrm{d} \boldsymbol{\omega} \, \mathrm{d}\boldsymbol{c}_j, & \\
  (i,j \in S), &
 \end{split} \label{eq:fast_collision_operator_cross_section}
\end{equation}
all possess the same structure for the fast processes listed in Table~\ref{tab:collisional_processes}. The integral in Eq.~(\ref{eq:fast_collision_operator_cross_section}) is short notation for a three-fold integral over velocity space, plus a surface integral over the unit sphere. We take the dependence of $f_i$ on $\boldsymbol{x}$, $\boldsymbol{c}_i$ and $t$ to be implicit. The variables $f_i^\prime, f_j^\prime$ represent the velocity distributions of pseudo-species $i$ and $j$ evaluated at the ``post-collision'' particle velocities $\boldsymbol{c}_i^\prime$ and $\boldsymbol{c}_j^\prime$ respectively (i.e. the right-hand side of the collision as written in Table~\ref{tab:collisional_processes}). Conversely, the unprimed $f_i$ represent the distribution evaluated at the ``pre-collision'' particle velocities (left-hand side) in the same table.

The collision operator in Eq.~(\ref{eq:fast_collision_operator_cross_section}) is made up of two competing terms: one involving the product $f_i \, f_j$, which accounts for depletion (negative sign) of $f_i$ due to collisions in the forward sense, and another one involving $f_i^\prime \, f_j^\prime$ accounts for simultaneous replenishment (positive sign) by inverse collisions. The term in parentheses is multiplied in Eq.~(\ref{eq:fast_collision_operator_cross_section}) by the magnitude of the pre-collision relative velocity $g = \left| \boldsymbol{c}_i - \boldsymbol{c}_j \right|$ and the differential scattering cross section $\sigma_{ij} = \sigma_{ij} \left( g, \boldsymbol{\omega} \right)$. 

The differential cross section may in general depend both on $g$ and on the orientation of the post-collision velocity $\boldsymbol{\omega} = \left( \boldsymbol{c}_i^\prime - \boldsymbol{c}_j^\prime \right) / \left| \boldsymbol{c}_i^\prime - \boldsymbol{c}_j^\prime \right|$. For the fast processes reversibility is enforced at the coarse-grain level and we postulate that the cross sections at both ``ends'' of the collision must verify the relation:
\begin{equation}
 \sigma_{ij} \left( g, \boldsymbol{\omega} \right) = \sigma_{ij} \left( g^\prime, \boldsymbol{\omega}^\prime \right), \qquad (i,j \in S), \label{eq:inverse_collisions}
\end{equation}
where $g^\prime = | \boldsymbol{c}_i^\prime - \boldsymbol{c}_j^\prime |$  and $\boldsymbol{\omega}^\prime = \left( \boldsymbol{c}_i - \boldsymbol{c}_j \right) / \left| \boldsymbol{c}_i - \boldsymbol{c}_j \right|$. This is what allows us to combine the contributions of depleting and replenishing collisions in Eq.~(\ref{eq:fast_collision_operator_cross_section}) into a single integral. Strictly speaking, Eq.~(\ref{eq:inverse_collisions}) will only hold for elastic collisions, i.e. those where $g = g^\prime$ and no change in internal energy states occurs. This is the case for the N-N collisions at the top of Table~\ref{tab:collisional_processes} and also for the other two fast processes we have defined. Since $\mathrm{N_2}(k)$-$\mathrm{N}$ and $\mathrm{N_2}(k)$-$\mathrm{N_2}(l)$ intra-bin scattering comprises all possible transitions between rovibrational levels within a given bin, they are not true elastic collisions. However, this energy is only exchanged within the bin, so we effectively treat them as if they were elastic collisions and in our coarse-grain model Eq.~(\ref{eq:inverse_collisions}) is assumed to hold true for all fast collision types.

Although the differential cross section $\sigma_{ij}$ may in general depend on both $g$ and $\boldsymbol{\omega}$, for the calculations discussed in Sec.~\ref{sec:normal_shock_bins}, we will neglect their dependence on the latter. This allows us to replace the differential cross sections in Eq.~(\ref{eq:fast_collision_operator_cross_section}) with their integral counterparts\footnote{In this work we refer to them as ``integral'' instead of ``total'' cross sections, because in our naming convention~\cite{torres18b} we reserve the latter to mean the sum over elastic, inelastic and reactive cross sections of a given collision pair.} $\sigma_{ij}^\mathrm{I}(g) = \int_{\mathcal{S}^2} \sigma_{ij} (g, \boldsymbol{\omega}) \, \mathrm{d} \boldsymbol{\omega}$ and employ the variable hard sphere (VHS) model~\cite{bird80a} for isotropic scattering in our DSMC calculations. As discussed in App.~\ref{app:collision_integrals}, the choice of scattering model has a direct effect on the transport properties of the corresponding Navier-Stokes calculations. We should note that employing transport coefficients based on the VHS model in CFD calculations of viscous flows is rather unusual, since much more accurate methods are available~\cite{capitelli00b, wright05a}. In fact, several researchers have gone the opposite route~\cite{kim08c, stephani12a, liechty19a} and ``calibrated'' the VHS, or similar cross sections in their DSMC codes with the state-of-the art transport collision integrals. In the present work, we base our transport properties on collision integrals derived from the VHS model (see App.~\ref{app:collision_integrals}) to ensure consistency with our DSMC calculations, thus making the comparisons in Sec.~\ref{sec:normal_shock_bins_navier_stokes_vs_dsmc} more straightforward.


\subsection{Slow collision operators for $\boldsymbol{\mathrm{N}}$ and $\boldsymbol{\mathrm{N_2}(k)}$} \label{sec:slow_collision_operators}

The slow collision operators account for all types of \emph{reactive} collisions in the broader sense of our coarse-grain state-to-state description. The general mathematical form of reactive collision terms has been derived in Sec.~4.2.5 of Giovangigli~\cite{giovangigli99a} and here we merely write down the particular cases applicable to the slow processes listed in Table~\ref{tab:collisional_processes}. 

Operator $\mathcal{C}_k (f)$ appears in all rows of Eq.~(\ref{eq:generalized_boltzmann_equations}) involving the pseudo-species $\mathrm{N_2}(k)$. It is itself composed of two separate terms, $\mathcal{C}_k (f) =\mathcal{C}_k^E (f) + \mathcal{C}_k^D (f)$. The first one accounts for the effect of excitation/deexcitation on $f_k$\footnote{Notice that we have included N-atom exchange reactions in this definition}:
\begin{equation}
 \begin{split}
  \mathcal{C}_k^E (f) = \sum_{\substack{l \in \mathcal{K}_\mathrm{N_2} \\ (l \ne k)}} \int\limits_{\mathcal{R}^3} \int\limits_{\mathcal{S}^2} \Bigl( f_l^\prime f_{\mathrm{N}}^\prime \frac{a_k}{a_{l}} - f_k \, f_{\mathrm{N}} \Bigr) g \, \sigma_{k, \mathrm{N}}^{l, \mathrm{N}} \, \mathrm{d} \boldsymbol{\omega} \, \mathrm{d} \boldsymbol{c}_\mathrm{N}, & \\
  k \in \mathcal{K}_\mathrm{N_2}. \quad  &
 \end{split} \label{eq:excitation_n2k_operator}
\end{equation}

Here, $\sigma_{k, \mathrm{N}}^{l, \mathrm{N}} = \sigma_{k, \mathrm{N}}^{l, \mathrm{N}} (g,\boldsymbol{\omega})$ is the differential cross section for the transition of an $\mathrm{N_2}(k)$+$\mathrm{N}$ pair into an $\mathrm{N_2}(l)$+$\mathrm{N}$ collision pair. The ratio of degeneracies $a_l / a_k$ corresponding to post- and pre-collision internal energy states $\mathrm{N_2}(k)$ and $\mathrm{N_2}(l)$ appears multiplying the post-collision distributions to account for detailed balance between forward (i.e. excitation) and the backward (i.e. deexcitation) reactions. For the excitation-deexcitation reaction, the detailed balance relation expressed in terms of forward and backward differential cross sections takes on the form:
\begin{equation}
 a_l \,  g^2 \, \sigma_{k, \mathrm{N}}^{l, \mathrm{N}} ( g, \boldsymbol{\omega} ) \, \mathrm{d} \boldsymbol{\omega}^\prime = a_k \, g^{\prime \, 2} \, \sigma_{l, \mathrm{N}}^{k, \mathrm{N}} ( g^\prime, \boldsymbol{\omega}^\prime ) \, \mathrm{d} \boldsymbol{\omega}, \, \begin{array}{l}k \in \mathcal{K}_\mathrm{N_2} \\ k \ne l \end{array}. \label{eq:microreversibitily_n2k_differential_xsec}
\end{equation}

Again, we will assume isotropic scattering for all such collisions and replace the differential cross section with their counterparts integrated over all deflection angles: 
\begin{equation}
 a_l \,  g^2 \, \sigma_{k \rightarrow l}^{E \mathrm{f}} \left( g \right) = a_k \,  g^{\prime \, 2} \, \sigma_{k \rightarrow l}^{E \mathrm{b}} \left( g^\prime \right), \quad k \ne l, \in \mathcal{K}_\mathrm{N_2}. \label{eq:microreversibitily_n2k_integrated_xsec}
\end{equation}

Here, $\sigma_{k \rightarrow l}^{E\mathrm{f}}(g) = \int_{\mathcal{S}^2} \sigma_{k, \mathrm{N}}^{l, \mathrm{N}} (g, \boldsymbol{\omega}) \, \mathrm{d} \boldsymbol{\omega}$ is the integrated excitation cross section evaluated at the ``pre-collision'' relative speed $g = | \boldsymbol{c}_k - \boldsymbol{c}_\mathrm{N} |$ and $\sigma_{k \rightarrow l}^{E \mathrm{b}}(g^\prime)$ represents the integrated cross section for deexcitation from bin $\mathrm{N_2}(l)$ to bin $\mathrm{N_2}(k)$ evaluated at the ``post-collision'' relative speed $g^\prime = | \boldsymbol{c}_l^\prime - \boldsymbol{c}_\mathrm{N}^\prime |$. Energy conservation implies that the relation $g^\prime = \sqrt{g^2 + 2 (E_k - E_l) / \mu_{\mathrm{N_2},\mathrm{N}}}$
must hold between pre- and post-collision pairs. Here, $\mu_{\mathrm{N_2},\mathrm{N}} = m_\mathrm{N_2} \, m_\mathrm{N} / (m_\mathrm{N_2} + m_\mathrm{N})$ is the reduced mass for the $\mathrm{N_2}$-N collision pair. Notice also that the summation in Eq.~(\ref{eq:excitation_n2k_operator}) excludes the term ($k = l$), because this corresponds $\mathrm{N_2}(k)$-$\mathrm{N}$ intra-bin scattering, which we consider belonging to the fast processes. 

The second term contributing to $\mathcal{C}_k (f)$ is due to dissociation-recombination reactions:
\begin{equation}
 \begin{split}  
  \mathcal{C}_k^D (f) = \int \Bigl( \tilde{f}_\mathrm{N} \hat{f}_\mathrm{N} \check{f}_\mathrm{N} \frac{\beta_{\mathrm{N}}^2}{\beta_k} - f_k \, f_\mathrm{N} \Bigr) \times \ldots \qquad \qquad & \\
  \times \mathcal{W}_{k, \mathrm{N}}^{3 \mathrm{N}} \, \mathrm{d} \tilde{\boldsymbol{c}}_\mathrm{N} \, \mathrm{d} \hat{\boldsymbol{c}}_\mathrm{N} \, \mathrm{d} \check{\boldsymbol{c}}_\mathrm{N} \, \mathrm{d} \boldsymbol{c}_\mathrm{N}, \quad k \in \mathcal{K}_\mathrm{N_2}. &
 \end{split} \label{eq:dissociation_n2k_operator}
\end{equation}

This expression is more complex than Eqs.~(\ref{eq:fast_collision_operator_cross_section}) and (\ref{eq:excitation_n2k_operator}), because it involves a three-body interaction (the three N atoms after dissociation). This is reflected in the triple product of ``post-collision'' distribution functions $f_\mathrm{N}$ appearing as part of the replenishing term in Eq.~(\ref{eq:dissociation_n2k_operator}). Notice that instead of being ``primed'', these three $f_\mathrm{N}$ are each identified by a unique overbar to distinguish them from one another. Equation~(\ref{eq:dissociation_n2k_operator}) now involves a 12-fold integral in velocity space. The factor $\mathcal{W}_{k,\mathrm{N}}^{3 \mathrm{N}}$ is referred to~\cite{alexeev94a, giovangigli99a} as the ``reaction probability'' for the dissociation-recombination reaction (in the forward sense), even though it has dimensions of $\mathrm{time}^8 \times \mathrm{length}^{-6}$. Unlike in Eqs.~(\ref{eq:fast_collision_operator_cross_section}) and (\ref{eq:excitation_n2k_operator}), it is not straightforward to write Eq.~(\ref{eq:dissociation_n2k_operator}) in terms of a differential, or integrated cross section. 


The factors $\beta_k$ and $\beta_\mathrm{N}$, which appear in Eq.~(\ref{eq:dissociation_n2k_operator}) multiplying the replenishing term are ``statistical weights'' of the colliding species:
\begin{equation}
 \beta_k = \frac{\mathrm{h_P^3}}{a_k \, m_\mathrm{N_2}^3}, \quad (k \in \mathcal{K}_\mathrm{N_2}) \quad \text{and} \quad \beta_\mathrm{N} = \frac{\mathrm{h_P^3}}{a_\mathrm{N} \, m_\mathrm{N}^3}, \label{eq:statistical_weights}
\end{equation}
where $\mathrm{h_P}$ is Planck's constant, $m_\mathrm{N_2}$, $m_\mathrm{N}$ are the molecular masses ($m_\mathrm{N_2} = 4.65\times 10^{-26} \, \mathrm{kg}$ for all $\mathrm{N_2}(k)$ and $m_\mathrm{N} = \frac{1}{2} m_\mathrm{N_2}$ for atomic nitrogen) and $a_k$, $a_\mathrm{N}$ again the degeneracies of pseudo-species $\mathrm{N}_2(k)$ and of N respectively. The ratio of statistical weights appears in Eq.~(\ref{eq:d3_n_operator}) to account for detailed balance between the forward (i.e. dissociation) and backward (i.e. recombination) reactions. Analogous to the case for excitation-deexcitation just discussed, the terms in Eqs.~(\ref{eq:dissociation_n2k_operator}) and (\ref{eq:d3_n_operator}) accounting for dissociation-recombination have been written exclusively in terms of the \emph{forward} probability $\mathcal{W}_{\mathrm{N}, k}^{\, 3 \mathrm{N}}$, i.e. in the \emph{left-to-right} sense as written in Table~\ref{tab:collisional_processes}). This is possible, because we have postulated the existence of a reversibility relation for this three-body interaction:
\begin{equation}
 \mathcal{W}_{k,\mathrm{N}}^{\, 3 \mathrm{N}} \, \beta_\mathrm{N}^3 = \mathcal{W}_{\, 3 \mathrm{N}}^{k,\mathrm{N}} \, \beta_k \, \beta_\mathrm{N}. \label{eq:n3_dissociation_microreversibility}
\end{equation}

The statistical weight of atomic nitrogen appears on both sides of Eq.~(\ref{eq:n3_dissociation_microreversibility}) with an exponent equal to its stoichiometric coefficient \emph{right} and \emph{left} of Eq.~(\ref{eq:n3_dissociation}), but simplifies once substituted into Eq.~(\ref{eq:dissociation_n2k_operator}). Notice also that Eq.~(\ref{eq:n3_dissociation_microreversibility}) implies that the dimensions of $\mathcal{W}_{\, 3 \mathrm{N}}^{k,\mathrm{N}}$ are now just $\mathrm{time}^{5}$.

Finally, when considering the Boltzmann equation for atomic nitrogen, $\mathcal{C}_\mathrm{N} (f)$ accounts for the effect of $\mathrm{N}+\mathrm{N_2}(k)$ dissociation-recombination on $f_\mathrm{N}$ and assumes the form:
\begin{equation}
 \begin{split}
  \mathcal{C}_\mathrm{N} (f) = \sum_{k \in \mathcal{K}_\mathrm{N_2}} \biggl\{ \int \Bigl( \bar{f}_\mathrm{N} \, \hat{f}_\mathrm{N} \, \check{f}_\mathrm{N} \frac{\beta_\mathrm{N}^2}{\beta_k} - f_\mathrm{N} \, f_k \Bigr) \times \ldots & \\
  \times \mathcal{W}_{k,\mathrm{N}}^{3 \mathrm{N}} \, \mathrm{d} \bar{\boldsymbol{c}}_\mathrm{N} \mathrm{d} \hat{\boldsymbol{c}}_\mathrm{N} \mathrm{d} \check{\boldsymbol{c}}_\mathrm{N} \mathrm{d} \boldsymbol{c}_k \ldots & \\
  - 3 \int \Bigl( f_\mathrm{N} \bar{f}_\mathrm{N} \hat{f}_\mathrm{N} \frac{\beta_\mathrm{N}^2}{\beta_k} - \check{f}_\mathrm{N} f_k \Bigr) \mathcal{W}_{k,\mathrm{N}}^{3 \mathrm{N}} \, \mathrm{d} \bar{\boldsymbol{c}}_\mathrm{N} \mathrm{d} \hat{\boldsymbol{c}}_\mathrm{N} \mathrm{d} \check{\boldsymbol{c}}_\mathrm{N} \mathrm{d} \boldsymbol{c}_k \biggr\}. & \label{eq:d3_n_operator}
 \end{split}
\end{equation}

Every element of the sum in Eq.~(\ref{eq:d3_n_operator}) is composed of two integrals. Both share the same structure as the one in Eq.~(\ref{eq:dissociation_n2k_operator}). The first one is focused on atomic nitrogen on the \emph{left} of Eq.~(\ref{eq:n3_dissociation}) and accounts for depletion of this species due to dissociation and its simultaneous replenishment due to recombination. The second integral does the same, but is focused on one of the three N-atoms on the \emph{right} of Eq.~(\ref{eq:n3_dissociation}). It accounts for depletion of any of the three N-atoms due to recombination and their simultaneous replenishment due to dissociation, hence the minus sign multiplying the integral. The factor 3 appears, because one must account cumulatively for the loss of the three nitrogen atoms on the right-hand side of Eq.~(\ref{eq:n3_dissociation}). 

Writing down Eq.~(\ref{eq:generalized_boltzmann_equations}) and the associated collision terms is a useful framework for deriving the macroscopic equations in Sec.~\ref{sec:hydrodynamic_description}. However, in this work we only solve the Boltzmann equation indirectly, by means of the particle-based DSMC method. In this approach the behavior of the collision terms has to be translated into a collision algorithm, which has been detailed previously in Ref.~\cite{torres18b}.


\subsection{Macroscopic flow variables in terms of velocity distributions} \label{sec:macroscopic_moments}

The set of kinetic equations represented by Eq.~(\ref{eq:generalized_boltzmann_equations}) can be solved (either indirectly using DSMC, or another suitable method) if well-posed initial and boundary conditions for the distribution functions of all mixture components are specified. From a mathematical viewpoint the solution is obtained once the distribution functions $f_i$ can be determined everywhere in phase space at any time of interest. However, from a practical viewpoint the solution only becomes useful after the distributions have been integrated over velocity space to yield their macroscopic moments. Here we recall the definitions of these flow field variables used in fluid dynamics in terms of moments of the distribution functions.

The mass density of every species is given by:
\begin{equation}
 \rho_i = m_i \int_{\mathcal{R}^3} f_i \, \mathrm{d} \boldsymbol{c}_i, \qquad i \in S, \label{eq:species_mass_density_moments}
\end{equation}
with individual species number densities following from $n_i = \rho_i / m_i $. Mixture number and mass densities are calculated as $n = \sum_{i \in S} \{ n_i \}$ and $\rho = \sum_{i \in S} \{ \rho_i \}$ respectively. The hydrodynamic velocity of the gas is given by:
\begin{equation}
 \boldsymbol{u} = \frac{1}{\rho} \sum_{i \in S} \left\lbrace m_i \int_{\mathcal{R}^3} \boldsymbol{c}_i \, f_i \, \mathrm{d} \boldsymbol{c}_i \right\rbrace, \label{eq:hydrodynamic_velocity_moments}
\end{equation}

Diffusion velocities of each species are given by:
\begin{equation}
 \boldsymbol{u}_i^\mathrm{d} = \frac{1}{n_i} \int_{\mathcal{R}^3} \boldsymbol{C}_i \, f_i \, \mathrm{d} \boldsymbol{C}_i, \qquad i \in S, \label{eq:diffusion_velocity_moments}
\end{equation}
where $\boldsymbol{C}_i = \boldsymbol{c}_i - \boldsymbol{u}$ represent the peculiar velocities of particles belonging to species $i \in S$. By definition, the diffusion velocities always verify the constraint $\sum_{i \in S} \{ \rho_i \, \boldsymbol{u}_i^\mathrm{d} \} = \boldsymbol{0}$. Of particular interest in Sec.~\ref{sec:normal_shock_bins_navier_stokes_vs_dsmc} is the diffusion velocity of $\mathrm{N_2}$, which is obtained as the mass-weighted average $\boldsymbol{u}_\mathrm{N_2}^\mathrm{d} = 1 / \rho_\mathrm{N_2} \sum_{k \in \mathcal{K}_\mathrm{N_2}} \{ \rho_k \, \boldsymbol{u}_k^\mathrm{d} \}$. The kinetic stress tensor is obtained as:
\begin{equation}
 \underline{\underline{\mathcal{P}}} = \sum_{i \in S} \left\lbrace m_i \int_{\mathcal{R}^3} \, \boldsymbol{C}_i \otimes \boldsymbol{C}_i \, f_i \, \mathrm{d} \boldsymbol{C}_i \right\rbrace, \label{eq:pressure_tensor_moments}
\end{equation}

The pressure tensor can be split into an isotropic and a remaining anisotropic contribution $\underline{\underline{\mathcal{P}}} = p \, \underline{\underline{I}} - \underline{\underline{\tau}}$, where $p$ is the hydrostatic pressure, $\underline{\underline{I}}$ stands for the unit tensor and $\underline{\underline{\tau}}$ is the viscous stress tensor. The hydrostatic pressure is calculated as $1/3$ of the trace of $\underline{\underline{\mathcal{P}}}$, e.g. in Cartesian coordinates $p = \frac{1}{3} \left( \mathcal{P}_{xx} + \mathcal{P}_{yy} + \mathcal{P}_{zz} \right)$. Next, we invoke the perfect gas law to introduce the translation temperature as $T = p / ( n \mathrm{k_B} )$. Since we are dealing with a dilute gas mixture, we may express the composition in terms of partial pressures $p_i = x_i \, p$, where $x_i = n_i / n$ are the species mole fractions. Alternatively, the mixture composition can be expressed in terms of mass fractions $y_i = \rho_i / \rho$. A separate temperature $T_\mathrm{int}$ can be defined for characterizing the internal energy content of $\mathrm{N_2}$, i.e. $n_\mathrm{N_2} E_\mathrm{N_2}^\mathrm{int} = \sum_{k \in \mathcal{K}_\mathrm{N_2}} \{ n_k E_k \}$. It is an implicit function of the number densities $n_k$, as explained Appendix~C of Ref.~\cite{torres18b}.

The total energy per unit volume in terms of the distribution is given by:
\begin{equation}
 \rho E = \sum_{i \in S} \left\lbrace \int_{\mathcal{R}^3} \left( \frac{1}{2} m_i \, \boldsymbol{C}_i \cdot \boldsymbol{C}_i + E_i \right) f_i \, \mathrm{d} \boldsymbol{C}_i \right\rbrace, \label{eq:total_energy_moments}
\end{equation}
where the $E_i$ represent the internal energies of each species $i \in S$. In our coarse-grained state-to-state description, they correspond to the bin-averaged energies $E_k$ for each internal state $\mathrm{N_2} (k), \, \forall \, k \in \mathcal{K}_\mathrm{N_2}$ and $E_\mathrm{N}$ to the 0-K energy of formation of atomic nitrogen. For consistency with our prior definitions~\cite{torres20a, torres18b, bellas20a}, we set $E_\mathrm{N} = D_0 / 2$, where $D_0 = 9.75 \, \mathrm{eV}$ is the heat of dissociation per $\mathrm{N_2}$-molecule from the ground rovibrational level as given by the NASA Ames N3 diatomic potential~\cite{jaffe18a}. Notice that the kinetic temperature $T$ and Eq.~(\ref{eq:total_energy_moments}) are related to one another through $\rho E = \frac{1}{2} \rho \, | \boldsymbol{u} |^2 + \frac{3}{2} \, n \,  \mathrm{k_B} T + n_\mathrm{N_2} E_\mathrm{N_2}^\mathrm{int} + n_\mathrm{N} E_\mathrm{N}$.

Finally, the mixture heat flux is the flux of kinetic and internal energy transported with every particle along each Cartesian direction:
\begin{equation}
 \boldsymbol{q} = \sum_{i \in S} \left\lbrace \int_{\mathcal{R}^3} \left( \frac{1}{2} \, m_i \, \boldsymbol{C}_i \cdot \boldsymbol{C}_i + E_i \right) \boldsymbol{C}_i \, f_i \, \mathrm{d} \boldsymbol{C}_i \right\rbrace, \label{eq:heat_flux_moments}
\end{equation}

For the exact expressions used to evaluate Eqs.~(\ref{eq:species_mass_density_moments})-(\ref{eq:heat_flux_moments}) in our DSMC calculations, refer to App.~\ref{app:macroscopic_moments}.



\section{Hydrodynamic description for coarse-grain model} \label{sec:hydrodynamic_description}

In this section we discuss the macroscopic balance equations used to model the flow at the hydrodynamic scale. They are derived from Eq.~(\ref{eq:generalized_boltzmann_equations}) by applying the Chapman-Enskog method.~\cite{giovangigli99a, ferziger72a, chapman70a} Here we give a quick overview of this procedure for our particular application.



\subsection{Chapman-Enskog method for coarse-grain model} \label{sec:chapman_enskog}

We introduce suitable reference quantities at the kinetic and macroscopic level to perform a dimensional order-of-magnitude analysis~\cite{graille09a} of Eq.~(\ref{eq:generalized_boltzmann_equations}). This allows us to re-write it in its non-dimensional form:  
\begin{equation}
 \tilde{\mathscr{D}} ( \tilde{f}_i ) = \frac{1}{\mathrm{Kn}} \left[ \tilde{\mathcal{J}}_i (\tilde{f}) + \frac{\sigma^\mathrm{slow}}{\sigma^\mathrm{fast}} \, \tilde{\mathcal{C}}_i (\tilde{f}) \right], \, i \in S, \label{eq:boltzmann_equation_nondimensional}
\end{equation}
where $\mathrm{Kn} = \lambda^0 / L^0$ is a pseudo-Knudsen number based on reference mean free path $\lambda^0$ and macroscopic length scale $L^0$. The scaling for arriving at the compressible Navier-Stokes equations is to select $\mathrm{Kn} \sim \varepsilon \ll 1$. The fast and slow processes in Eq.~(\ref{eq:boltzmann_equation_nondimensional}) are assumed to occur at time scales different enough to require separate reference cross sections and the Maxwellian reaction regime~\cite{giovangigli99a} is obtained assuming that $\sigma^\mathrm{slow} \sim \varepsilon^2 \sigma^\mathrm{fast}$. Applying this scaling is a choice, which ultimately determines the structure of the resulting hydrodynamic equations. Expressed in terms of the small parameter $\varepsilon$ and reverting back to dimensional variables for convenience, we will thus seek solutions to Eq.~(\ref{eq:generalized_boltzmann_equations}) in the continuum limit of the form:
\begin{equation}
 \mathscr{D}_i \left( f_i \right) = \frac{1}{\varepsilon} \mathcal{J}_i (f) + \varepsilon \, \mathcal{C}_i (f), \qquad (i \in S) \label{eq:boltzmann_equation_small_parameter}
\end{equation}

Performing an Enskog expansion around the zero-order velocity distributions $f_i^0$ in terms of the small parameter $\varepsilon$: $f_i = f_i^0 \, ( 1 + \varepsilon \, \phi_i + \varepsilon^2 \, \phi_i^{(2)} + \dots )$, where $\phi_i$ and $\phi_i^{(2)}$ are perturbation functions, and substituting back into Eq.~({\ref{eq:boltzmann_equation_small_parameter}}) yields:
\begin{equation}
 \begin{split}
  & \mathscr{D}_i ( f_i^0 ) + \varepsilon \, \mathscr{D}_i ( f_i^0 \phi_i ) + \dots = \frac{1}{\varepsilon} \mathcal{J}_i (f^0) - f_i^0 \mathscr{F}_i ( \phi ) \\
  & + \varepsilon \left( - f_i^0 \, \mathscr{F}_i ( \phi^{(2)} ) + \mathcal{J}_i (f^0 \phi) + \mathcal{C}_i (f^0) \right) + \dots \label{eq:boltzmann_equation_enskog}
 \end{split}
\end{equation}
where $\mathscr{F}_i ( \phi)= - \sum\limits_{j \in S} \{ \mathcal{J}_{ij} ( f_i^0 \phi_i , f_j^0 ) + \mathcal{J}_{ij} ( f_i^0, f_j^0 \phi_j ) \} / f_i^0$ is the linearized fast collision operator.

Solving Eq.~(\ref{eq:boltzmann_equation_enskog}) at order $\varepsilon^{-1}$ (corresponding to the fastest time scale) yields the equilibrium, or Maxwell-Boltzmann distribution. Here we have defined the macroscopic moments: species mass density $\rho_i = \int_{\mathcal{R}^3} m_i \, f_i^0 \, \mathrm{d} \boldsymbol{c}_i, \, (i \in S)$, mixture momentum density $\rho \boldsymbol{u} = \sum_{i \in S} \{ \int_{\mathcal{R}^3} m_i \, \boldsymbol{c}_i f_i^0 \, \mathrm{d} \boldsymbol{c}_i \}$ and total energy density $\rho E = \sum_{i \in S} \{ \int_{\mathcal{R}^3} \left( \frac{1}{2} m_i \, \boldsymbol{c}_i \cdot \boldsymbol{c}_i + E_i \right) f_i^0 \, \mathrm{d} \boldsymbol{c}_i \}$ exclusively in terms of the zero-order velocity distribution. In terms of the macroscopic flow variables it takes on the form:
\begin{equation}
 f_i^0 = \left( \frac{m_i}{2 \pi \mathrm{k_B} T} \right)^{3/2} n_i \exp \left( - \frac{m_i | \boldsymbol{c}_i - \boldsymbol{u} |^2 }{2 \mathrm{k_B} T} \right) \label{eq:maxwellian}
\end{equation}

The equality $f_i^{0 \prime} f_j^{0 \prime} = f_i^{0} f_j^{0}$ allows us now to write the linearized fast operator as: $\mathscr{F}_i ( \phi ) = \sum_{j \in S} \{ \int_{\mathcal{R}^3} f_j^0 ( \phi_i + \phi_j - \phi_i^\prime - \phi_j^\prime) g \, \sigma_{ij} \, \mathrm{d} \boldsymbol{\omega} \, \mathrm{d} \boldsymbol{c}_j \}$.

Averaging Eq.~(\ref{eq:boltzmann_equation_enskog}) at order $\varepsilon^0$ over pseudo-species mass, momentum and energy leads to the Euler equations for the non-reacting gas mixture:
\begin{align}
 \partial_t ( \rho_i ) & + \nabla_{\boldsymbol{x}} \cdot \bigl( \rho_i \boldsymbol{u} \bigr) = 0, \qquad i \in S \label{eq:euler_mass_balance} \\
 \nonumber \\[-1em]
 \partial_t ( \rho \boldsymbol{u} ) & + \nabla_{\boldsymbol{x}} \cdot \bigl( \rho \boldsymbol{u} \otimes \boldsymbol{u} + p \, \underline{\underline{I}} \bigr) = \boldsymbol{0} \label{eq:euler_momentum_balance} \\
 \nonumber \\[-1em]
 \partial_t ( \rho E ) & + \nabla_{\boldsymbol{x}} \cdot \bigl( \rho \boldsymbol{u} \bigl( E  + p/\rho \bigr) \bigr) = 0 \label{eq:euler_total_energy_balance}
\end{align}

Notice that due to the choice of constraints, the definitions of macroscopic moments in the Chapman-Enskog solution slightly differ from those introduced in Sec.~\ref{sec:macroscopic_moments} at the kinetic scale. However, out of convenience here we will use the same symbols for both definitions. Definitions for $\rho_\mathrm{N_2}$, etc. and corresponding number densities follow the same pattern as in Sec.~\ref{sec:macroscopic_moments}. Note also that, given the scaling in Eq.~(\ref{eq:boltzmann_equation_small_parameter}), the slow collision operators do not contribute to the solution at order $\varepsilon^0$, and thus no chemical source terms appear on the right hand side of Eq.~(\ref{eq:euler_mass_balance}).



\subsection{Macroscopic balance (Navier-Stokes) equations for coarse-grain system including viscous and chemical source terms}
\label{sec:macroscopic_balance_equation}

With $f_i^0$ known, we go back to solving Eq.~(\ref{eq:boltzmann_equation_enskog}) at order $\varepsilon^0$ for the first-order perturbations $\phi = (\phi_i)_{i \in S}$:
\begin{equation}
 \mathscr{F}_i ( \phi ) = \Psi_i, \qquad i \in S \label{eq:epsilon_0}
\end{equation}

Uniqueness of the solution is ensured through the constraint that the perturbations do not contribute to the macroscopic moments, i.e.: $\int_{\mathcal{R}^3} m_i \, f_i^0 \phi_i \, \mathrm{d} \boldsymbol{c}_i = 0 \, (i \in S)$, $\sum_{i \in S} \{ \int_{\mathcal{R}^3} m_i \boldsymbol{c}_i \, f_i^0 \phi_i \, \mathrm{d} \boldsymbol{c}_i \} = \boldsymbol{0}$, and $\sum_{i \in S} \{ \int_{\mathcal{R}^3} ( \frac{1}{2} m_i \, \boldsymbol{c}_i \cdot \boldsymbol{c}_i \, + E_i ) f_i^0 \phi_i \, \mathrm{d} \boldsymbol{c}_i \} = 0$.

Next, we evaluate the right hand side of Eq.~(\ref{eq:epsilon_0}) $\Psi_i = - \mathscr{D}_i ( \ln f_i^0 )$. With help of Eq.~(\ref{eq:maxwellian}), we express all resulting time derivatives of macroscopic flow variables in terms of spatial gradients by re-arranging Eqs.~(\ref{eq:euler_mass_balance})-(\ref{eq:euler_total_energy_balance}). The result is a linear combination of the transport forces~\cite{giovangigli99a}, i.e. gradients in flow velocity, species partial pressure and temperature:
\begin{equation}
 \begin{split}
  \Psi_i = - \boldsymbol{\Psi}_i^\eta : \nabla_{\boldsymbol{x}} \, \boldsymbol{u} & - \sum_{j \in S} \boldsymbol{\Psi}_i^{D_j} \cdot \boldsymbol{d}_j \\
  & - \boldsymbol{\Psi}_i^{\widehat{\lambda}} \cdot \nabla_{\boldsymbol{x}} \left( \frac{1}{\mathrm{k_B} T} \right), \quad i \in S,
 \end{split}
\end{equation}
where we have defined the driving forces for diffusion of species $j$ as: $\boldsymbol{d}_j = ( \nabla_{\boldsymbol{x}} \, p_j ) / p$. The individual contributions to $\Psi_i$ are given by:
\begin{align}
 \boldsymbol{\Psi}_i^\eta = & \frac{m_i}{\mathrm{k_B} T} \left( \boldsymbol{C}_i \otimes \boldsymbol{C}_i - \tfrac{1}{3} \boldsymbol{C}_i \cdot \boldsymbol{C}_i \, \underline{\underline{I}} \right), \quad i \in S, \\
 \boldsymbol{\Psi}_i^{D_j} = & \frac{1}{p_i} \left( \delta_{ij} - y_i \right) \, \boldsymbol{C}_i, \qquad \qquad \quad i, j \in S, \\
 \boldsymbol{\Psi}_i^{\widehat{\lambda}} = & \left( \tfrac{5}{2} \mathrm{k_B} T - \tfrac{1}{2} m_i \, \boldsymbol{C}_i \cdot \boldsymbol{C}_i \right) \boldsymbol{C}_i, \qquad i \in S 
\end{align}

It can be shown that the unique solution to Eq.~(\ref{eq:epsilon_0}) is a linear combination of the transport fluxes:
\begin{equation}
 \begin{split}
  \phi_i = - \boldsymbol{\phi_i}^\eta : \nabla_{\boldsymbol{x}} \, \boldsymbol{u} & - \sum_{j \in S} \boldsymbol{\phi}_i^{D_j} \cdot \boldsymbol{d}_j \\
  & - \boldsymbol{\phi}_i^{\widehat{\lambda}} \cdot \nabla_{\boldsymbol{x}} \left( \frac{1}{\mathrm{k_B} T} \right), \quad i \in S,
 \end{split}
\end{equation}

The tensorial functions $\boldsymbol{\phi}^\eta = ( \boldsymbol{\phi}_i^\eta )_{i \in S}$ and vectorial functions $\boldsymbol{\phi}^{D_j} = ( \boldsymbol{\phi}_i^{D_j} )_{(i,j) \in S}$ and $\boldsymbol{\phi}^{\widehat{\lambda}} = (\boldsymbol{\phi}_i^{\widehat{\lambda}})_{i \in S}$ are solutions to linearized Boltzmann equations decoupled for each driving force contribution (see Eq.~(4.6.24) of Giovangigli~\cite{giovangigli99a}) 
\begin{equation}
 \mathscr{F}_i ( \boldsymbol{\phi}^\mu ) = \Psi_i^\mu, \qquad i \in S, \label{eq:epsilon_0-mu}
\end{equation}
with the superscript $\mu\in\{\eta,D_j,(j\in S), \widehat{\lambda}\}$. Constraints are imposed as $\int_{\mathcal{R}^3} m_i \, f_i^0 \boldsymbol{\phi}_i^\mu \, \mathrm{d} \boldsymbol{c}_i = 0 \,  (i \in S)$, $\sum_{i \in S} \{ \int_{\mathcal{R}^3} m_i \boldsymbol{c}_i \, f_i^0 \boldsymbol{\phi}_i^\mu \, \mathrm{d} \boldsymbol{c}_i \} = \boldsymbol{0}$, and $\sum_{i \in S} \{ \int_{\mathcal{R}^3} ( \frac{1}{2} m_i \, \boldsymbol{c}_i \cdot \boldsymbol{c}_i \, + E_i ) f_i^0 \boldsymbol{\phi}_i^\mu \, \mathrm{d} \boldsymbol{c}_i \} = 0$. 

In the continuum, or hydrodynamic limit the complete governing equations are finally obtained by averaging Eq.~(\ref{eq:boltzmann_equation_enskog}) at order $\varepsilon^1$ over pseudo-species mass, total momentum and energy:
\begin{align}
 \partial_t ( \rho_i ) & + \nabla_{\boldsymbol{x}} \cdot \bigl( \rho_i \boldsymbol{u} + \boldsymbol{j}_i \bigr) = \omega_i, \qquad i \in S, \label{eq:species_mass_balance} \\
 \nonumber \\[-1em]
 \partial_t ( \rho \boldsymbol{u} ) & + \nabla_{\boldsymbol{x}} \cdot \bigl( \rho \boldsymbol{u} \otimes \boldsymbol{u} + p \, \underline{\underline{I}} - \underline{\underline{\tau}} \bigr) = \boldsymbol{0}, \label{eq:momentum_balance} \\
 \nonumber \\[-1em]
 \partial_t ( \rho E ) & + \nabla_{\boldsymbol{x}} \cdot \bigl( \rho \boldsymbol{u} \bigl( E  + p/\rho \bigr) - \underline{\underline{\tau}} \cdot \boldsymbol{u} + \boldsymbol{q} \bigr) = 0. \label{eq:total_energy_balance}
\end{align}

Here, Eq.~(\ref{eq:species_mass_balance}) represents the set of continuity equations for every pseudo-species $i \in S$. The structure of the chemical source terms on the right hand side is discussed in more detail in Sec.~\ref{sec:navier_stokes_chemistry_source_terms}. 
The transport fluxes for pseudo-species mass, momentum and energy appearing in Eqs.~(\ref{eq:species_mass_balance})-(\ref{eq:total_energy_balance}) are given in the Chapman-Enskog approximation by:
\begin{align}
 \boldsymbol{j}_i & = \int_{\mathcal{R}^3} m_i \, \boldsymbol{C}_i \, f_i^0 \phi_i \, \mathrm{d} \boldsymbol{C}_i, \qquad i \in S, \label{eq:diffusion_flux_perturbation} \\
 \underline{\underline{\tau}} & = - \sum_{i \in S} \left\lbrace \int_{\mathcal{R}^3} m_i \, \boldsymbol{C}_i \otimes \boldsymbol{C}_i \, f_i^0 \phi_i \, \mathrm{d} \boldsymbol{C}_i \right\rbrace, \label{eq:stress_tensor_perturbation} \\
 \boldsymbol{q} & = \sum_{i \in S} \left\lbrace \int_{\mathcal{R}^3} \left( \frac{1}{2} \boldsymbol{C}_i \cdot \boldsymbol{C}_i + E_i \right) f_i^0 \phi_i \, \mathrm{d} \boldsymbol{C}_i \right\rbrace, \label{eq:heat_flux_perturbation}
\end{align}
respectively. We discuss the manner in which these fluxes are evaluated in Sec.~\ref{sec:navier_stokes_transport}.



\subsection{Transport fluxes} \label{sec:navier_stokes_transport}

Solving the kinetic equations \eqref{eq:epsilon_0} leads to expressions for Eqs.~(\ref{eq:diffusion_flux_perturbation})-(\ref{eq:heat_flux_perturbation}) in terms of spatial gradients of flow field variables and transport coefficients.  The transport properties can be obtained through the solution of linear systems arising from Galerkin approximations (see Sec.~4.6.5 and 4.7 of Giovangigli~\cite{giovangigli99a} for details). This ultimately provides closure for the viscous terms in the Navier-Stokes equations. 

The diffusion fluxes appearing in Eq.~(\ref{eq:species_mass_balance}) are found as a solution to the system of Stefan-Maxwell equations of multi-component diffusion:
\begin{equation}
 \sum_{j \in S} \left\lbrace \Delta_{ij} \frac{\boldsymbol{j}_j}{\rho_j} \right\rbrace = - ( \boldsymbol{d}_i + \chi_i \, \nabla_{\boldsymbol{x}} \ln T), \qquad i \in S \label{eq:stefan_maxwell_eqn}  
\end{equation}
subject to the constraint $\sum_{j \in S} \{ \boldsymbol{j}_j \} = \boldsymbol{0}$ to ensure mass conservation. The Stefan-Maxwell matrix entries in Eq.~(\ref{eq:stefan_maxwell_eqn}) are:
\begin{align}
 \Delta_{ij} & = - \frac{x_i x_j}{\mathcal{D}_{ij}}, \qquad i,j \in S, \quad i \ne j, \\
 \Delta_{ii} & = \sum_{\substack{j \in S\\ i \ne j}} \frac{x_i x_j}{\mathcal{D}_{ij}}, \qquad i \in S.
\end{align}

In the absence of external force fields, all remaining driving forces for diffusion appear on the right hand side of Eq.~(\ref{eq:stefan_maxwell_eqn}). The previously introduced linearly dependent driving forces can be decomposed as $\boldsymbol{d}_i = ( \nabla_{\boldsymbol{x}} \, p_i ) / p = \nabla_{\boldsymbol{x}} \, x_i + \left( x_i - y_i \right) \nabla_{\boldsymbol{x}} \ln p$, which means they account for diffusion due to gradients of mole fraction and pressure (baro-diffusion). The remaining term in Eq.~(\ref{eq:stefan_maxwell_eqn}) represents thermo-diffusion (Soret effect), induced by temperature gradients. 
Formally, all three gradients can induce species mass transfer, but in the Navier-Stokes calculations of Sec.~\ref{sec:normal_shock_bins_navier_stokes_vs_dsmc} only mole fraction gradients were taken into account. In order to evaluate the entries of the Stefan-Maxwell matrix, one must supply the binary diffusion coefficients $\mathcal{D}_{ij} (p,T) \, \forall \, (i \ne j), (i,j \in S)$ and the thermal diffusion ratios $\chi_i$. 
The expression for $\chi_i = \chi_i \left( p_j \, \forall (j \in S), T \right)$ is given in Chapter 5 of Ref.~\cite{giovangigli99a} and in App.~\ref{app:transport_systems}. Following the structure of the matrix for the thermal conductivity transport system, it can be shown that their sign is not defined, but that $\sum_{i\in S} \{ \chi_i \}  = 0$ must hold~\cite{ern94a}. Alternatively, the diffusion fluxes can be expressed in terms of multi-component diffusion coefficients
$\boldsymbol{j}_i= -\sum_{j\in S} \{ D_{ij}(\boldsymbol{d}_j + \chi_j \, \nabla_{\boldsymbol{x}} \ln T)/\rho_i \}, \, (i \in S)$. The diffusion matrix is semi-positive definite, $D_{ij} \ge 0, i\neq j$, $D_{ii}>0$, $(i,j \in S)$, and is the pseudo-inverse of the Stefan-Maxwell matrix appearing in Eq.~(\ref{eq:stefan_maxwell_eqn}).

The viscous stress tensor $\underline{\underline{\tau}}$ appearing in Eqs.~(\ref{eq:momentum_balance}) and (\ref{eq:total_energy_balance}) takes on the form:
\begin{equation}
 \underline{\underline{\tau}} = 2 \, \eta \, \underline{\underline{\mathsf{S}}}, \label{eq:viscous_stress_tensor_ns}
\end{equation}
where $\eta$ is the mixture shear viscosity and $\underline{\underline{\mathsf{S}}} = \frac{1}{2} \left[ \nabla_{\boldsymbol{x}} \boldsymbol{u} + ( \nabla_{\boldsymbol{x}} \boldsymbol{u} )^T - \frac{2}{3} \, ( \nabla_{\boldsymbol{x}} \cdot \boldsymbol{u} ) \, \underline{\underline{I}} \right]$ is the traceless symmetric velocity gradient tensor. Notice that when compared with Eq.~(4.6.43) of Giovangigli~\cite{giovangigli99a}, Eq.~(\ref{eq:viscous_stress_tensor_ns}) lacks a reaction pressure term, since our scaling of Eq.~(\ref{eq:boltzmann_equation_small_parameter}) places us in the Maxwellian reaction regime. Furthermore, a bulk viscosity term is also missing, because it is not needed within the state-to-state description. The expression for $\eta = \eta \left( p_i \, \forall (i \in S), T \right)$ is given in Chapter 5 of Ref.~\cite{giovangigli99a} and in App.~\ref{app:transport_systems}. Following the structure of the matrix for the viscosity transport system, it can be shown that $\eta > 0$~\cite{ern94a}.

Finally, the heat flux vector in Eq.~(\ref{eq:total_energy_balance}) takes on the form:
\begin{equation}
 \boldsymbol{q} = - \lambda \, \nabla_{\boldsymbol{x}} T + \sum_{i \in S} \left\lbrace h_i \, \boldsymbol{j}_i \right\rbrace + p \, \sum_{i \in S} \left\lbrace \chi_i \, \boldsymbol{j}_i / \rho_i \right\rbrace \label{eq:heat_flux_ns}
\end{equation}

The first term on the right hand side is the contribution due to heat conduction $\boldsymbol{q}^\mathrm{cond}$. It is the product of the mixture thermal conductivity $\lambda$ and the temperature gradient. The second term $\boldsymbol{q}^\mathrm{diff}$ accounts for heat transfer by diffusion of enthalpy of each mixture component, i.e. $h_i = \frac{5}{2} \mathrm{k_B} T  + E_i$.  The expression for the thermal conductivity $\lambda = \lambda \left( p_i \, \forall (i \in S), T \right)$ is given in Chapter 5 of Ref.~\cite{giovangigli99a} and in App.~\ref{app:transport_systems}. An alternative formulation for the heat flux is to use the partial thermal conductivity $\widehat{\lambda}$ and the thermal diffusion coefficients $\theta_i, \, (i\in S)$. Both formulations are equivalent, but the one chosen here is advantageous to study the entropy production in Sec.~\ref{sec:thermodynamic_entropy_equation_derivation}. Following the structure of the matrix for the thermal conductivity transport system, it can be shown that $\lambda > 0$, provided that some conditions on the collision integral data are met~\cite{ern94a}. Note that within the state-to-state formalism there is no need to consider Eucken's correction to the thermal conductivity~\cite{ferziger72a}, because transfer of internal energy is implicitly taken into account through the chemical production rates $\omega_i, \, i \in S$. 
The third term formally accounts for heat transfer induced by concentration gradients (Dufour effect). It is the complement to the Soret effect appearing in Eq.~(\ref{eq:stefan_maxwell_eqn}). However, note that it is also being neglected in the Navier-Stokes calculations presented in Sec.~\ref{sec:normal_shock_bins_navier_stokes_vs_dsmc}. 

The necessary routines for the solution of the transport systems have been implemented in the Mutation++~\cite{scoggins20a} thermodynamic and transport library, which is tightly coupled to the Navier-Stokes flow solver used to generate the results of Sec.~\ref{sec:normal_shock_bins_navier_stokes} and \ref{sec:normal_shock_bins_navier_stokes_vs_dsmc}.



\subsection{Chemistry source terms} \label{sec:navier_stokes_chemistry_source_terms}

The terms on the right hand side of Eq.~(\ref{eq:species_mass_balance}) represent the mass production terms for atomic nitrogen and every $\mathrm{N_2}(k)$ respectively. For the latter, both excitation-deexcitation and dissociation-recombination reactions contribute to the source term: $\omega_k = \omega_k^E + \omega_k^D$. These two contributions are obtained by averaging Eqs.~(\ref{eq:excitation_n2k_operator}) and (\ref{eq:dissociation_n2k_operator}) (evaluated at the local Maxwellians $f^0$) under the constraint of pseudo-species mass conservation. This yields $\omega_k^E = m_\mathrm{N_2} \, \int \mathcal{C}_k^E (f^0) \, \mathrm{d} \boldsymbol{c}_k$ and $\omega_k^D = m_\mathrm{N_2} \, \int \mathcal{C}_k^D (f^0) \, \mathrm{d} \boldsymbol{c}_k$ respectively. Normalized with the respective molecular masses, these terms take on the following form:
\begin{equation}
 \frac{\omega_k^E}{m_\mathrm{N_2}} = \sum_{\substack{l \in \mathcal{K}_\mathrm{N_2} \\ (k \ne l)}} \Bigl\{ \Bigl( - k_{k \rightarrow l}^{E \mathrm{f}} \, n_k + k_{k \rightarrow l}^{E \mathrm{b}} n_l \Bigr) n_\mathrm{N} \Bigr\}, \quad k \in \mathcal{K}_\mathrm{N_2} \label{eq:n2k_excitation_mass_production_rate}
\end{equation}
for excitation-deexcitation and: 
\begin{equation}
 \frac{\omega_k^D}{m_\mathrm{N_2}} = \Bigl( - k_{k}^{D \mathrm{f}} \, n_k + k_{k}^{D \mathrm{b}} \, n_\mathrm{N}^2 \Bigr) n_\mathrm{N}, \qquad k \in \mathcal{K}_\mathrm{N_2} \label{eq:n2k_dissociation_mass_production_rate}
\end{equation}
for dissociation-recombination. For atomic nitrogen only the dissociation-recombination reactions contribute to the source term. Taking the moments of Eq.~(\ref{eq:d3_n_operator}) in analogous manner yields $\omega_\mathrm{N} = m_\mathrm{N} \, \int \mathcal{C}_\mathrm{N}(f^0) \, \mathrm{d} \boldsymbol{c}_\mathrm{N}$, and can be simplified to:
\begin{equation}
 \frac{\omega_\mathrm{N}}{m_\mathrm{N}} = - 2 \sum_{k \in \mathcal{K}_\mathrm{N_2}} \Bigl\{ \omega_k^D \Bigr\} \label{eq:atomic_n_mass_production_rate}
\end{equation}

In previous work~\cite{torres20a} the coarse-grain reaction cross sections $\sigma_{k \rightarrow l}^{E\mathrm{f}} (g)$ and $\sigma_k^{D\mathrm{f}} (g)$ were fitted to an analytical form consistent with Arrhenius-type expressions for the corresponding rate coefficients $k_{k \rightarrow l}^{E \mathrm{f}} (T)$ and $k_{k}^{D \mathrm{f}} (T)$ appearing in Eqs.~(\ref{eq:n2k_excitation_mass_production_rate})-(\ref{eq:atomic_n_mass_production_rate}). Special care was taken to ensure consistency between the kinetic and hydrodynamic description. This meant that the reversibility relations postulated to exist between \emph{forward} and \emph{backward} cross sections/probabilities as discussed in Sec.~\ref{sec:boltzmann_equation} have their counterparts at the hydrodynamic scale. For further context refer to Sec.~2.4.2 of Giovangigli~\cite{giovangigli99a} and in particular Remark 2.4.1 therein. The final result is that the backward rate coefficient for excitation/deexcitation processes in Eq.~(\ref{eq:n2k_excitation_mass_production_rate}) must be obtained from:
\begin{equation}
 k_{k \rightarrow l}^{E \mathrm{b}} = k_{k \rightarrow l}^{E \mathrm{f}} \, Z_k / Z_l, \qquad (k \ne l \in \mathcal{K}_\mathrm{N_2}) \label{eq:deexcitation_rate_coeff}
\end{equation}
whereas the recombination rate coefficient appearing in Eqs.~(\ref{eq:n2k_dissociation_mass_production_rate}) and (\ref{eq:atomic_n_mass_production_rate}) is obtained as: 
\begin{equation}
 k_{k}^{D \mathrm{b}} = k_{k}^{D \mathrm{f}} \, Z_k / Z_\mathrm{N}^2, \qquad (k \in \mathcal{K}_\mathrm{N_2}). \label{eq:recombination_rate_coeff}
\end{equation}

Here, the partition function per unit volume of each pseudo-species $i$ has the form: $Z_i (T) = ( 2 \pi \, m_i \mathrm{k_B} T / \mathrm{h_P^2} )^{3/2} \, a_i \exp [ - E_i / ( \mathrm{k_B} T) ]$.



\subsection{Entropy equation and sign of the chemical entropy production term} \label{sec:thermodynamic_entropy_equation_derivation}

A macroscopic balance equation for the entropy per unit volume based on thermodynamic considerations is derived in Chapter~2.6 of Giovangigli~\cite{giovangigli99a}. In the form applicable to our case it reads:
\begin{equation}
 \partial_t \left( \rho s \right) + \nabla_{\boldsymbol{x}} \cdot \Bigl( \rho \boldsymbol{u} \, s  + \boldsymbol{j}^S \Bigr) = \Upsilon, \label{eq:thermodynamic_entropy_equation}
\end{equation}
where terms on the left-hand side represent the (1) local time rate of change of entropy, (2) the advection and (3) diffusion of entropy in physical space. The term $\boldsymbol{j}^S = ( \boldsymbol{q} - \sum_{i \in S} \{ \boldsymbol{j}_i \, g_i\} ) / T$ represents the diffusive flux of entropy for the gas mixture. It contains the product of diffusion fluxes of every mixture component with their respective Gibbs free energy per unit mass: $g_i = \mathrm{k_B} T / m_i \ln \left( n_i / Z_i \right)$. 

On the right hand side of Eq.~(\ref{eq:thermodynamic_entropy_equation}) the volumetric entropy production rate can be split up into $\Upsilon = \Upsilon_\mathrm{tran} + \Upsilon_\mathrm{chem}$, i.e. entropy production due to (a) transport phenomena and (b) chemical reactions~\footnote{Recall that in the coarse-grained approach inelastic transitions between internal energy states of a molecule are also treated as chemical reactions}. General expressions for both terms have been derived by Giovangigli,~\cite{giovangigli99a} and here we recall only the terms relevant for our fluid model. The first production term can be written as:
\begin{equation}
 \begin{split}
  & \Upsilon_\mathrm{tran} = \frac{\lambda}{T^2} \, \nabla_{\boldsymbol{x}} T \cdot \nabla_{\boldsymbol{x}} T + \frac{2 \, \eta}{T} \, \underline{\underline{\mathsf{S}}} : \underline{\underline{\mathsf{S}}} \, \ldots \\
  & + \frac{p}{T} \sum_{i,j \in S} D_{ij} \, (\boldsymbol{d}_i + \chi_i \nabla_{\boldsymbol{x}} \ln T ) \cdot (\boldsymbol{d}_j + \chi_j \nabla_{\boldsymbol{x}} \ln T ). \label{eq:chemical_entropy_production_transport}
 \end{split}
\end{equation}

Given the structure of the first two terms on the right hand side of Eq.~(\ref{eq:chemical_entropy_production_transport}) and the fact that $\eta, \lambda > 0$ it can be easily seen that they must always be non-negative. The third term contains as factors $D_{ij}$ the components of the multi-component diffusion matrix. Its properties guarantee that the associated entropy production term will always remain non-negative. 
Thus, $\Upsilon_\mathrm{tran} \ge 0$ must hold for any physically realizable flow.

Now, for the particular set of reactions given by Eqs.~(\ref{eq:n3_excitation}) and (\ref{eq:n3_dissociation}), it is worthwhile to have a closer look at the entropy production due to chemical reactions: $\Upsilon_\mathrm{chem} = - ( \sum_{i \in S} \{ g_i \, \omega_i \} ) / T$. It is a function on the Gibbs free energies per unit mass and the chemical source terms appearing on the right hand side of Eq.~(\ref{eq:species_mass_balance}). 

Following the general procedure outlined by  Giovangigli~\cite{giovangigli99a}, it is possible to show that $\Upsilon_\mathrm{chem} \ge 0$ for all cases, in accordance with the second law of thermodynamics. The key to demonstrating this lies in re-writing the chemical production terms in the \emph{symmetric} form (Sec.~4.6.6 of Giovangigli~\cite{giovangigli99a}), where the rate coefficients for the excitation-deexcitation and dissociation-recombination reaction become $k_{E(k \rightarrow l)}^{s} = [ k_{k \rightarrow l}^{E \mathrm{f}} \, k_{k \rightarrow l}^{E \mathrm{b}} \, Z_k \, Z_l \, Z_\mathrm{N}^2 ]^{1/2}$ and $k_{D(k)}^{s} = [ k_{k}^{D \mathrm{f}} \, k_{k}^{D \mathrm{b}} \, Z_k \, Z_\mathrm{N}^4 ]^{1/2}$ respectively. Consistency between these production rates in symmetric form and the original notation of Eqs.~(\ref{eq:n2k_excitation_mass_production_rate})-(\ref{eq:atomic_n_mass_production_rate}) is contingent upon the elementary reactions expressed by Eqs.~(\ref{eq:n3_excitation}) and (\ref{eq:n3_dissociation}) verifying detailed balance. This, in turn, implies that the backward rate coefficients for excitation-deexcitation and dissociation-recombination must be computed according to Eqs.~(\ref{eq:deexcitation_rate_coeff}) and (\ref{eq:recombination_rate_coeff}) respectively. After some algebraic manipulation, one arrives at the final form:
\begin{equation}
 \begin{split}
  \frac{\Upsilon_\mathrm{chem}}{\mathrm{k_B}} & = \sum_{\substack{k,l \in \mathcal{K}_\mathrm{N_2} \\ (l > k)}} \biggl\{ k_{E(k \rightarrow l)}^s \ln \biggl( \frac{A}{B} \biggr) \left( A - B \right) \biggr\} \\
  & + \sum_{k \in \mathcal{K}_\mathrm{N_2}} \biggl\{ k_{D(k)}^{s} \ln \biggl( \frac{A}{C} \biggr) \left( A - C \right) \biggr\} \label{eq:chemical_entropy_production_gibbs_final}
 \end{split}
\end{equation}
for the entropy production due to chemical reactions. Here, we have defined the relations $\ln \left( A \right) = \left( g_k \, m_\mathrm{N_2} + g_\mathrm{N} \, m_\mathrm{N} \right) / \mathrm{k_B} T$, $\ln \left( B \right) = \left( \bar{g}_l \, m_\mathrm{N_2} + g_\mathrm{N} \, m_\mathrm{N} \right) / \mathrm{k_B} T$ and $\ln \left( C \right) = \left( 3 \, g_\mathrm{N} \, m_\mathrm{N} \right) / \mathrm{k_B} T$.

Regardless of the signs of $A$,$B$ and $C$, all the elements of the sums in Eq.~(\ref{eq:chemical_entropy_production_gibbs_final}) must be non-negative. Since the rate coefficients themselves are always non-negative, this means that $\Upsilon_\mathrm{chem} \ge 0$ in all instances. Satisfying this condition for all terms contributing to $\Upsilon$ in Eq.~(\ref{eq:thermodynamic_entropy_equation}) is crucial for constructing a fluid model fully consistent with the second law of thermodynamics.



\section{Internal energy excitation and dissociation across normal shock wave} \label{sec:normal_shock_bins}

In this section we present simulation results for a steady, normal shock wave. We apply three distinct numerical approaches and compare them in terms of their degree of physical fidelity. In order to formulate a discretized version of the macroscopic balance equations amenable to numerical solution, we re-write Eqs.~(\ref{eq:species_mass_balance})-(\ref{eq:total_energy_balance}) for the unsteady, one-dimensional case in the form:
\begin{equation}
 \frac{\partial \mathbf{U}}{\partial t} + \frac{\partial \mathbf{F}}{\partial x} - \frac{\partial \mathbf{F}^\mathrm{d}}{\partial x} = \mathbf{S}, \label{eq:navier_stokes_conservative}
\end{equation}
where $\mathbf{U} = \left( \rho_i \; (i \in S), \, \rho u_x, \, \rho E \right)^T$ is the vector of conserved variables, $\mathbf{F} = \left( \rho_i u_x \; (i \in S), \, \rho u_x^2 + p, \, \rho u_x \, ( E + p / \rho) \right)^T$ is the inviscid flux vector and $\mathbf{F}^\mathrm{d} = \left( j_{x, i} \; (i \in S), \, -\tau_{xx}, \, -\tau_{xx} u_x + q_x \right)^T$ is the vector of diffusive fluxes. On the right-hand side of Eq.~(\ref{eq:navier_stokes_conservative}), $\mathbf{S} = \left( \omega_i \, (i \in S), \, 0, \, 0 \right)^T$ represents the source term vector. Further manipulation of Eq.~(\ref{eq:navier_stokes_conservative}) yields the appropriate discretized equations solved numerically in Sec.~\ref{sec:normal_shock_bins_inviscid}, \ref{sec:normal_shock_bins_navier_stokes} and \ref{sec:normal_shock_bins_navier_stokes_vs_dsmc}.

In Sec.~\ref{sec:normal_shock_bins_inviscid} we first obtain steady-state solutions to Eq.~(\ref{eq:navier_stokes_conservative}) in the inviscid limit using the master equation approach. When coupled with the Rankine-Hugoniot jump relations for a chemically frozen free-stream, this approach is fully equivalent to solving the steady-state Euler equations across the normal shock. The resulting flow fields are influenced primarily by the detailed chemistry terms on the right hand side. Out of all our calculations, these are the cheapest from a computational standpoint. Thus, they can be easily repeated for a range of bin resolutions and allow us to study the convergence of the coarse-grain chemistry model toward the full rovibrational state-to-state solution. The inviscid post-shock profiles are obtained by marching forward in space from the initial discontinuity, so the calculations can be carried out without a-priori knowledge of the extent of the post-shock relaxation region. However, this information becomes crucial to set up the computational domains for the Finite Volume (FV) calculations described in Sec.~\ref{sec:normal_shock_bins_navier_stokes}. We first obtain FV solutions to the Euler equations, including the chemical source terms. We verify that the FV Euler solution agrees with those of Sec.~\ref{sec:normal_shock_bins_inviscid} and make sure that we limit numerical dissipation to the minimum necessary to capture the shock within a few cells. This makes us confident that any additional diffusive phenomena observed in the full Navier-Stokes flow fields are entirely caused by the viscous flux terms in Eq.~(\ref{eq:navier_stokes_conservative}). Finally, in Sec.~\ref{sec:normal_shock_bins_dsmc} we describe the process of obtaining the normal shock flow fields using DSMC and them compare them to the equivalent Navier-Stokes FV flow fields in Sec.~\ref{sec:normal_shock_bins_navier_stokes_vs_dsmc}. 




\subsection{Hydrodynamic inviscid solution based on Finite difference ODE method} \label{sec:normal_shock_bins_inviscid}

To obtain a first estimate of the thermo-chemical non-equilibrium region, we simulate the normal shock following a steady-state, one-dimensional inviscid approach. When such conditions are assumed, the time derivatives $\partial \mathbf{U} / \partial t$ and the diffusive transport fluxes $\mathbf{F}^\mathrm{d}$ in Eq.~(\ref{eq:navier_stokes_conservative}) all vanish. This makes it possible to re-cast the original set of equations into an ordinary differential equation (ODE) system:
\begin{equation}
 \frac{\mathrm{d} \mathbf{P}}{\mathrm{d} x} = \mathbf{Q} \left( \mathbf{P} \right), \label{eq:shocking_ode_system}
\end{equation}
where the solution vector is now given by $\mathbf{P} = \left( y_i \, (i \in S), u, T \right)^T$ and the $y_i$ are the mass fractions of atomic nitrogen plus each internal energy bin of $\mathrm{N_2}$. The right hand side of Eq.~(\ref{eq:shocking_ode_system}) is given by $\mathbf{Q} \left( \mathbf{P} \right) = ( \partial \mathbf{F} / \partial \mathbf{P} )^{-1} \mathbf{S}$. The system can be solved as an initial value problem~\cite{gear71a} marching along the $x$-axis under the condition that a suitable initial state $\mathbf{P} (x = 0)$  is provided. The solver used is equipped with mesh adaptation techniques implemented in the LSODE package~\cite{radhakrishnan93a} and the code used in this study has been applied to similar problems in the past~\cite{munafo14c, munafo14d, panesi14a}. 

\begin{figure}

  \begin{minipage}{0.8\columnwidth}
   \subfloat[Mixture density and partial density of $\mathrm{N_2}$ $\mathrm{[kg/m^3] \times 10^3}$]{\includegraphics[width=\columnwidth]{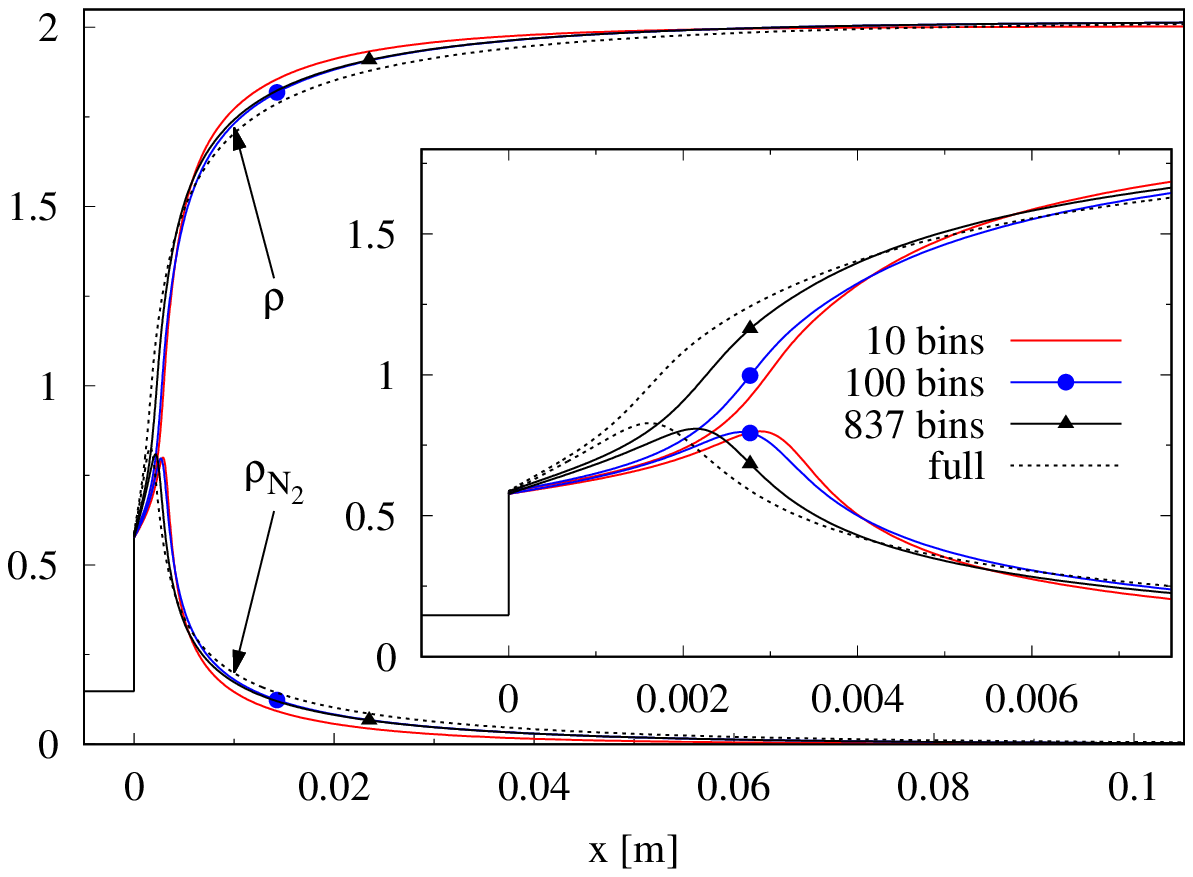}\label{fig:shocking_F90_densities_10kmsec_linear}}
  \end{minipage}

  \begin{minipage}{0.8\columnwidth}
   \subfloat[Mixture kinetic temperature $T$ and internal temperature of $\mathrm{N_2}$ $T_\mathrm{int} \mathrm{[K]}$ behind the shock]{\includegraphics[width=\columnwidth]{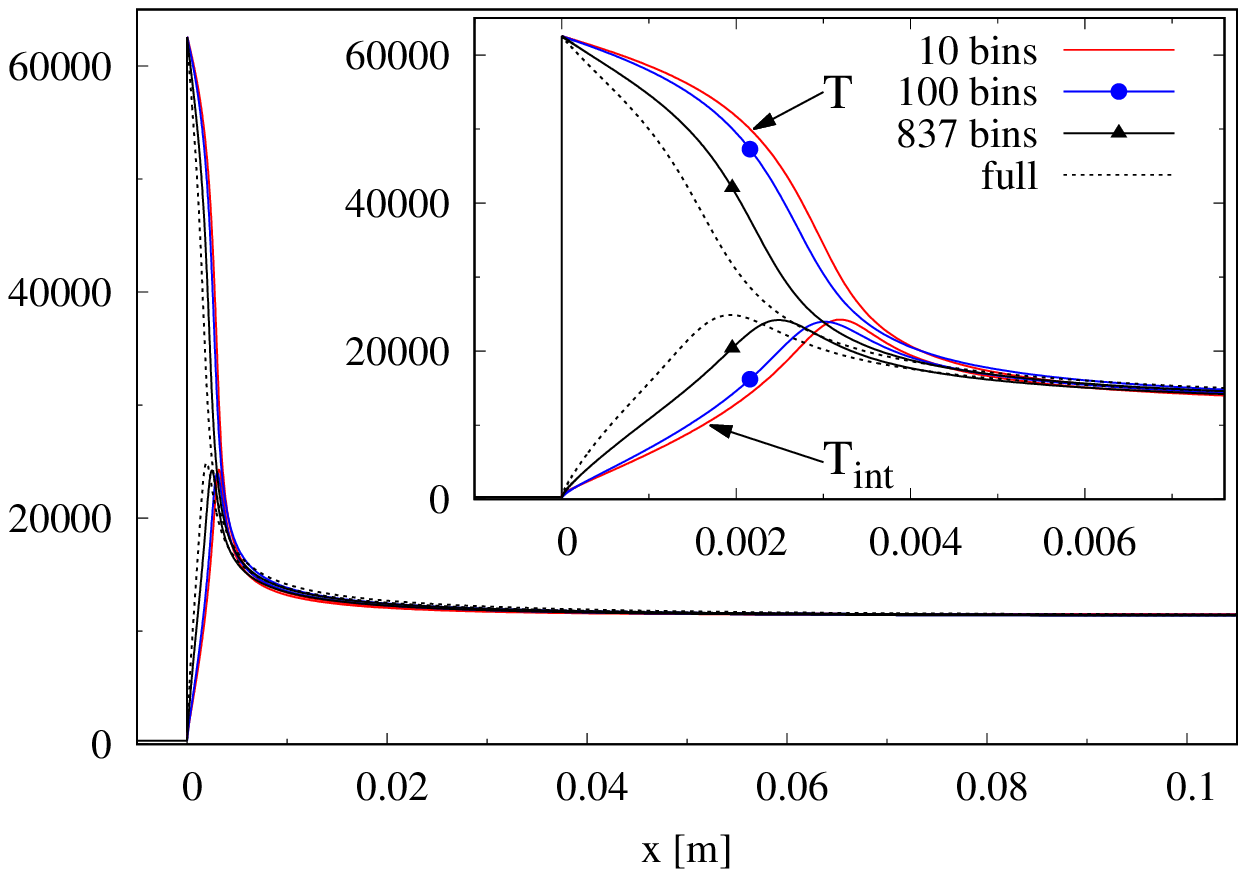}\label{fig:shocking_F90_temperatures_10kmsec_linear}}
  \end{minipage}
  
  \caption{Inviscid shock at $u_1 = 10 \, \mathrm{km \cdot s^{-1}}$.}
 \label{fig:shocking_F90_10kmsec_linear}
\end{figure}

\begin{figure}

 \begin{minipage}{0.8\columnwidth}
  \subfloat[Mixture density and partial density of $\mathrm{N_2}$ $\mathrm{[kg/m^3] \times 10^3}$]{\includegraphics[width=\columnwidth]{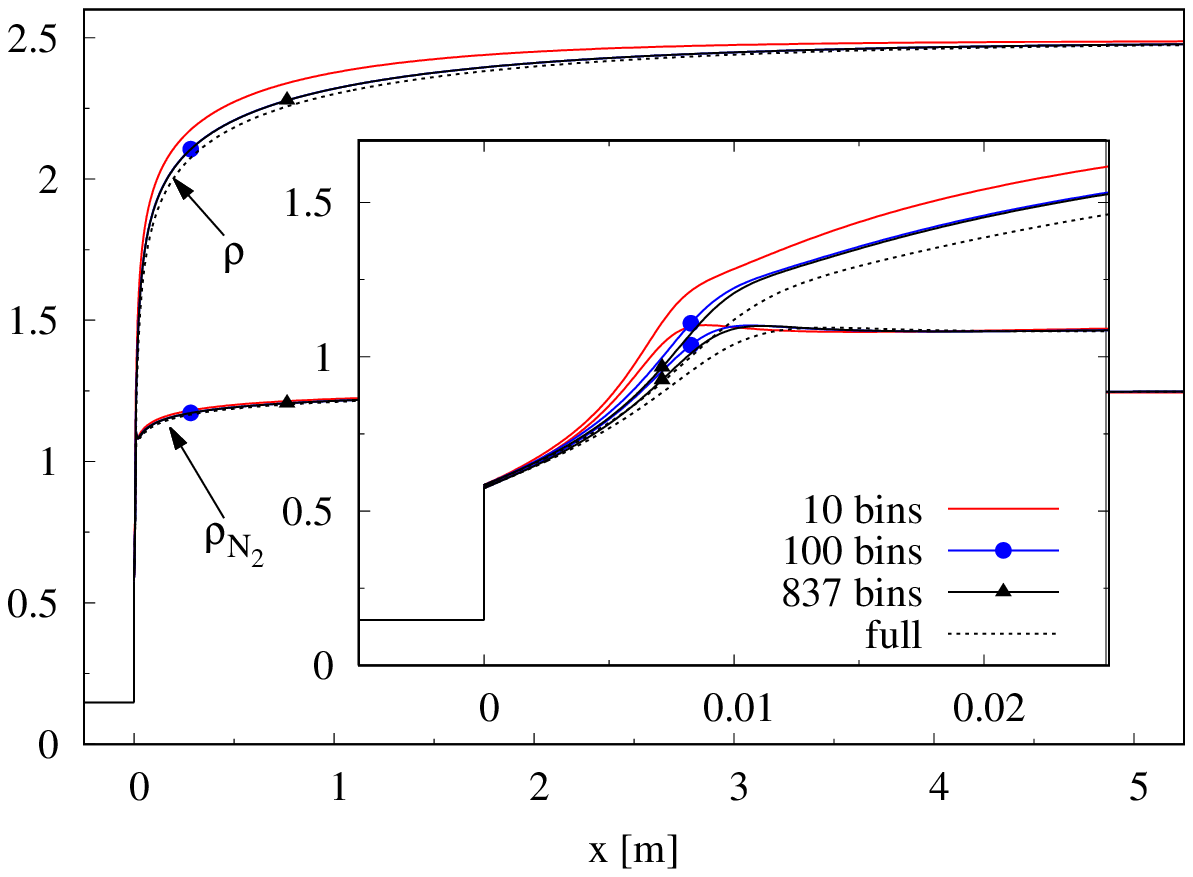}\label{fig:shocking_F90_densities_7kmsec_linear}}
 \end{minipage}
 
 \begin{minipage}{0.8\columnwidth}
  \subfloat[Mixture kinetic temperature $T$ and internal temperature of $\mathrm{N_2}$ $T_\mathrm{int} \mathrm{[K]}$ behind the shock]{\includegraphics[width=\columnwidth]{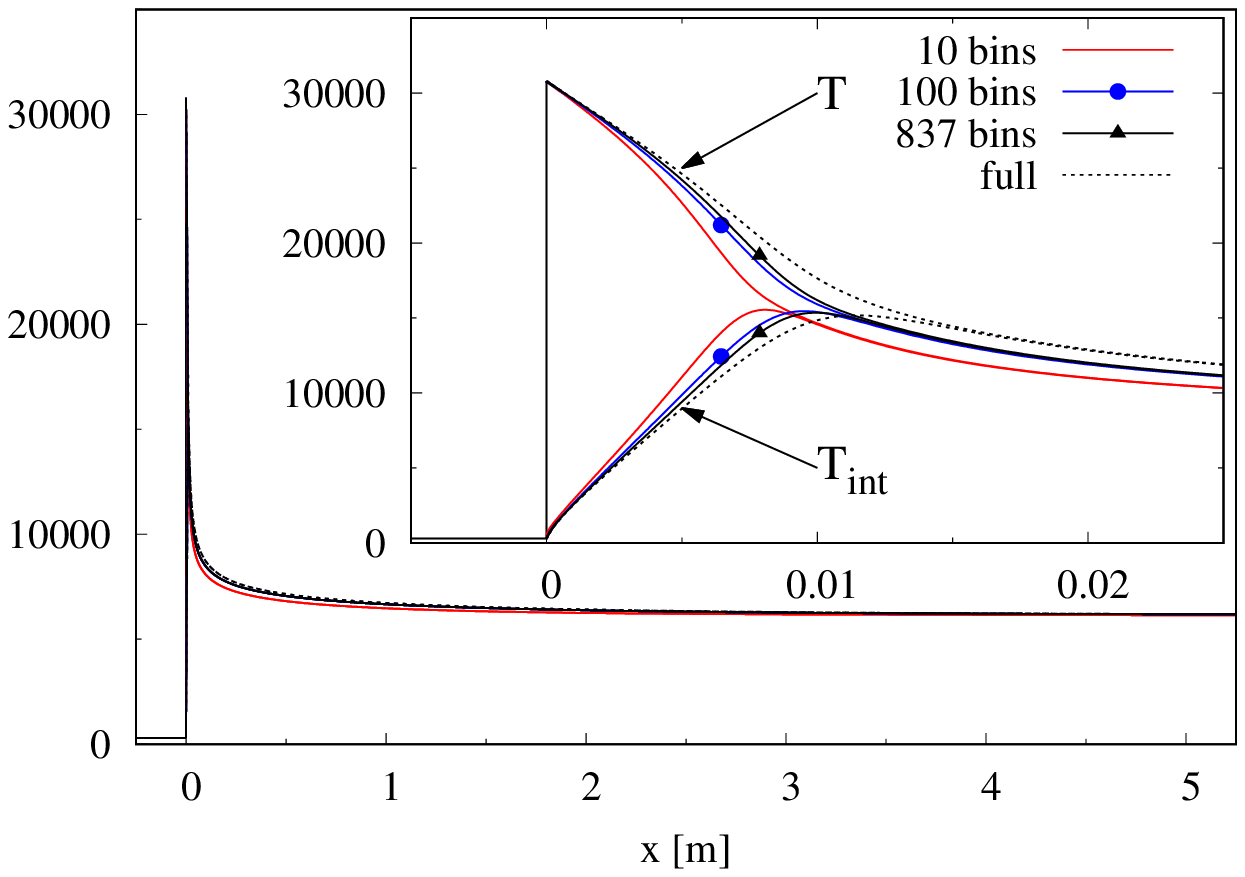}\label{fig:shocking_F90_temperatures_7kmsec_linear}}
 \end{minipage}
  
 \caption{Inviscid shock at $u_1 = 7 \, \mathrm{km \cdot s^{-1}}$.}
 \label{fig:shocking_F90_7kmsec_linear}
\end{figure}

Two different supersonic free-stream conditions are considered. For the \emph{high-speed} case, we impose a free-stream velocity of $u_1 = 10 \, \mathrm{km \cdot s^{-1}}$, while for the \emph{low-speed} case we use $u_1 = 7 \, \mathrm{km \cdot s^{-1}}$. All other parameters, such as free-stream temperature, pressure and composition are the same for both cases. The higher-speed conditions are listed in Table~\ref{tab:normal_shock_bins_10kmsec_bc}, where they are labeled as (1) pre-shock. In the ODE approach the shock is not captured by the numerical method. It is instead replaced by a sudden jump in flow conditions at $x = 0$, which only affects the translational mode. Therefore, the analytical Rankine-Hugoniot jump relations with specific heat ratio $\gamma = 5/3$ are used to predict the non-equilibrium post-shock state (state (1a) in Table~\ref{tab:normal_shock_bins_10kmsec_bc}). While the kinetic temperature reaches $T = 62550 \, \mathrm{K}$ behind the discontinuity, the internal temperature and composition remain \emph{frozen} at the free-stream values. Thus, the initial bin mass fractions $y_k, (k \in \mathcal{K}_\mathrm{N_2})$ in Eq.~(\ref{eq:shocking_ode_system}) are made to follow a Boltzmann distribution at $T_\mathrm{int} = 300 \, \mathrm{K}$. The ODE algorithm then marches along $x$ starting from state (1a). Notice that the free stream contains a non-zero amount of atomic nitrogen, even though the gas in equilibrium at $300 \, \mathrm{K}$ should only consist of $\mathrm{N_2}$-molecules. We add a small amount of N to the free-stream gas to trigger internal energy exchange and dissociation processes, since only reactions induced by N-$\mathrm{N_2}$ collisions are taken into account by the chemical source terms of Eqs.~(\ref{eq:n2k_excitation_mass_production_rate})-(\ref{eq:atomic_n_mass_production_rate}). The pre- and post-shock conditions for the low-speed case are listed in Table~\ref{tab:normal_shock_bins_7kmsec_bc}. Due to the lower post-shock temperature, the gas does not dissociate to the same degree as at the high-speed conditions and about 1/3 of the post-shock gas remains in the form of molecular nitrogen. We carry out four separate simulations at the high- and low-speed conditions respectively. The first simulations provide reference solutions with the original Ames database. These results are labeled ``full'' in Tables~\ref{tab:normal_shock_bins_10kmsec_bc} and \ref{tab:normal_shock_bins_7kmsec_bc} and in Figs.~\ref{fig:shocking_F90_temperatures_10kmsec_linear} and \ref{fig:shocking_F90_temperatures_7kmsec_linear} respectively.
We then compare the reference curves with calculations in which the full database has been replaced with with reduced-size equivalents based on the URVC bin model~\cite{magin12a, torres20a}. In Tables~\ref{tab:normal_shock_bins_10kmsec_bc} and \ref{tab:normal_shock_bins_7kmsec_bc}, under label (2) we list the post-shock equilibrium state reached by the simulations when using 837, 100 and 10 bins respectively and compare them the ones obtained with the full database and its 9390 energy levels. As the number of bins is reduced from 837 down to 10, the post-shock equilibrium conditions begin to diverge from the ones predicted by the full model. However, even for the 10-bin system, the deviations in the post-shock equilibrium state are only of a few percent. We obtain such close agreement with the full model, because we are using energy bins of variable, instead of constant width. In previous work~\cite{torres18b} we were able to show that switching to variably-sized bins allows us to closely match the thermodynamic properties of the full model with a much smaller number of URVC bins. In particular, using more bins of smaller width to group together the lowest-energy rovibrational levels is advantageous to accurately capture the internal energy content of the cold free stream.


\begin{table}
 \centering
 \caption{Normal shock wave at $u_1 = 10 \, \mathrm{km \cdot s^{-1}}$: Upstream and downstream boundary conditions as a function of bin number} \label{tab:normal_shock_bins_10kmsec_bc}
 \begin{tabular}{l c c c c c c}
            & $p$ & $T$ & $T_\mathrm{int}$ & $\rho \times 10^3$ &  $u$ & $x_\mathrm{N}$ \\
            & [Pa] & [K] & [K] & [$\mathrm{kg / m^{3}}$] & [$\mathrm{m / s}$] & \\ \hline
  (1) pre-shock: & 13.3 & 300 & 300 & 0.1473 & 10000 & 0.02813 \\ \hline
  \multicolumn{7}{l}{(1a) post-shock frozen:}         \\
   & 11040 & 62550 & 300 & 0.5864 & 2511 & 0.02813 \\ \hline
  \multicolumn{7}{l}{(2) post-shock equilibrium:} \\
  full & 13665 & 11422 & 11422 & 2.0161 & 730.5 & 0.9998 \\
  837 bins & 13665 & 11422 & 11422 & 2.0161 & 730.5 & 0.9998 \\
  100 bins & 13665 & 11422 & 11422 & 2.0161 & 730.5 & 0.9998 \\
  10 bins & 13658 & 11493 & 11493 & 2.0024 & 735.5 & 0.9998 
 \end{tabular}
\end{table}


\begin{table}
 \centering
 \caption{Normal shock wave at $u_1 = 7 \, \mathrm{km \cdot s^{-1}}$: Upstream and downstream boundary conditions as a function of bin number} \label{tab:normal_shock_bins_7kmsec_bc}
 \begin{tabular}{l c c c c c c}
            & $p$ & $T$ & $T_\mathrm{int}$ & $\rho \times 10^3$ &  $u$ & $x_\mathrm{N}$ \\
            & [Pa] & [K] & [K] & [$\mathrm{kg / m^{3}}$] & [$\mathrm{m / s}$] &                \\ \hline
  (1) pre-shock: & 13.3 & 300 & 300 & 0.1473 & 7000 & 0.02813 \\ \hline
  \multicolumn{7}{l}{(1a) post-shock frozen:} \\
   & 5409.1 & 30784 & 300 & 0.5837 & 1766 & 0.02813 \\ \hline
  \multicolumn{7}{l}{(2) post-shock equilibrium:} \\
   full & 6802.3 & 6158.1 & 6158.1 & 2.4858 & 414.7 & 0.6642 \\
   837 bins & 6802.3 & 6158.1 & 6158.1 & 2.4858 & 414.7 & 0.6642 \\
   100 bins & 6802.3 & 6157.9 & 6157.9 & 2.4859 & 414.7 & 0.6642 \\
   10 bins & 6802.8 & 6141.2 & 6141.2 & 2.4886 & 414.3 & 0.6665 
 \end{tabular}
\end{table}


Mass density and temperature profiles for the high- and low-speed cases are plotted in Figs.~\ref{fig:shocking_F90_10kmsec_linear} and \ref{fig:shocking_F90_7kmsec_linear} respectively. The initial discontinuity, where the gas suddenly transitions from the free-stream conditions to the frozen post-shock conditions, is visible at $x=0$. Recall that the ODE system is only solved starting from the frozen post-shock conditions, i.e. state (1a) in Tables~\ref{tab:normal_shock_bins_10kmsec_bc} and \ref{tab:normal_shock_bins_7kmsec_bc}, and the method does not capture the shock front itself. Close-ups immediately downstream of the discontinuity are shown as insets in all four sub-figures. All plots follow the same labeling conventions. The reference solution is shown as dashed black lines, while results obtained with the URVC binning approach are plotted as continuous lines: 837 bins (black triangle on black line), 100 bins (blue circle on blue line) and 10 bins (red line). 

In Fig.~\ref{fig:shocking_F90_densities_10kmsec_linear} we plot profiles of mixture density $\rho$ (continuous lines) and molecular nitrogen $\rho_\mathrm{N_2}$ (dotted lines) for the high-speed case. The behavior in all four cases is very similar and the main differences are confined to the region immediately behind the shock front. Each one of the four $\rho_\mathrm{N_2}$-profiles reaches its maximum several millimeter downstream of the discontinuity, before dissociation begins to consume the remaining molecular nitrogen. The reference solution for the full N3 system exhibits the quickest response to the shock, whereas the coarse-grained systems lag behind. The response becomes slower with decreasing number of bins. In Fig.~\ref{fig:shocking_F90_temperatures_10kmsec_linear} we examine the corresponding temperature profiles. The kinetic temperature $T$ quickly decreases from its initial value of $62550 \, \mathrm{K}$ to about $30000 \, \mathrm{K}$ in the first $2-3 \, \mathrm{mm}$ behind the discontinuity. Simultaneously, the internal temperature rises from its free-stream value of $300 \, \mathrm{K}$ to a maximum of about $25000 \, \mathrm{K}$ in the same distance, before slowly decreasing again. Both temperatures then slowly approach each other as the gas continues to cool due to the effect of $\mathrm{N_2}$-dissociation. The relaxation of translational and internal energy proceeds quickest in the reference solution (dashed lines) and becomes progressively slower for the coarse-grain cases with decreasing number of bins. The internal temperatures reported in Figs.~\ref{fig:shocking_F90_temperatures_10kmsec_linear} and \ref{fig:shocking_F90_temperatures_7kmsec_linear} are the result of post-processing the internal state populations behind the shock. For the full reference solution, $T_\mathrm{int}$ is based on the rovibrational level populations (refer to Eqs.~(23) and (24) in Panesi et al.~\cite{panesi13a}). For the coarse-grained systems $T_\mathrm{int}$ is based on the bin populations and obtained in an analogous manner, following the procedure of App.~C of Ref.~\cite{torres18b}.  Thanks to the variably-spaced bin formulation, the macroscopic post-shock equilibrium state (i.e. temperature, composition) reached by all simulations closely matches the reference solution. As was shown by Munaf\`o et al~\cite{munafo14c, munafo14d} for the same flow conditions, the internal energy level populations exhibit strong departure from Boltzmann distributions and internal energy relaxation and dissociation effectively proceed at a common time scale.


In Fig.~\ref{fig:shocking_F90_densities_7kmsec_linear} we now show the density profiles for the low-speed case. Again, all four systems follow the same general behavior. Whereas in the high-speed case practically all molecular nitrogen eventually dissociated behind of the shock front, at these lower-speed conditions the $\mathrm{N_2}$-profiles remain fairly flat further downstream. However, the trend is now reversed, in the sense that the 10-bin system is the quickest to react to the shock, whereas the response becomes slower as the number of bins is increased all the way up to the full system. Figure~\ref{fig:shocking_F90_temperatures_7kmsec_linear} shows the corresponding temperatures for the low-speed case. With a length of approximately $5 \, \mathrm{m}$, the post-shock non-equilibrium region is now almost two orders of magnitude greater than in Fig.~\ref{fig:shocking_F90_temperatures_10kmsec_linear}. A closer look suggests that at these lower-speed conditions internal energy relaxation and cooling due to $\mathrm{N_2}$-dissociation proceed at distinct time scales. For the full reference solution, $T$ and $T_\mathrm{int}$ reach a common value of $\approx 15000 \, \mathrm{K}$ about $1 \, \mathrm{cm}$ downstream of the discontinuity, while the N mole fraction at this point has barely surpassed 20\% (not shown). Beyond $x = 1.5 \, \mathrm{cm}$ the remainder of the dissociation then effectively proceeds at a common temperature. With regard to the coarse-grained model solutions, another difference relative to the high-speed conditions is apparent. Whereas in Fig.~\ref{fig:shocking_F90_temperatures_10kmsec_linear} the reference solution showed the quickest initial relaxation, in Fig.~\ref{fig:shocking_F90_temperatures_7kmsec_linear} the full system is now the slowest of all four cases. In fact the ``convergence'' of the coarse-grained profiles with increasing bin number toward the reference solution occurs in the opposite sense relative to the high-speed case. 


By studying these two flow conditions with the inviscid ODE method we found that the relaxation region for the high-speed case extends for about $10 \, \mathrm{cm}$ and for the low-speed case roughly $5 \, \mathrm{m}$ from the discontinuity. This helps us size the domain and to adjust the computational parameters for the Navier-Stokes and DSMC calculations discussed in Secs.\ref{sec:normal_shock_bins_navier_stokes} and \ref{sec:normal_shock_bins_dsmc}. Furthermore, we see that the coarse-grained model has an influence on the evolution of the gas state in the post-shock region and these profiles diverge to some degree from the reference solution. As would be expected, the closest agreement with the full system is observed for the cases with the largest number of bins (837), while the biggest differences are observed for the 10-bin cases. However, these deviations become less severe further downstream of the initial discontinuity.



\subsection{Normal shock solution Euler vs. Navier-Stokes using Finite Volume method} \label{sec:normal_shock_bins_navier_stokes}

Based on the findings of Sec.~\ref{sec:normal_shock_bins_inviscid}, we simulate the normal shock by solving the Euler and Navier-Stokes equations on a one-dimensional domain with the finite volume (FV) method~\cite{hirsch88a}. Equations~(\ref{eq:navier_stokes_conservative}) are discretized in space and advanced in time using the implicit Backward-Euler method~\cite{gear71a}. The numerical inviscid fluxes at cell interfaces are computed using Roe's approximate Riemann solver~\cite{roe81a}. The particular form of Roe's dissipation matrix for the set of variables in Eq.~(\ref{eq:navier_stokes_conservative}) is discussed in detail elsewhere~\cite{munafo14d}. The purpose of this study is two-fold. First we compare the FV Euler result to the inviscid ODE results of Sec.~\ref{sec:normal_shock_bins_inviscid} to confirm that, when solving them on a sufficiently refined FV grid, we obtain the same answer as in Fig.~\ref{fig:shocking_F90_temperatures_10kmsec_linear}. 
Then we show how the shock structure changes once the viscous and diffusive terms of the Navier-Stokes equations are taken into account. For the sake of conciseness, in this section we only compare results for the high-speed case using the 10-bin coarse-grained system. However, the findings also apply to the low-speed flow condition and other bin numbers studied. Additional FV Navier-Stokes results will then be shown in Sec.~\ref{sec:normal_shock_bins_navier_stokes_vs_dsmc}, where we compare to equivalent DSMC simulations. All viscous shock solutions are obtained in a two-step approach. First, an Euler FV calculation is performed until reaching the inviscid steady-state solution. The simulation is carried out in the shock's frame of reference, where its steady-state structure develops over time around an initial discontinuity in flow parameters. The portion of the flow field left of the discontinuity is initialized to the pre-shock equilibrium state, whereas to its right the post-shock equilibrium state is imposed (recall Tables~\ref{tab:normal_shock_bins_10kmsec_bc} and \ref{tab:normal_shock_bins_7kmsec_bc} for the equilibrium conditions imposed in the high- and low-speed cases respectively). The final steady-state Euler solution is then re-used as initial condition for the subsequent Navier-Stokes simulation on the same grid. 
For both flow conditions a one-dimensional FV mesh with variable spacing is used. The region near the initial discontinuity is highly resolved, with a grid spacing of $\Delta x = 2 \times 10^{-5} \, \mathrm{m}$. 
Such severe refinement was performed only to minimize the effect of numerical diffusion near the shock front and lies well below the mean free path of $\lambda \approx 10^{-3} \, \mathrm{m}$ estimated at the same location. From this central region the grid is gradually coarsened in both the upstream and downstream directions to reduce computational cost, while ensuring numerical stability in the FV scheme. 

\begin{figure}
 \raggedright
 \includegraphics[width=\columnwidth]{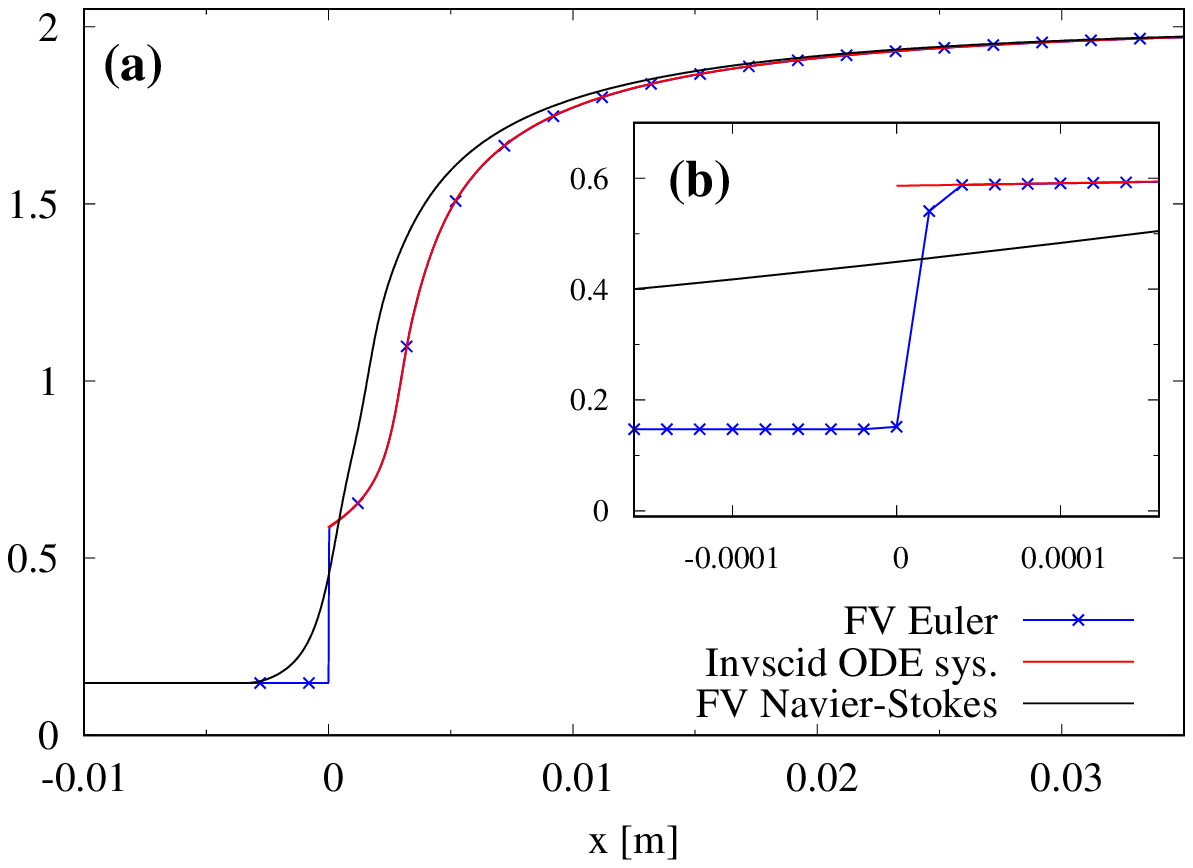}
 
 \caption{Gas density $\rho \times 10^3$ $[\mathrm{kg/m^3}]$ for shock at $u_1 = 10 \, \mathrm{km \cdot s^{-1}}$ with 10 bins. FVM solutions for Euler vs. Navier-Stokes and inviscid ODE approach.}
 \label{fig:fv_comparison_density_10v}
\end{figure}

\begin{figure}
 \raggedright
 \includegraphics[width=\columnwidth]{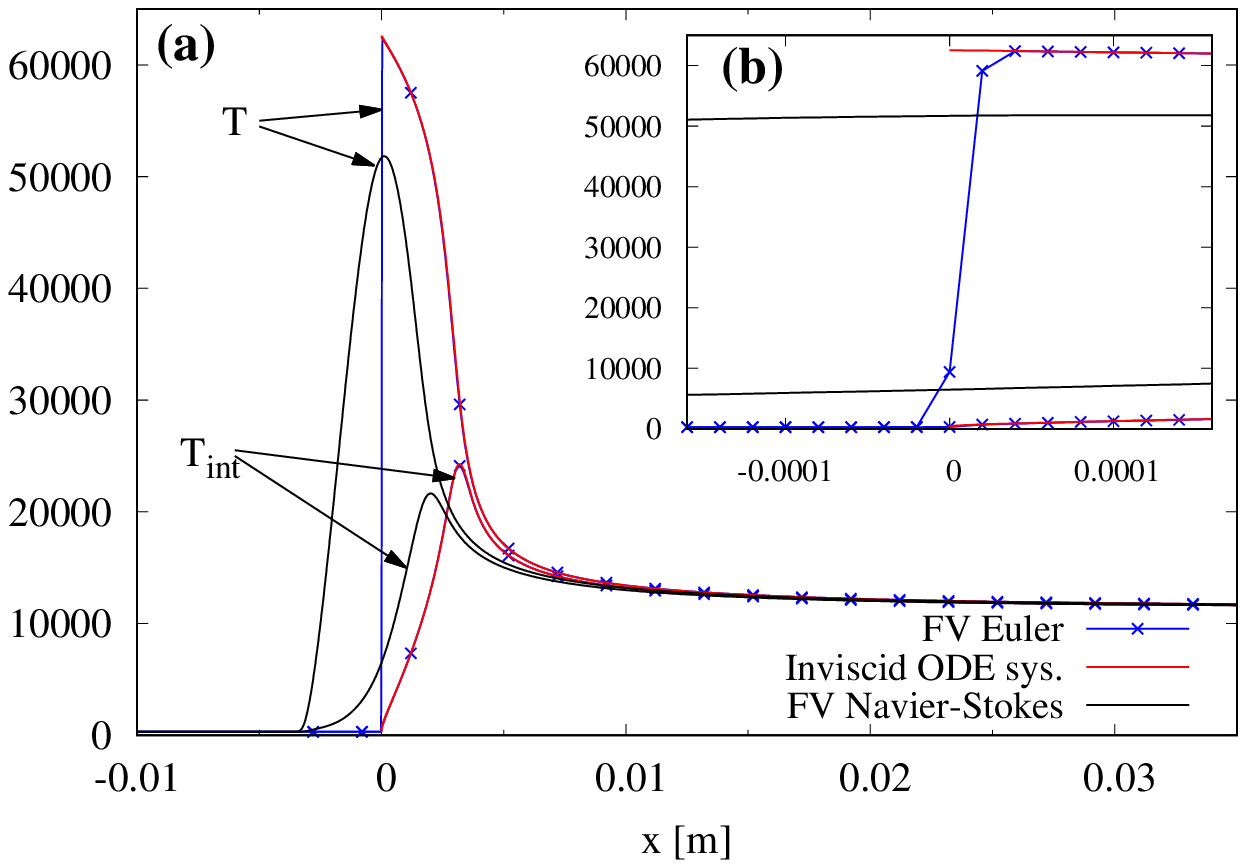}
 
 \caption{Kinetic and internal temperatures [K] for shock at $u_1 = 10 \, \mathrm{km \cdot s^{-1}}$ with 10 bins. FVM solutions for Euler vs. Navier-Stokes and inviscid ODE approach.}
 \label{fig:fv_comparison_temperatures_10v}
\end{figure}

Figures~\ref{fig:fv_comparison_density_10v} and \ref{fig:fv_comparison_temperatures_10v} show a comparison between the FV Euler (x-symbols on blue lines), Navier-Stokes (black lines) and inviscid ODE flow field of Sec.~\ref{sec:normal_shock_bins_inviscid} (red lines). All profiles shown are for the high-speed condition using the 10-bin coarse-grained system. Density profiles are shown first in Fig.~\ref{fig:fv_comparison_density_10v}. The origin of the $x$-axis lies at the location of the initial discontinuity for the Euler cases. Due to numerical diffusion in the FV approach this discontinuity is captured over an extent of 2-3 cells (see close-up in Fig.~\ref{fig:fv_comparison_density_10v}(b)). However, the grid has been carefully refined in the vicinity to ensure that this adverse numerical effect remains minimal. This is confirmed by the excellent agreement of the FV-Euler and inviscid ODE density profiles over the remainder of Fig.~\ref{fig:fv_comparison_density_10v}(a): past the discontinuity both the FV Euler and ODE solution curves lie on top of each other. Once the diffusive terms in the Navier-Stokes equations are taken into account, the discontinuity at $x=0$ disappears and is replaced by a smooth transition from pre-shock to post-shock density. Differences between the inviscid and viscous solutions are appreciable within about $\pm 0.01 \, \mathrm{m}$ of the initial discontinuity. The corresponding temperature profiles are shown in Fig.~\ref{fig:fv_comparison_temperatures_10v}. Excellent agreement between the FV-Euler and inviscid ODE solutions is observed to within 2 cells of the discontinuity (see close-up in Fig.~\ref{fig:fv_comparison_temperatures_10v}(b)). The jump in kinetic temperature is captured well by the FV method, as is its peak value for the inviscid case. Again, viscous effects act to smooth out these flow features and diffuse the shock front upstream. In the Navier-Stokes profile the gas temperature begins to depart from its pre-shock value about $0.003 \, \mathrm{m}$ ahead of the initial discontinuity and reaches a lower maximum ($T_\mathrm{max} \approx 51800 \, \mathrm{K}$ for Navier-Stokes vs. $62550 \, \mathrm{K}$ for Euler). The internal temperature profile is also affected by the inclusion of diffusive transport. The peak in the viscous profile ($T_\mathrm{int, max} \approx 21600 \, \mathrm{K}$) lies slightly upstream compared to the maximum of $24200 \, \mathrm{K}$ for the inviscid case. Consistent with the density profiles, differences in the viscous and inviscid temperature fields are only significant up to about $\pm 0.01 \, \mathrm{m}$ away from the initial discontinuity.

This comparison only covered flow quantities, which exhibit sharp discontinuities in their inviscid FV profiles. It showed that the Euler FV solutions are consistent with the inviscid ODE approach of Sec.~\ref{sec:normal_shock_bins_inviscid} and not polluted by numerical diffusion. This guarantees that any diffusive effects observed in the Navier-Stokes profiles reported in Sec.~\ref{sec:normal_shock_bins_navier_stokes_vs_dsmc} are physical in nature, i.e. exclusively due to the actual molecular diffusion terms in the Navier-Stokes equations.


\subsection{Normal shock solution with DSMC} \label{sec:normal_shock_bins_dsmc}

In this section we describe how the normal shock for both the high- and low-speed conditions was simulated using the DSMC method~\cite{bird94a}. The macroscopic flow profiles with DSMC are then compared with corresponding Navier-Stokes solutions in Sec.~\ref{sec:normal_shock_bins_navier_stokes_vs_dsmc}. Since DSMC can be used to indirectly solve the Boltzmann equation~\cite{wagner92a}, it allows us to resolve the shock structure with the highest level of detail. The VKI DSMC code used for this purpose is able to simulate one-dimensional steady and unsteady flows. Coarse-grained URVC cross sections~\cite{torres20a} for the N-$\mathrm{N_2}$ system are used and implementation details concerning the inelastic and reactive collision routines are discussed elsewhere.~\cite{torres18b}.

As was the case in Secs.~\ref{sec:normal_shock_bins_inviscid} and \ref{sec:normal_shock_bins_navier_stokes}, here we simulate the steady, one-dimensional flow across a normal shock. However, the precise manner in which the DSMC solution is obtained differs for the high- and low-speed cases. For the former, we simulate the flow in the shock's frame of reference. Both extremes of the domain are treated as open stream boundaries~\cite{bird94a}. In the VKI DSMC code~\cite{torres17a} we use the surface reservoir technique~\cite{tysanner05a} to generate the correct number and distribution of particles each time step at the upstream and downstream boundaries. The supersonic upstream gas enters from the left and, after traversing the standing shock wave, leaves the domain toward the right, where particles conforming to the post-shock equilibrium conditions are injected. The boundary conditions, expressed in terms of the equilibrium macroscopic flow parameters, are listed in Table~\ref{tab:normal_shock_bins_10kmsec_bc}. The velocity distributions at both boundaries conform to Maxwellians with the respective average velocities $\boldsymbol{u}_1 = \left( u_1, 0, 0 \right)^T$ and $\boldsymbol{u}_2 = \left( u_2, 0, 0 \right)^T$ and equilibrium temperatures $T_1$ and $T_2$. The particles representing molecular nitrogen entering at the left and right boundaries populate the rovibrational bins according to Boltzmann distributions at the pre- and post-shock equilibrium temperatures respectively. Given the degree of dissociation in the post-shock region, the number of $\mathrm{N_2}$-particles injected through the downstream boundary is negligible. As before, a trace amount of atomic nitrogen is added to the upstream gas to trigger inelastic $\mathrm{N}$-$\mathrm{N_2}(k)$ processes. 

To ensure that the shock front builds up at a well-defined location within the domain, we generate initial particles corresponding to the pre-shock equilibrium state (1) in the region left of the initial discontinuity and particles corresponding to post-shock equilibrium state (2) to the right of this location. This becomes the point where the supersonic free stream is ``tripped'' into transitioning to the post-shock equilibrium state and marks the initial location of the standing shock. As the simulation progresses, this discontinuity is smoothed out by particle transport. Once this phase is complete, the steady-state flow parameters are gathered from the DSMC particles and further refined through time-averaging. The location of the initial discontinuity is somewhat arbitrary, but if it is placed too close to either boundary, random walk may push the shock front out of the domain before steady-state macro-parameters can be extracted. Given that our primary goal is to observe as much of the relaxation region behind the shock, we place it as close as is reasonable to the left boundary. By setting $L_u = 3 \, \mathrm{cm}$ (see Table~\ref{tab:normal_shock_bins_simulation_parameters}), we make sure to leave ample space (i.e. $6\,000$ cells) between the inlet and the location of the initial discontinuity. Notice that for the high-speed condition only parameters for the 10-bin and 100-bin systems are listed in Table~\ref{tab:normal_shock_bins_simulation_parameters}. Due to the greater computational cost of the DSMC method compared to the ODE approach of Sec.~\ref{sec:normal_shock_bins_inviscid} and the Navier-Stokes calculations of Sec.~\ref{sec:normal_shock_bins_navier_stokes}, no DSMC simulations for the higher-resolution 837-bin case and the full database were carried out. In both high-speed simulations, the DSMC particle weight is set to ensure that at least 20 particles are present in every upstream cell. Due to the rise in density across the shock, there are $\approx 540$ particles per cell in the downstream region. For the 100-bin case the domain length is reduced to $L_\mathrm{u} = 3 \, \mathrm{cm}$ and $L_\mathrm{d} = 10 \, \mathrm{cm}$ respectively. This reduction is justified, as we are still able to capture the full relaxation region, while significantly reducing the computational expense.

Two complementary measures are taken to reduce the statistical noise inherent in DSMC flow fields. For the two high-speed cases in Table~\ref{tab:normal_shock_bins_simulation_parameters} we perform 64 simulations (using independent random number seeds) and ensemble-average the results. Thus, they become equivalent to a single simulation using 1280 particles per cell in the upstream- and about 34500 particles per cell in the downstream region. Past the transient phase (which lasts between $600\,000$ and $700\,000$ time steps) steady-state flow field samples are gathered over another $50\,000$ time steps. During this phase, instantaneous samples are taken every 10 time steps and added to a cumulative steady-state sample. 


\begin{table}[htb]
 \centering
 \caption{Normal shock wave with DSMC: domain and simulation parameters}
 \label{tab:normal_shock_bins_simulation_parameters}

 \begin{tabular}{r c c c c c}
  Case & & \multicolumn{2}{c}{high-speed} & & low-speed \\
  System & & 10 bins & 100 bins & $\quad$ & 10 bins \\ \hline
  & & & & \\[-1em]
  DSMC cell size $\Delta x$ & [$\mathrm{\mu m}$] & 5 & 5 & & 1.5 \\
  \\[-1em] \hline
  \\[-1em]
  Domain length & [cm] & 20 & 13 & & 9 \\
  upstream $L_\mathrm{u}$ & [cm] & 3 & 3 & & - \\
  downstream $L_\mathrm{d}$ & [cm] & 17 & 10 & & - \\ \hline
  & & & & & \\[-1em]
  DSMC cells & & $40\,000$ & $26\,000$ & & $60\,000$ \\
  upstream & & $6\,000$ & $6\,000$ & & - \\
  downstream & & $34\,000$ & $20\,000$ & & - \\
  \\[-1em] \hline
  \\[-1em]
  Total simulator & & & & & \\
  particles (million) & $\approx$ & 18.5 & 11 & & 16 \\ \hline
  \\[-1em]
  Particle weight & & \multicolumn{2}{c}{$8.02762 \times 10^{14}$} & & $2.4083 \times 10^{14}$ \\ \hline
  \\[-1em]
  DSMC $\Delta t$ & [ns] & 0.5 & 0.5 & & 0.2 \\
  DSMC steps & & & & & \\
  transient & & $600\,000$ & $700\,000$ & & $600\,000$ \\
  time avg. & & $50\,000$ & $50\,000$ & & $300\,000$ \\
            & & \multicolumn{2}{c}{(every 10 steps)} & & (every 1000)
 \end{tabular}
\end{table}


The flow field for the low-speed condition could not be obtained in the shock's frame of reference. Given the available computational resources, the domain size necessary to contain the entire steady-state shock profile would have become prohibitively large. Based on Fig.~\ref{fig:shocking_F90_temperatures_7kmsec_linear}, such a domain would have to extend at least $5 \, \mathrm{m}$ downstream of the shock front. While for the high-speed case we could comfortably contain the entire shock within $40\,000$ collision cells, this was not feasible for the low-speed case. Fortunately, for our purposes it is not necessary to simulate the entire post-shock relaxation region with DSMC. As was seen for the high-speed case, most of the diffusive effects are only appreciable within a narrow region surrounding the shock front. By concentrating on this portion we managed to significantly reduce the domain size. To accomplish this, we resort to the approach described by Strand and Goldstein~\cite{strand13a}, where the normal shock is treated as inherently unsteady. The supersonic free stream is fed into the domain on the left boundary, while a specular wall reflects all particles on the boundary to the right. This stagnates the incoming flow and generates a shock wave moving from right to left into the undisturbed gas upstream. Unlike in the previous set-up, the reference frame is now attached to the post-shock equilibrium gas, implying that $u_2^\prime = 0$. Therefore, in order to obtain the desired post-shock thermodynamic conditions of Table~\ref{tab:normal_shock_bins_7kmsec_bc} in our simulation, we adjust the inflow velocity to $u_1^\prime =  u_1 - u_2$. Once the shock front has left the near-wall region, it begins to take on its steady-state structure and travels upstream at approximately $u_\mathrm{shock} = -u_2$. At this point macroscopic flow parameters can be sampled and individual samples time-averaged to reduce statistical noise. Since the shock is continuously moving upstream, these instantaneous samples have to be displaced to a common origin before time-averaging. Again, we resort to the procedure described in Ref.~\cite{strand13a} to define a common reference location for all profiles.

At the low-speed condition the higher post-shock density and lower temperature (see Table~\ref{tab:normal_shock_bins_7kmsec_bc}) impose more stringent constraints on the collision cell- and time step size. Thus, in the rightmost column of Table~\ref{tab:normal_shock_bins_simulation_parameters}, several simulation parameters were adjusted accordingly. Just as for the high-speed condition, ensemble-averaging over 64 independent simulations is used to reduce the statistical scatter in the instantaneous samples.



\subsection{Comparison Navier Stokes vs. DSMC} \label{sec:normal_shock_bins_navier_stokes_vs_dsmc}

\begin{figure}
 
 \begin{minipage}{0.8\columnwidth}
  \subfloat[Density $\rho \times 10^{3} \, \mathrm{[kg/m^3]}$ (solid lines) and partial density of molecular nitrogen $\rho_\mathrm{N_2} \times 10^{3}$ (dotted lines)]{\includegraphics[width=\columnwidth]{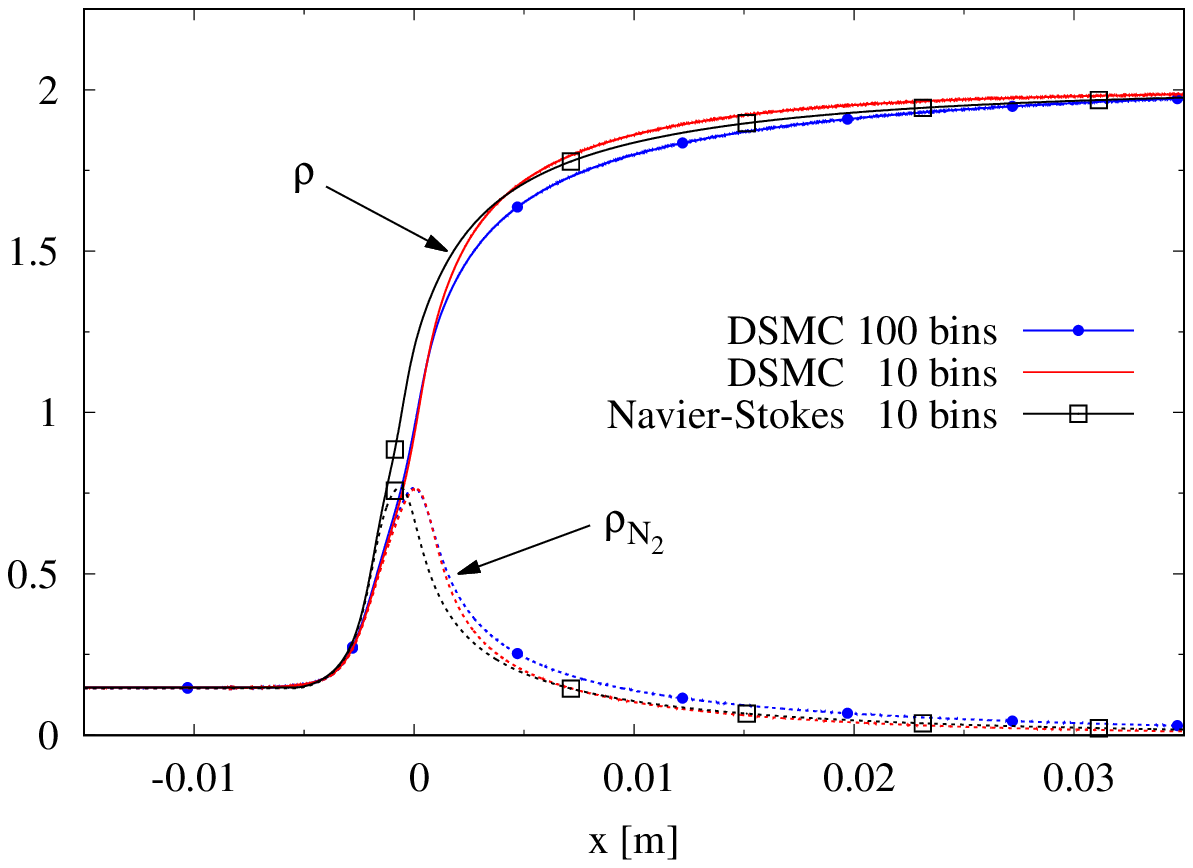}\label{fig:comparison_density_10v}}
 \end{minipage}

 \begin{minipage}{0.8\columnwidth}
  \subfloat[Gas kinetic temperature $T \, \mathrm{[K]}$ and internal temperature of $\mathrm{N_2}$-molecules $T_\mathrm{int} \, \mathrm{[K]}$]{\includegraphics[width=\columnwidth]{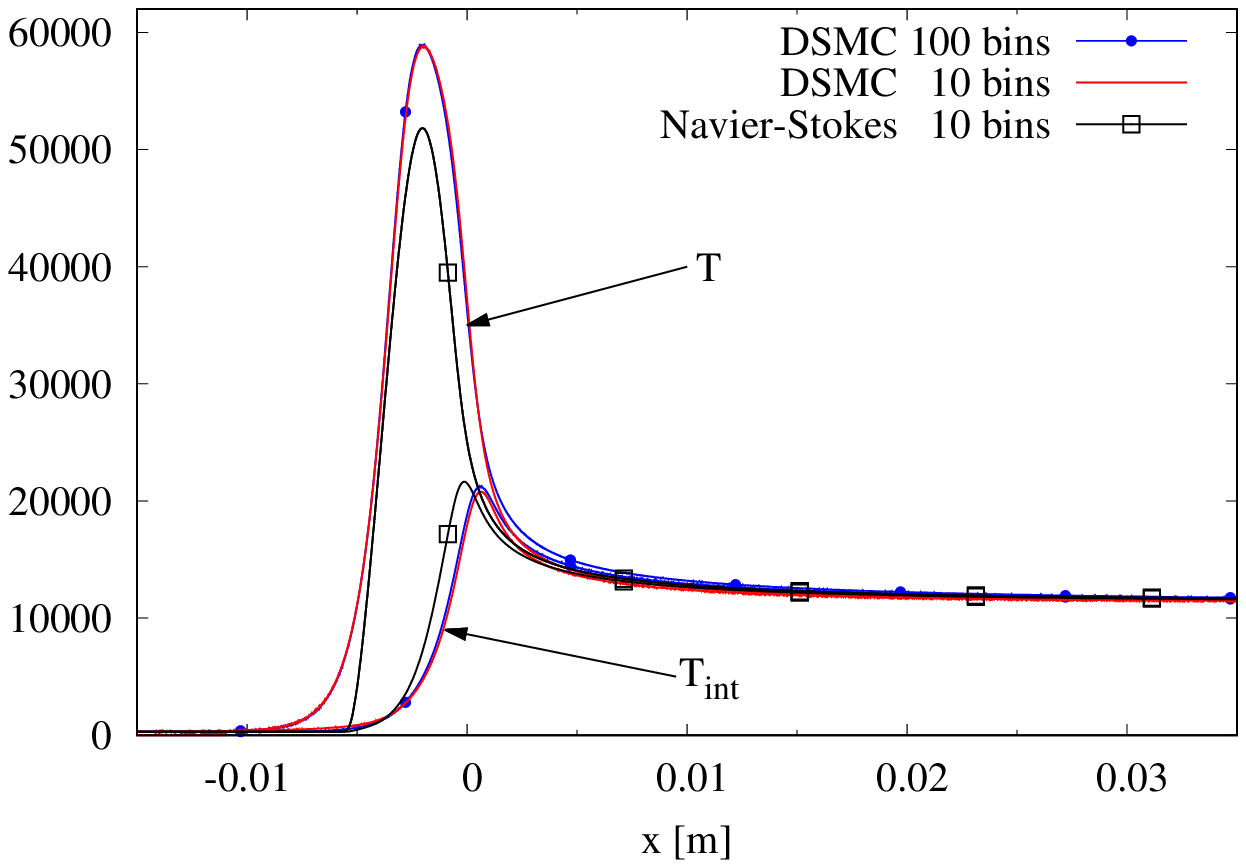}\label{fig:comparison_temperatures_10v}}
 \end{minipage}
 
 \caption{Gas density and temperature profiles for high-speed condition ($u_1 = 10 \, \mathrm{km \cdot s^{-1}}$). DSMC with 100 bins (dot on blue lines) vs. DSMC with 10 bins (red lines) vs. Navier-Stokes with 10 bins (unfilled squares on black lines).}
 \label{fig:comparison_01_10kmsec}
\end{figure}

\begin{figure}
 
 \begin{minipage}[t]{0.5\columnwidth}
  \subfloat[$x$-component of mass diffusion flux for $\mathrm{N_2} \, \mathrm{[kg \cdot m/s]}$]{\includegraphics[width=\columnwidth]{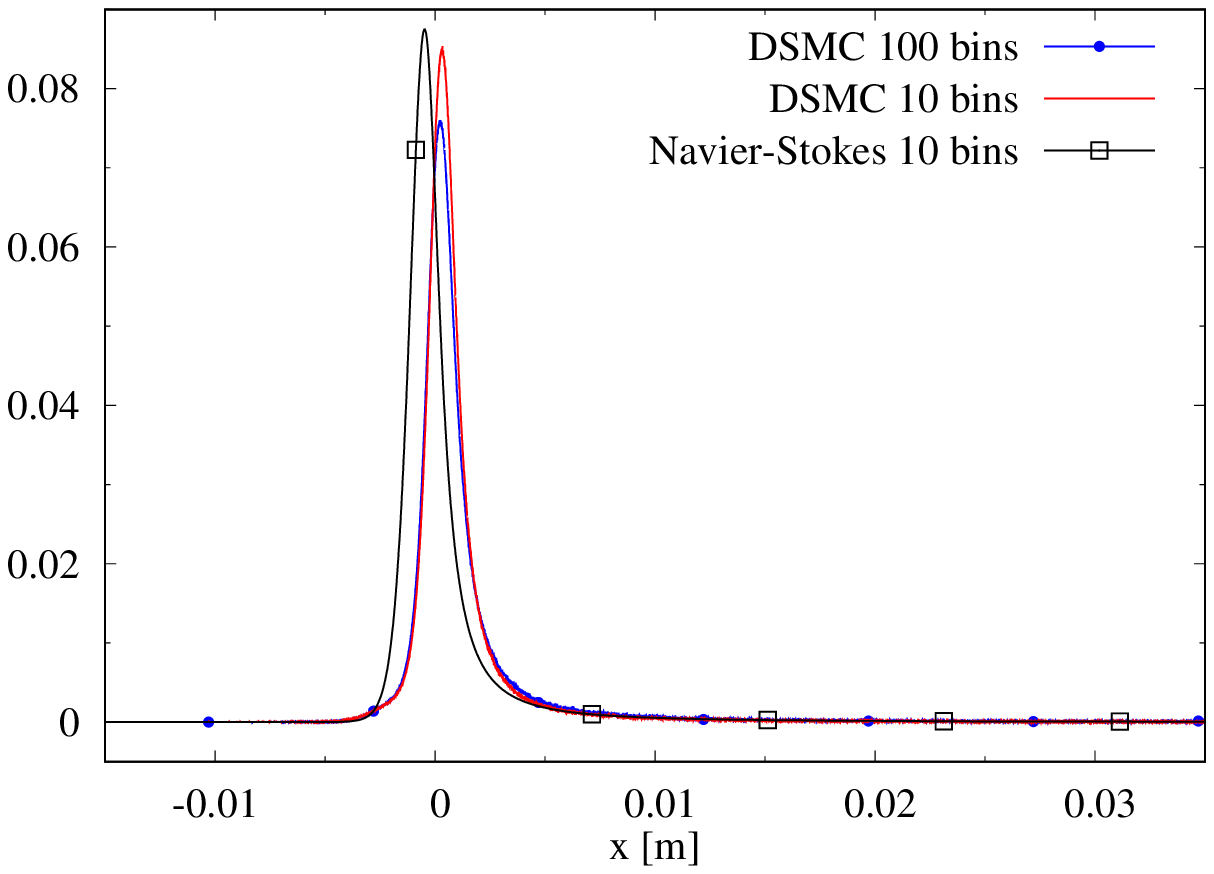}\label{fig:comparison_diffusion_fluxes_N2_10v}}
 \end{minipage}~
 \begin{minipage}[t]{0.5\columnwidth}
  \subfloat[Normal components of viscous stress tensor $\mathrm{[kPa]}$]{\includegraphics[width=\columnwidth]{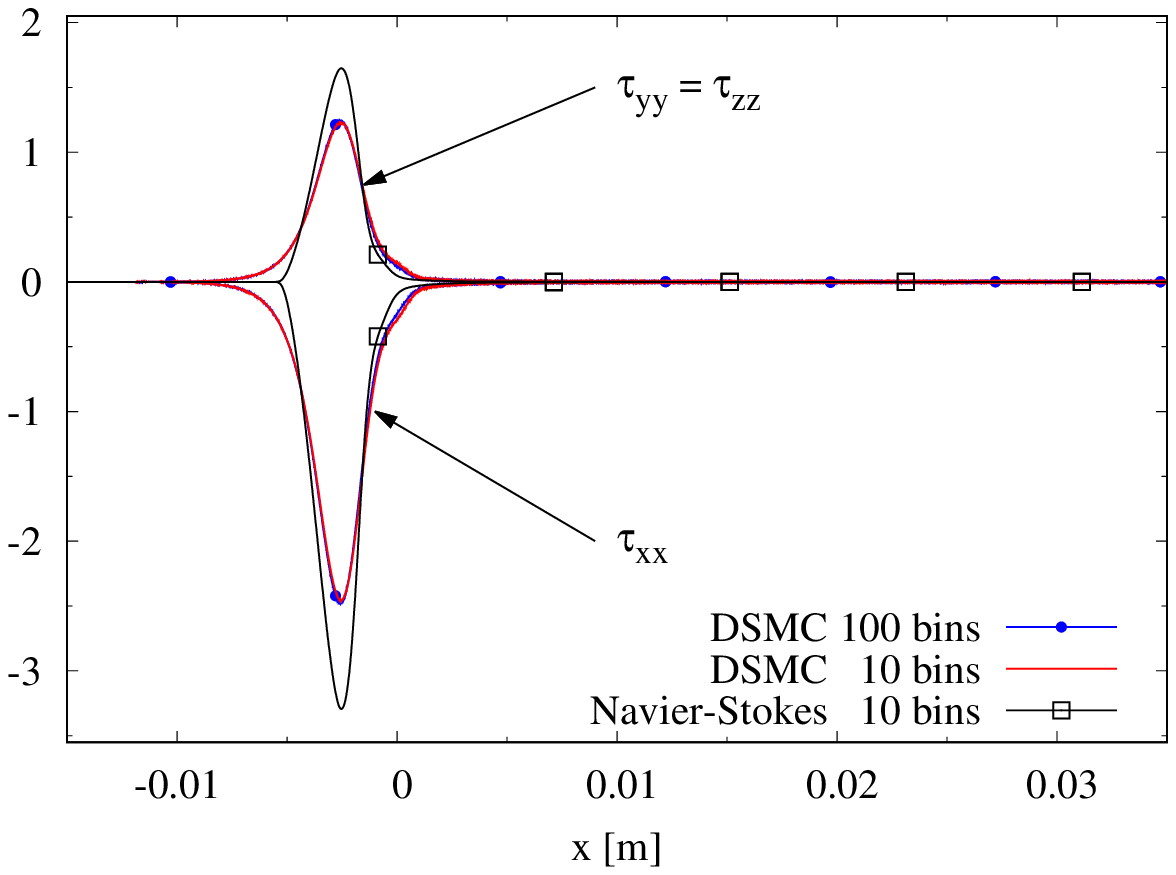}\label{fig:comparison_viscous_stresses_10v}}
 \end{minipage}
 
 \begin{minipage}{0.5\columnwidth}
  \subfloat[$x$-component of heat flux $\mathrm{[MW/m^2]}$. Navier-Stokes profile split into contributions due to conduction (dash-dotted lines) and diffusion of enthalpy (dotted lines)]{\includegraphics[width=1.0\columnwidth]{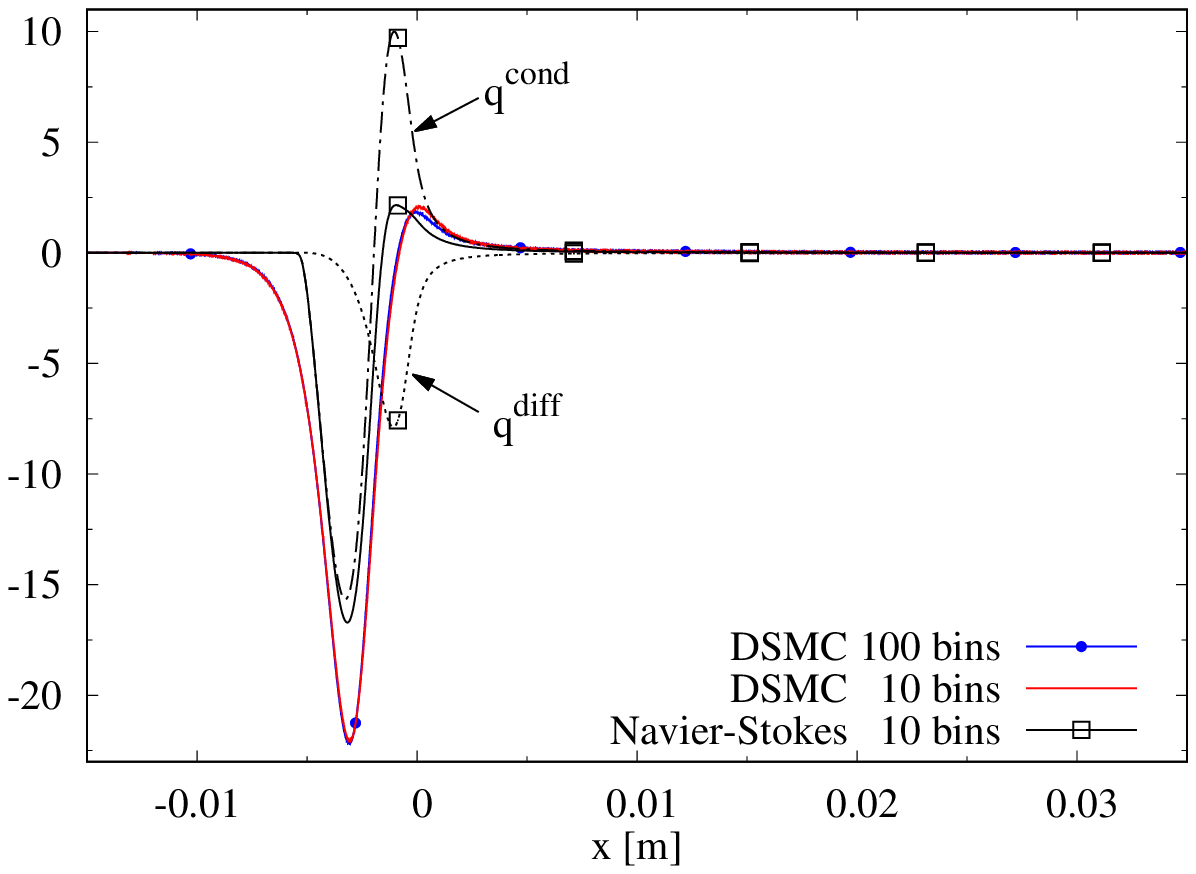}\label{fig:comparison_heat_fluxes_10v}}
 \end{minipage}
 
 \caption{Diffusive transport fluxes for high-speed condition ($u_1 = 10 \, \mathrm{km \cdot s^{-1}}$). DSMC with 100 bins (filled circle on blue lines) vs. DSMC with 10 bins (red lines) vs. Navier-Stokes with 10 bins (unfilled squares on black lines).}
 \label{fig:comparison_02_10kmsec}
\end{figure}

We now examine the flow fields obtained through the methods described in Secs.~\ref{sec:normal_shock_bins_navier_stokes} and \ref{sec:normal_shock_bins_dsmc}. First, in Figs.~\ref{fig:comparison_01_10kmsec} and \ref{fig:comparison_02_10kmsec} we compare DSMC profiles obtained using the 100-bin (blue dot on blue line) and 10-bin (red line) systems to Navier-Stokes profiles with the 10-bin system (black square on black line) at the high-speed conditions.

We start with the gas density profiles in Fig.~\ref{fig:comparison_density_10v}. The DSMC and Navier-Stokes curves have been translated on the $x$-axis, such that the initial rise in density occurs at the same location for all three profiles. The location of the origin is arbitrary, but the same convention is used consistently in all flow parameter plots in Figs.~\ref{fig:comparison_01_10kmsec} and \ref{fig:comparison_02_10kmsec}. Focusing on the 10-bin system, both the DSMC and Navier-Stokes density profiles show close agreement, except for a weak increase of the density slope in the DSMC result at $x \approx 0$, which is absent from the Navier-Stokes curve. The Navier-Stokes density profile exhibits a quicker and more uniform initial rise, before intersecting the DSMC profile at $x \approx 0.003 \, \mathrm{m}$.

Next, in Fig.~\ref{fig:comparison_temperatures_10v} we compare the corresponding kinetic and internal temperatures. Here, the differences between both methods are more apparent. The maximum $T$-value obtained with DSMC (10 bins) is $T_\mathrm{max} \approx 58800 \, \mathrm{K}$, which lies roughly $7000 \, \mathrm{K}$ above the corresponding peak for Navier-Stokes. Incidentally, both maxima lie very close to one another, at $x \approx -0.002 \, \mathrm{m}$. By contrast, the maximum $T_\mathrm{int}$-values for all three curves are much closer to one another, with the Navier-Stokes curve slightly leading the DSMC profiles. The most noticeable difference is that both $T$-curves for DSMC begin to rise farther upstream and more gradually than the Navier-Stokes profile. Back in Fig.~\ref{fig:comparison_density_10v}, we also plot the partial density of $\mathrm{N_2}$ using dotted lines. As was observed in Fig.~\ref{fig:shocking_F90_densities_10kmsec_linear} for the inviscid case, there is an initial rise in $\rho_\mathrm{N_2}$ across the shock, before dissociation kicks in and gradually consumes the molecular nitrogen further downstream. At these high temperatures, the post-shock gas is almost entirely made up of atoms. Here, Navier-Stokes predicts dissociation occurring slightly ahead of the corresponding DSMC (10 bins) curve. This is consistent with the lower kinetic temperature observed for Navier-Stokes in Fig.~\ref{fig:comparison_temperatures_10v}.


We now move on to Fig.~\ref{fig:comparison_02_10kmsec} and the comparison of flow parameters associated with diffusive transport at the high-speed condition. In Fig.~\ref{fig:comparison_diffusion_fluxes_N2_10v} we first show the mass diffusion flux of $\mathrm{N_2}$ along the $x$-direction. For the two DSMC curves and the single Navier-Stokes result $j_{x, \mathrm{N_2}}$ is calculated as the mass-weighted average over all internal energy bins, i.e.: $j_{x,\mathrm{N_2}} = \sum_{k \in \mathcal{K}_\mathrm{N_2}} \{ \rho_k \, u_k^\mathrm{d} \}$. The corresponding mass diffusion flux of atomic nitrogen: $j_{x,\mathrm{N}} = \rho_\mathrm{N} \, u_\mathrm{N}^\mathrm{d}$ (not shown) is equal in magnitude, but opposite in sign. The peak of $j_{x, \mathrm{N_2}}$ captured by the DSMC and Navier-Stokes methods with 10 bins agrees to within less than 5\%, although in the Navier-Stokes profile this maximum appears slightly ahead of the DSMC curve. For the 100-bin DSMC case, the peak diffusion flux lies about 10\% below the corresponding 10-bin DSMC value, but at almost exactly the same $x$-location. Next, in Fig.~\ref{fig:comparison_viscous_stresses_10v} we plot the three normal components of the viscous stress tensor. For our one-dimensional flow configuration only the velocity derivative $\partial u_x / \partial x$ becomes non-zero across the shock. As a consequence, the only components of $\underline{\underline{\tau}}$ in Eq.~(\ref{eq:momentum_balance}), which take on non-zero values turn out to be $\tau_{xx} = \frac{4}{3} \, \eta \, (\partial u_x / \partial x)$ and $\tau_{yy} = \tau_{zz} = - \frac{2}{3} \, \eta \, (\partial u_x / \partial x)$. Both DSMC and the Navier-Stokes profiles reach their maxima at essentially the same $x$-location. The DSMC stress profiles are slightly more spread out than their Navier-Stokes counterparts. The ratio $\tau_{xx, \mathrm{max}} / \tau_{yy, \mathrm{max}}$ yields exactly $-2$ for the Navier-Stokes profiles, in accordance with the analytical expressions for $\tau_{xx}$ and $\tau_{yy}$. The same ratio of $-2$ is maintained for the DSMC profiles, although the peak viscous stresses obtained with Navier-Stokes lie about 34\% above the corresponding DSMC values. As can be seen by comparing the two DSMC profiles, the number of bins has practically no effect on the shape of the viscous stress profiles.
Finally, in Fig.~\ref{fig:comparison_heat_fluxes_10v} we compare $q_x$, i.e. the heat flux component along the flow direction. Both DSMC and the Navier-Stokes profiles exhibit their peak negative values (due to heat being transferred upstream across the shock front) at roughly the same $x$-location. However, the maximum flux for DSMC is nearly $-22.1 \, \mathrm{MW/m^2}$, while for the Navier-Stokes result it only reaches $-16.7 \, \mathrm{MW/m^2}$. As was the case for the kinetic temperature in Fig.~\ref{fig:comparison_temperatures_10v}, the DSMC heat flux profiles are noticeably more diffuse and begin to deviate from zero much sooner upstream than their Navier-Stokes counterpart. A second smaller, but positive peak appears in all three $q_x$-profiles further downstream. Thus, some amount of heat is also being transferred from the shock front in the downstream direction.

It is interesting to note that the location of this second, positive peak in $q_x$ nearly coincides with the maximum in $j_{x,\mathrm{N_2}}$ reported in Fig.~\ref{fig:comparison_diffusion_fluxes_N2_10v} for all three calculations. One might thus assume that ``diffusion of enthalpy'' plays a significant role in shaping the heat flux profile in this region. In order to answer this question we have decomposed the Navier-Stokes (solid black lines) result into $q_x^\mathrm{cond}$, i.e. its contributions due to heat conduction (dash-dotted line) and $q_x^\mathrm{diff}$, i.e. its contribution due to diffusion of enthalpy (dotted line). It turns out that the second peak observed in the $q_x$-profile is the net result of a sizable conductive heat flux in the downstream direction and a nearly as large diffusive heat flux in the opposite sense. With about $10 \, \mathrm{MW / m^2}$ the peak of $q_x^\mathrm{cond}$ in the downstream direction is about 2/3 in magnitude of the amount being transferred upstream. Simultaneously, this effect is almost completely compensated for by the $q_x^\mathrm{diff}$-contribution in the opposite sense, which reaches a peak value of nearly $-8  \, \mathrm{MW / m^2}$. 

No such decomposition is shown for the DSMC results in Fig.~\ref{fig:comparison_heat_fluxes_10v}. Indeed it would be tricky to achieve a rigorous separation into the aforementioned $q^\mathrm{cond}$ and $q^\mathrm{diff}$ terms for the DSMC profiles. In DSMC the macroscopic heat flux emerges as the net result of advection of kinetic and internal energy attached to each individual molecule and atom (see App.~\ref{app:macroscopic_moments} for the definitions used in our calculations). The DSMC heat flux profiles naturally account for all contributions due to conduction, diffusion of enthalpy and heat transfer induced by concentration gradients (Dufour effect). However, since transport coefficients, such as thermal conductivity $\lambda$ and species-dependent thermal diffusion ratio $\chi_i$ have no meaning at the gas-kinetic scale, a rigorous separation into individual contributions is not possible.

The overall close agreement between the DSMC and Navier-Stokes profiles in Figs.~\ref{fig:comparison_01_10kmsec} and \ref{fig:comparison_02_10kmsec} is somewhat surprising. Given the strong deceleration, the molecular velocity distributions across the shock obtained with DSMC will deviate significantly from the Chapman-Enskog distribution, on which the Navier-Stokes solution is based. Thus, one might have expected a greater difference between both results. 
Another noteworthy aspect is that, apart from minor differences in the mixture and partial density profiles, the 10-bin and 100-bin DSMC flow fields exhibit almost the same behavior. This is in contrast with what was observed in Fig.~\ref{fig:shocking_F90_temperatures_10kmsec_linear} for the inviscid case, where the temperature profiles are very sensitive to the number of bins employed. Although an exhaustive study was not conducted, this suggests that diffusive phenomena significantly reduce differences due to bin number originally observed in the inviscid profiles.




\begin{figure}
 
 \begin{minipage}{0.8\columnwidth}
  \subfloat[Density $\rho \times 10^{3} \, \mathrm{[kg/m^3]}$ (solid lines) and partial density of molecular nitrogen $\rho_\mathrm{N_2} \times 10^{3}$ (dotted lines)]{\includegraphics[width=\columnwidth]{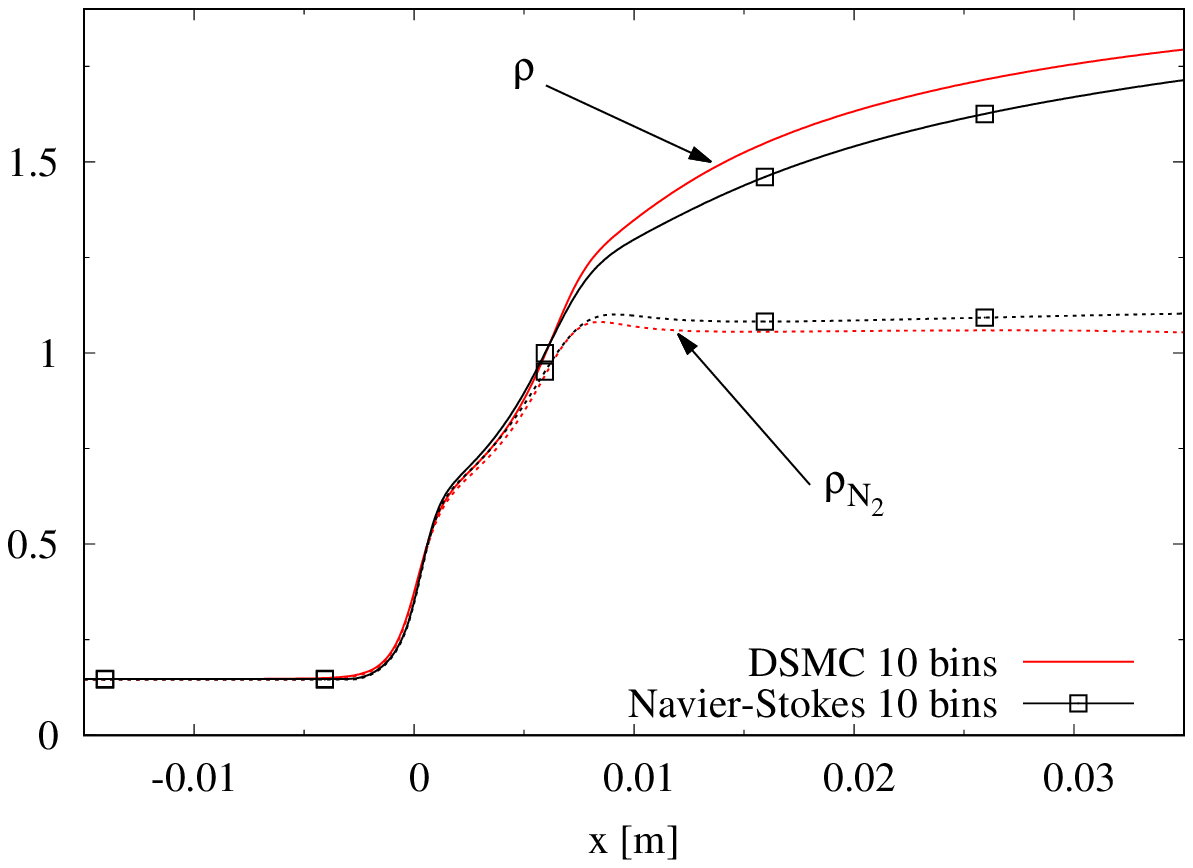}\label{fig:comparison_density_10v_7kmsec}}
 \end{minipage}
 
 \begin{minipage}{0.8\columnwidth}
  \subfloat[Gas kinetic temperature and internal temperature of $\mathrm{N_2}$ molecules $\mathrm{[K]}$]{\includegraphics[width=\columnwidth]{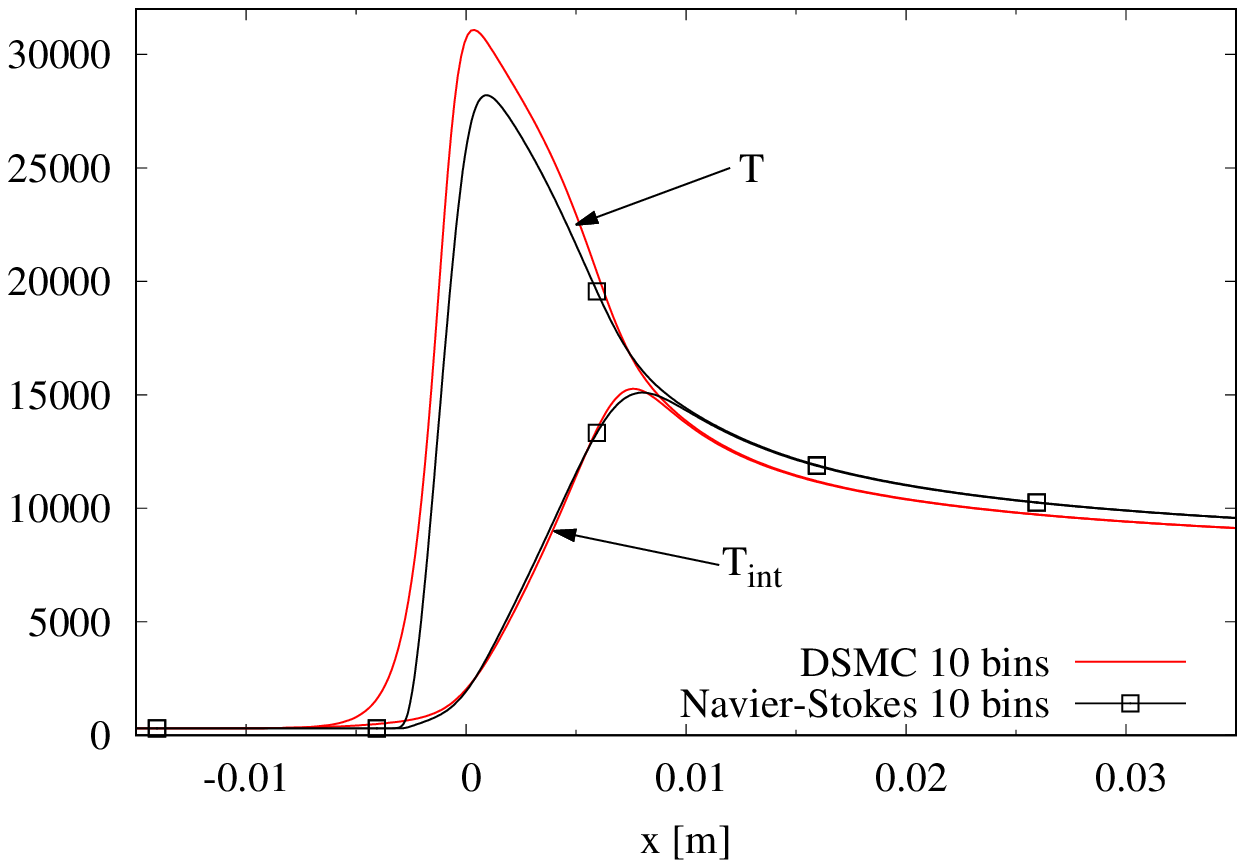}\label{fig:comparison_temperatures_10v_7kmsec}}
 \end{minipage}
 
 \caption{Gas density and temperature profiles for low-speed condition ($u_1 = 7 \, \mathrm{km \cdot s^{-1}}$). DSMC with 10 bins (red lines) vs. Navier-Stokes with 10 bins (squares on black lines).}
 \label{fig:comparison_01_7kmsec}
\end{figure}

\begin{figure}
 
 \begin{minipage}[t]{0.5\columnwidth}
  \subfloat[$x$-component of mass diffusion flux for $\mathrm{N_2} \, \mathrm{[kg \cdot m/s]}$]{\includegraphics[width=\columnwidth]{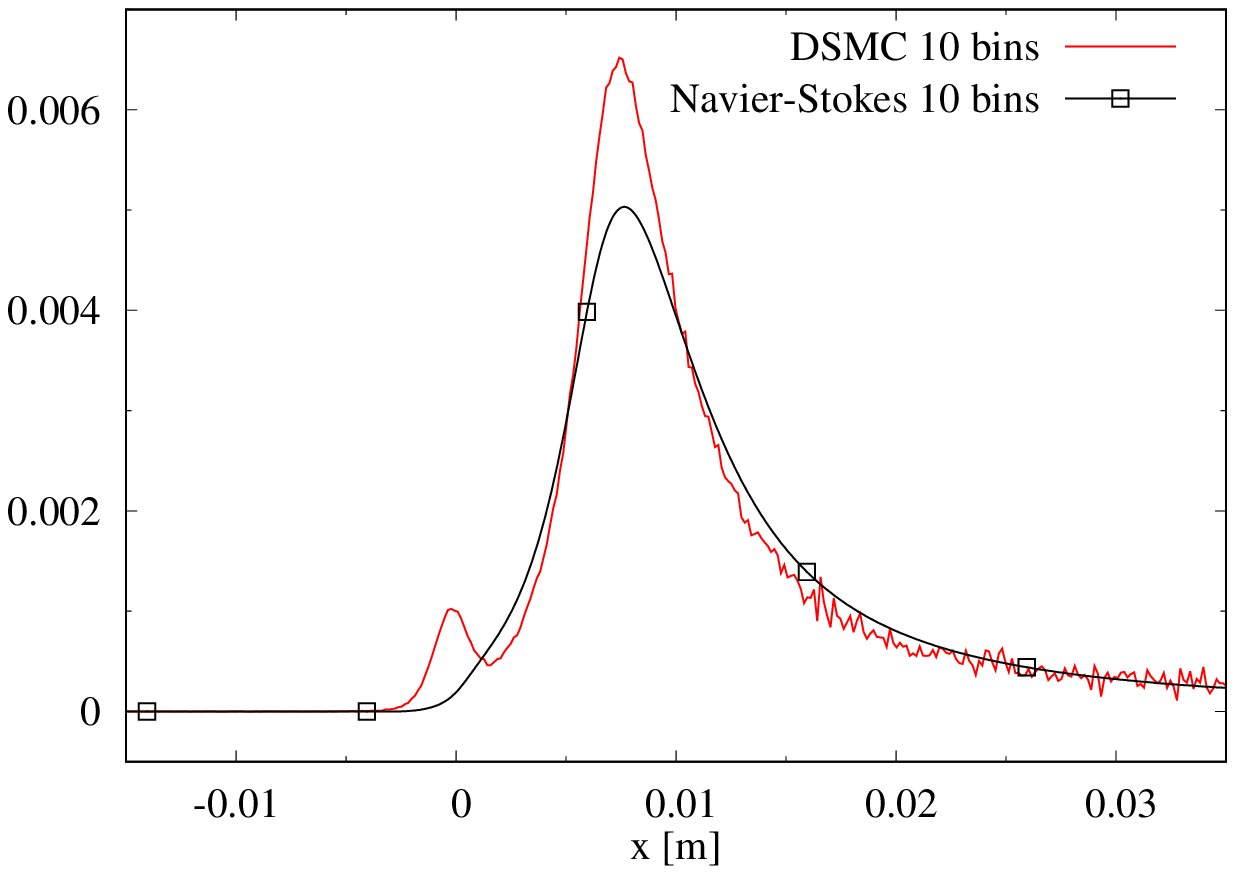}\label{fig:comparison_diffusion_fluxes_N2_10v_7kmsec}}
 \end{minipage}~ 
 \begin{minipage}[t]{0.5\columnwidth}
  \subfloat[Normal components of viscous stress tensor $\mathrm{[kPa]}$]{\includegraphics[width=\columnwidth]{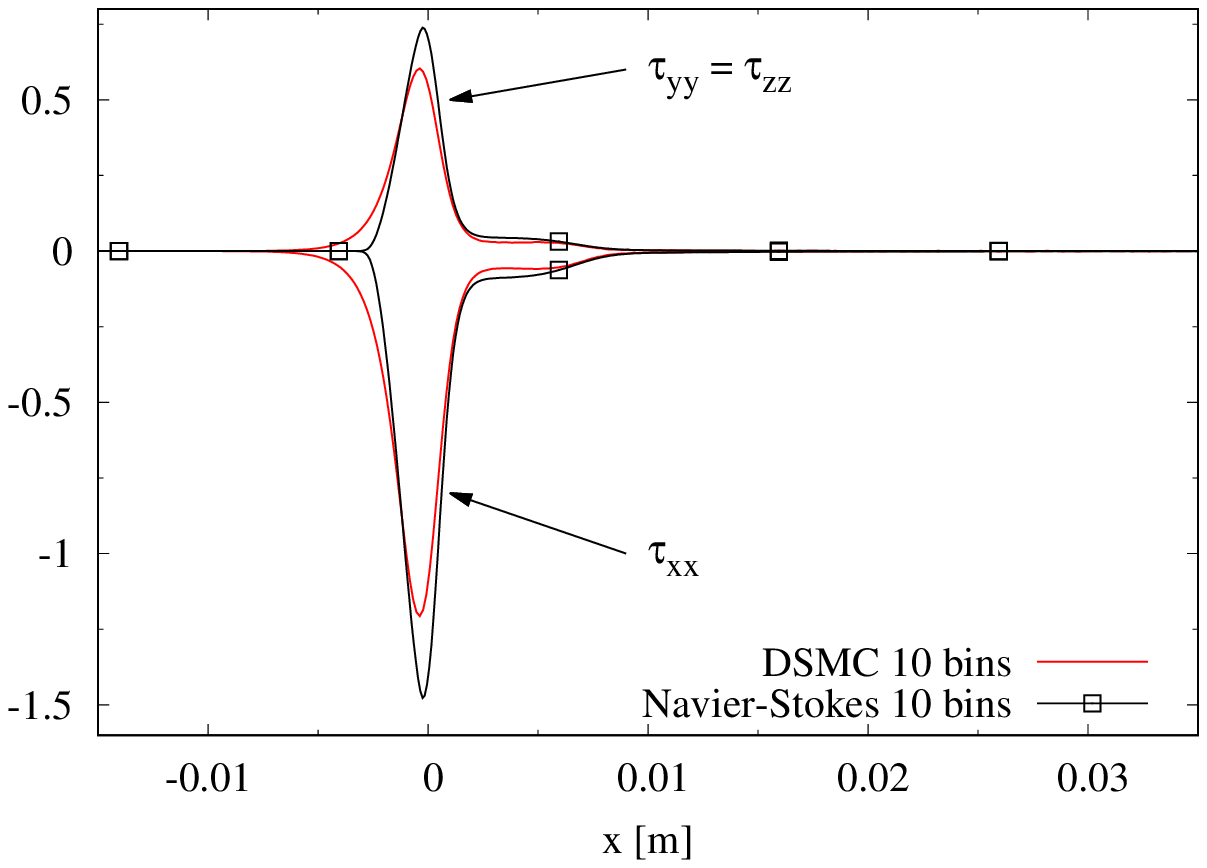}\label{fig:comparison_viscous_stresses_10v_7kmsec}}
 \end{minipage}

 \begin{minipage}{0.5\columnwidth}
  \subfloat[$x$-component of heat flux $\mathrm{[MW/m^2]}$. Navier-Stokes profile split into contributions due to conduction (dash-dotted lines) and diffusion of enthalpy (dotted lines)]{\includegraphics[width=1.0\columnwidth]{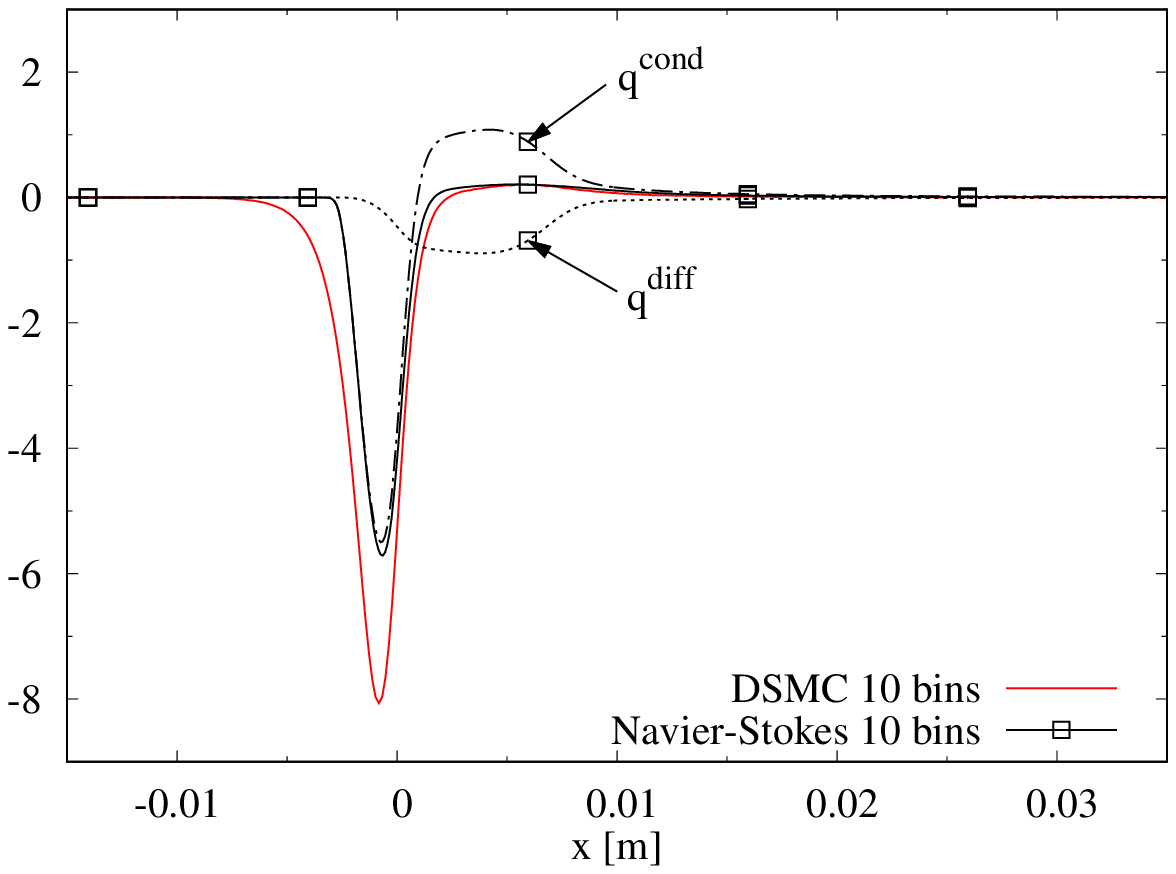}\label{fig:comparison_heat_fluxes_10v_7kmsec}}
 \end{minipage}

  \caption{Diffusive transport fluxes for low-speed condition ($u_1 = 7 \, \mathrm{km \cdot s^{-1}}$). DSMC with 10 bins (red lines) vs. Navier-Stokes with 10 bins (squares on black lines)}
  \label{fig:comparison_02_7kmsec}
\end{figure}

In Figs.~\ref{fig:comparison_01_7kmsec} and \ref{fig:comparison_02_7kmsec}, we now compare DSMC (red lines) and Navier-Stokes results (unfilled squares on black lines) for the low-speed case. Here we focus exclusively on the 10-bin system. Recall from Sec.~\ref{sec:normal_shock_bins_inviscid} that at $7 \, \mathrm{km \cdot s^{-1}}$ the post-shock chemical nonequilibrium region extends much farther downstream than at $10 \, \mathrm{km \cdot s^{-1}}$. However, here we focus on the region immediately surrounding the shock front, where the strongest  thermo-chemical nonequilibrium is observed. Thus, density, temperature and in particular mixture composition do not fully reach their post-shock equilibrium values in the $x$-range plotted. However, the moments associated with viscous and diffusive phenomena adjust much more quickly and are fully contained within the region shown.

In Fig.~\ref{fig:comparison_density_10v_7kmsec} we begin by plotting density profiles. As was done for the high-speed case, the DSMC and Navier-Stokes profiles have been aligned such that the initial rise in density occurs at a common $x$-location. For both the DSMC and Navier-Stokes calculations the overall gas density $\rho$ is represented by solid lines, whereas $\rho_\mathrm{N_2}$ is shown using dotted lines. One can see two distinct ``bumps'' in both $\rho$-profiles, with the first one appearing at the same $x$-location with both methods. Near the second bump further downstream, the two $\rho$-curves begin to diverge, and beyond this point the DSMC profile remains slightly above the corresponding Navier-Stokes curve. Up until the second bump in the $\rho$-profiles dissociation plays only a minor role. But past this point the amount of atomic nitrogen begins to rapidly increase, while $\rho_\mathrm{N_2}$ remains almost constant. In Fig.~\ref{fig:comparison_temperatures_10v_7kmsec} we plot the corresponding temperature profiles. As was seen for the high-speed case in Fig.~\ref{fig:comparison_temperatures_10v}, the peaks in kinetic temperature $T$ appear at almost the same $x$-location for both DSMC and Navier-Stokes. Of course, given the significantly lower total enthalpy of the flow, the peak $T$-values are much lower than for the high-speed case. At $T_\mathrm{max} \approx 31100 \, \mathrm{K}$, DSMC predicts a somewhat higher peak value than Navier-Stokes, where a maximum of $\approx 28200 \, \mathrm{K}$ is reached. Similar to the high-speed case, the kinetic temperature profile from DSMC in Fig.~\ref{fig:comparison_temperatures_10v_7kmsec} is more diffuse and exhibits a more gradual initial rise than the Navier-Stokes curve. The location of the $T_\mathrm{int}$-maximum appears almost exactly at the same $x$-location and both values differ by less than 2\% (DSMC: $T_\mathrm{int} \approx 15300 \, \mathrm{K}$ vs. Navier-Stokes: $T_\mathrm{int} \approx 15100 \, \mathrm{K}$). Slightly different behavior is seen downstream of this point, with the common DSMC temperature decreasing somewhat faster than in the Navier-Stokes profile. It is worth noting that both methods predict the highest kinetic temperature about $0.005 \, \mathrm{m}$ upstream of the point where significant amounts of N-atoms begin to be produced. In fact, for both methods the location in Fig.~\ref{fig:comparison_density_10v_7kmsec} where the $\rho$ and $\rho_\mathrm{N_2}$ profiles begin to diverge coincides with the peak in $T_\mathrm{int}$ observed in Fig.~\ref{fig:comparison_temperatures_10v_7kmsec}, and beyond which the translational and internal temperatures reach a common value. This suggests that at these lower-speed conditions a noticeable ``incubation length'' for dissociation exists and that dissociation primarily occurs under near-equilibrium conditions downstream of the shock front. Overall, DSMC predicts slightly quicker dissociation of $\mathrm{N_2}$ than the Navier-Stokes calculation. This can be seen by comparing the density profiles in Fig.~\ref{fig:comparison_density_10v_7kmsec}. The behavior of the temperature profiles in Fig.~\ref{fig:comparison_temperatures_10v_7kmsec} is consistent with this fact. Since in the DSMC calculation a slightly larger number of endothermic dissociation reactions remove a greater amount of energy from the translational and internal modes, the DSMC temperature stays below the Navier-Stokes profile past the initial shock front.



Next, in Fig.~\ref{fig:comparison_02_7kmsec} we compare the flow parameters associated with diffusive transport for the low-speed case. First, in Fig.~\ref{fig:comparison_diffusion_fluxes_N2_10v_7kmsec} we examine the diffusion fluxes of $\mathrm{N_2}$ along the $x$-direction. Here, slightly different behavior between DSMC and the Navier-Stokes profiles are apparent. The diffusion flux for $\mathrm{N_2}$ obtained with DSMC exhibits two distinct peaks, one at $x \approx -0.001 \, \mathrm{m}$ and another closer to $x = 0.0075 \, \mathrm{m}$. This behavior is exactly mirrored for $\mathrm{N}$, although with opposite sign (not shown). By contrast, in the Navier-Stokes solution the first peak does not appear at all. Furthermore, the maxima in predicted $j_{x, \mathrm{N_2}}$ lie at about $0.0065 \, \mathrm{kg \cdot m/s}$ for DSMC vs. $0.005 \, \mathrm{kg \cdot m/s}$ for Navier-Stokes. 

In Fig.~\ref{fig:comparison_viscous_stresses_10v_7kmsec} we plot the three normal components of the viscous stress tensor for the low-speed case. The magnitudes of these stresses are approximately half of those for the high-speed case, but follow the same general behavior. Both for DSMC and Navier-Stokes we retrieve precisely $\tau_{xx, \mathrm{max}} / \tau_{yy, \mathrm{max}} = -2$, but the ratio between the peak values is now $[\tau_\mathrm{xx, \mathrm{max}}]_\mathrm{NS} / [\tau_\mathrm{xx, \mathrm{max}}]_\mathrm{DSMC} = 1.23$. In a slight departure from the high-speed case, the normal stresses do not immediately return to zero downstream of their peaks. Instead, a small plateau forms in both the DSMC and Navier-Stokes profiles.

Finally, in Fig.~\ref{fig:comparison_heat_fluxes_10v_7kmsec} we compare the heat flux profiles for the low-speed shock. The peak heat flux for DSMC was observed to be $-8.06 \, \mathrm{MW/m^2}$, whereas it was $-5.70 \, \mathrm{MW/m^2}$ in the Navier-Stokes result. This amounts to a ratio $[q_\mathrm{max}]_\mathrm{NS} / [q_\mathrm{max}]_\mathrm{DSMC} = 0.708$, as opposed to $0.756$ for the high-speed case. As was the case for the high-speed case, the DSMC and Navier-Stokes profiles agree in general shape, but differ somewhat in the location and magnitude of their maxima. As had been observed for the high-speed case, the initial departure from zero begins further upstream and is more gradual in DSMC than in the Navier-Stokes profile. Past the initial negative peak in $q_x$, both profiles exhibit a second, slightly positive overshoot downstream of the shock front. This peak, or plateau is much less pronounced and more spread out than in the high-speed case. Again, in Fig.~\ref{fig:comparison_heat_fluxes_10v_7kmsec} we have split up the Navier-Stokes profile into contributions due to heat conduction (dash-dotted line) and diffusion of enthalpy (dotted line) to assess the relative contributions of both transfer mechanisms. It can be seen that in the plateau region heat conduction in the downstream direction is almost exactly compensated for by diffusion of enthalpy in the opposite sense. The magnitudes of these fluxes are less significant when compared to the high-speed case, but the general effect is still present at this condition.



\section{Conclusions} \label{sec:conclusions}

We have presented the procedure to build a coarse-grain fluid model incorporating internal energy exchange and nonequilibrium chemistry fully consistent with the gas-kinetic description. The resulting hydrodynamic equations are equipped with dissipative transport and chemical source terms that are rigorously derived from the collision operators of the underlying kinetic equation. 

We have used a state-to-state approach, which allows for detailed description of inelastic processes in a gas mixture. A set of coarse-grain cross sections and corresponding rate coefficients derived from the NASA Ames \emph{\emph{ab initio}} database for the $\mathrm{N_2} (v,J)$-N system was employed to model internal energy exchange and dissociation-recombination reactions. The uniform rovibrational collisional (URVC) bin model was used to reduce this database to a manageable size for flow calculations. The simplicity of the URVC model makes it possible to impose reversibility relations between forward and backward elementary reactions at the coarse-grain level. These relations are expressed in terms of cross section pairs at the kinetic scale and equivalent rate coefficient pairs at the hydrodynamic scale. By means of the Chapman-Enskog method we have obtained expressions for the diffusive and viscous transport terms in the Navier-Stokes equations that are consistent with the elastic collision operators of the Boltzmann equation. All associated transport properties are calculated from the corresponding scattering cross sections. These two features of the coarse-grain model allow for the unambiguous formulation of the entropy production rates due to viscous transport and chemistry, which in turn ensures that the second law of thermodynamics is respected by the fluid equations. Demonstrating strict non-negativity of the entropy production terms is a sanity check on our derivations and in fact a fundamental requirement for any well-posed coarse-grain fluid model.

We have implemented both the fluid-scale and kinetic-scale coarse grain model in dedicated flow solvers. In order to compare their behavior, we have performed simulations of normal shock waves in nitrogen exhibiting strong thermo-chemical nonequilibrium. Flow fields at two different shock speeds were obtained, each through three numerical approaches of increasing fidelity: (1) a steady, one-dimensional inviscid flow solution obtained by coupling the master equations for detailed chemistry to momentum and energy balances along the flow direction, (2) a one-dimensional viscous flow solution to the Navier-Stokes equations by means of the Finite Volume method and (3) a gas-kinetic-scale solution using the direct simulation Monte Carlo (DSMC) method.

Our calculations reveal rather close agreement between the Navier-Stokes and DMSC predictions. This is somewhat surprising given the free-stream Mach numbers we studied ($\mathrm{Ma}_\infty \approx 28$ and $20$ respectively). At such extreme conditions one could have expected that the inability of the Navier-Stokes solutions to fully reproduce the strong translational nonequilibrium effects across the shock (i.e. bi-modal velocity distributions) would cause them to more noticeably deviate from the DSMC results. However, even though the macroscopic flow properties predicted with DSMC are more diffuse than in the Navier-Stokes calculations, all major features appear in both solutions at nearly the same $x$-location and are comparable in magnitude. With regard to resolving the shock structure with Navier-Stokes, our study reinforces the notion that employing transport properties consistent with the corresponding scattering cross sections is fundamental to obtaining close agreement with kinetic-scale solutions. Furthermore, our DMSC calculations suggest that the sensitivity of the flow field to the number of bins used in our coarse-grain model is greatly attenuated when viscous and diffusive transport effects are included. 

It should be recalled that the URVC bin model we employ assumes a constant average energy for all rovibrational levels and freezes their relative populations within a bin. This is rather restrictive and clearly not the ideal reduction strategy. On the other hand, these constraints make deriving the associated fluid equations rather simple from a mathematical viewpoint. At this stage it is not clear whether equivalent asymptotic solutions can be derived in the same manner for other existing coarse-grain models. In particular, if each bin is assumed to have an associated temperature (e.g. Boltzmann bins), it is not straightforward to translate the model into the Chapman-Enskog framework, because of the need for compatibility of this ansatz with associated scaling of the collision operators. A rigorous treatment of the internal energy for multi-temperature gases  is still an open problem for the Chapman-Enskog method. Other types of closures for transport phenomena, such as as the Maximum Entropy closure,~\cite{muller93a, levermore96a} may be more natural.

\begin{acknowledgments}
 The authors would like to thank Dr. Federico Bariselli for his contributions to the improvement of the URVC bin model and to Dr. Alessandro Munaf\`o for the codes used to generate the hydrodynamic solutions. We would also like to thank Dr. R. L. Jaffe and Dr. D. W. Schwenke from NASA Ames Research Center for access to the kinetic database used in this work.
\end{acknowledgments}

\section*{Data availability statement}

The data that support the findings of this study are available from the corresponding author upon reasonable request




\appendix


\section{Obtaining macroscopic moments of the velocity distribution in DSMC} \label{app:macroscopic_moments}


In Sec.~\ref{sec:macroscopic_moments} we gave the definitions for all macroscopic moments of the velocity distributions relevant to the comparisons of Sec.~\ref{sec:normal_shock_bins_navier_stokes_vs_dsmc}. When the numerical solution to Eq.~(\ref{eq:generalized_boltzmann_equations}) is obtained through a classical discretization in phase space, such as by Muaf\`o et al.~\cite{munafo14b}, a discretized version of the distribution function for each mixture component is obtained at every $\boldsymbol{x}$-location. This can then be numerically integrated over velocity space to yield the macroscopic moments. By contrast, in the DSMC method the distribution function is implicitly represented by a finite number of simulated particles taking on random velocities, while physical space is discretized into an array of contiguous cells. Therefore, one directly estimates the local macroscopic moments by averaging over the ensemble of DSMC particles in each cell. In this section we give the equivalent expressions for macroscopic flow field variables resulting from the DSMC simulations discussed in Secs.~\ref{sec:normal_shock_bins_dsmc} and \ref{sec:normal_shock_bins_navier_stokes_vs_dsmc}.

The mass densities defined by Eq.~(\ref{eq:species_mass_density_moments}) in a given DSMC cell are calculated as:
\begin{equation}
 \rho_i = W_p \, m_i \, [N_i]_\mathrm{cell} / V_\mathrm{cell}, \qquad i \in S \label{eq:species_number_density_dsmc}
\end{equation}
where $W_p = N^\mathrm{real} / N^\mathrm{sim}$ is the ``particle weight'' relating real gas molecules to DSMC simulated particles, $[N_i]_\mathrm{cell}$ is the number of particles of species $i \in S$ in the given DSMC cell and $V_\mathrm{cell}$ is the cell volume. Translating the definition of Eq.~(\ref{eq:hydrodynamic_velocity_moments}) into the DSMC convention, its Cartesian velocity components for a given cell are calculated as: 
\begin{equation}
 u_\nu = \frac{1}{\rho} \sum_{i \in S} \biggl\{ \rho_i \langle c_\nu \rangle^i \biggr\}, \qquad \nu = \{ 1,2,3 \}, \label{eq:hydrodynamic_velocity_dsmc}
\end{equation}
where $\langle c_\nu \rangle^i = \sum_\mathrm{cell} \left\{ c_\nu / [N_i]_\mathrm{cell} \right\}$ represents the cell-average of all particle velocity components along Cartesian direction $\nu$ belonging to species $i$. Average velocity components for each mixture component in DSMC are simply $u_\nu^i = \langle c_\nu \rangle^i$ for $\nu = \{ 1,2,3 \}$. The diffusion velocities for every mixture component are obtained as $\boldsymbol{u}_i^\mathrm{d} = \boldsymbol{u}_i - \boldsymbol{u}$. The kinetic pressure tensor was defined in Eq.~(\ref{eq:pressure_tensor_moments}) as the mass-weighted sum over all mixture components' second-order velocity moments. In DSMC its Cartesian components are calculated as:
\begin{equation}
 \begin{split}
  \mathcal{P}_{\nu \eta} = \sum\limits_{i \in S} \biggl\{ \rho_i \Bigl( & \langle c_\nu \, c_\eta \rangle^i + u_\nu \, u_\eta - \langle c_\nu \rangle^i \, u_\eta \ldots \\
  & - u_\nu \, \langle c_\eta \rangle^i \Bigr) \biggr\}, \qquad \nu,\eta = \{ 1, 2, 3 \},
  \label{eq:kinetic_stress_tensor_dsmc}
 \end{split}
\end{equation}
where $\langle c_\nu \, c_\eta \rangle^i = \sum_\mathrm{cell} \left\{ c_\nu \, c_\eta / [N_i]_\mathrm{cell} \right\}$ represent the cell-averaged products of the Cartesian velocity components of all particles belonging to species $i \in S$. Finally, in DSMC the Cartesian components of the mixture heat flux defined in Eq.~(\ref{eq:heat_flux_moments}) are calculated as:
\begin{equation}
 \begin{split}
  & q_\nu = \sum_{i \in S} \biggl\{ \frac{\rho_i}{2} \Bigl[ \langle \left| \boldsymbol{c} \right|^2 c_\nu \rangle^i - \langle \left| \boldsymbol{c} \right|^2 \rangle^i u_\nu + \left| \boldsymbol{u} \right|^2 \left( \langle c_\nu \rangle^i - u_\nu \right) \ldots \\
  & + \sum_{\eta = 1}^{3} \left\lbrace 2 u_\nu \, u_\eta \langle c_\eta \rangle^i - 2 \langle c_\nu \, c_\eta \rangle^i u_\eta \right\rbrace \Bigr] + n_i E_i \left( \langle c_\nu \rangle^i - u_\nu \right) \biggr\}, \\
  & \qquad \qquad \qquad \nu = \{ 1,2,3 \} \label{eq:heat_flux_dsmc}
 \end{split}
\end{equation}
where $\langle \left| \boldsymbol{c} \right|^2 c_\nu \rangle^i = \sum_\mathrm{cell} \{ (c_1^2 + c_2^2 + c_3^2) \, c_\nu / [N_i]_\mathrm{cell} \}$ and $\langle \left| \boldsymbol{c} \right|^2 \rangle^i = \sum_\mathrm{cell} \{ (c_1^2 + c_2^2 + c_3^2) / [N_i]_\mathrm{cell} \}$, are again understood to be cell-averages taken over all particles belonging to species $i \in S$.

Thus, in DSMC the macroscopic moments can be entirely reconstructed from instantaneous, or time-accumulated samples of the quantities $[N_i]_\mathrm{cell}$, $\langle c_\nu \rangle^i$, $\langle c_\nu \, c_\eta \rangle^i$, $\langle \left| \boldsymbol{c} \right|^2 c_\nu \rangle^i$ and $\langle \left| \boldsymbol{c} \right|^2 \rangle^i$. Using Eqs.~(\ref{eq:species_number_density_dsmc})-(\ref{eq:heat_flux_dsmc}) is especially convenient in DSMC, because it makes it possible to calculate all flow variables based only on samples gathered in the laboratory frame of reference, as opposed to a frame moving with to the local flow velocity. This avoids the need to calculate peculiar velocities for each particle and makes it possible to defer the calculation of the flow velocity and other moments depending on it to a separate post-processing stage.



\section{Transport linear systems} \label{app:transport_systems}

Linear systems of size $\mathcal{N}_\mathrm{s}$ must be solved to compute the mixture transport coefficients. For shear viscosity one must solve $\sum_{j \in S} \{ G_{ij}^\eta \alpha_j^\eta \} = x_i, \, \forall \, (i \in S)$, with the viscosity matrix given by:
\begin{align}
 G_{ii}^\eta = & \sum_{\substack{j \in S \\ i \ne j}} \left\lbrace \frac{x_i x_j}{n \mathcal{D}_{ij}} \frac{1}{m_i + m_j} \left[ 1 + \frac{3}{5} \frac{m_j}{m_k} A_{ij} \right] \right\rbrace + \frac{x_i^2}{\eta_i}, \nonumber \\
 & \qquad \qquad \qquad \qquad \qquad \qquad \qquad \quad i \in S \\
 G_{ij}^\eta = & \frac{x_i x_j}{n \mathcal{D}_{ij}} \frac{1}{m_i + m_j} \left[ \frac{3}{5} A_{ij} - 1 \right], \quad \begin{array}{c}
                                                                                                                        (i, j) \in S\\
                                                                                                                        i \ne j
                                                                                                                       \end{array}
\end{align}
and the right hand side given by the species mole fractions. The matrix entries in turn depend on the binary diffusion coefficients $\mathcal{D}_{ij} = 3/16 \sqrt{2 \pi \mathrm{k_B} T / \mu_{ij}} / ( n \, \bar{Q}_{ij}^{(1,1)} )$, the collision integral ratios $A_{ij} = \bar{Q}_{ij}^{(2,2)} / \bar{Q}_{ij}^{(1,1)}$ and the viscosity coefficients for each pure species $\eta_i = 5 / 16 \sqrt{\pi m_i \mathrm{k_B} T} / \bar{Q}_{ii}^{(2,2)}$. The mixture shear viscosity is then obtained as $\eta = \sum_{j \in S} \{ x_j \alpha_j^\eta \}$. 

In analogous manner, the system for thermal conductivity is written as $\sum_{j \in S} \{ G_{ij}^\lambda \alpha_j^\lambda \} = x_i, \, \forall \, (i \in S)$, with the entries of the thermal conductivity matrix given by:
\begin{align}
 G_{ii}^\lambda & = \frac{1}{\mathrm{k_B}} \sum_{\substack{j \in S \\ i \ne j}} \biggl\{ \frac{x_i x_j}{n \mathcal{D}_{ij}} \frac{m_i m_j}{(m_i + m_j)^2} \biggl[ \frac{30}{25} \frac{m_i}{m_j} + \frac{m_j}{m_i}  \nonumber \\
 - \frac{12}{25} & \frac{m_j}{m_i} \, B_{ij} + \frac{16}{25} A_{ij} \biggr] \biggr\} + \frac{4}{15 \, \mathrm{k_B}} \frac{x_i^2 m_i}{\eta_i}, \qquad i \in S  \label{eq:G_lambda_ii} \\
 G_{ij}^\lambda & = \frac{1}{\mathrm{k_B}} \frac{x_i x_j}{n \mathcal{D}_{ij}} \frac{m_i m_j}{(m_i + m_j)^2} \biggl[ \frac{16}{25} A_{ij} + \frac{12}{25} B_{ij} - \frac{11}{5} \biggr], \nonumber \\
 & \qquad \qquad \qquad \qquad \qquad (i, j) \in S, \quad i \ne j. \label{eq:G_lambda_ij}
\end{align}

In addition to $\mathcal{D}_{ij}$ and $A_{ij}$, Eqs.~(\ref{eq:G_lambda_ii}) and (\ref{eq:G_lambda_ij}) also depend on the collision integral ratios $B_{ij} = ( 5 \, \bar{Q}_{ij}^{(1,2)} - 4 \, \bar{Q}_{ij}^{(1,3)} ) / \bar{Q}_{ij}^{(1,1)}$. The mixture thermal conductivity is then obtained as $\lambda = \sum_{j \in S} \{ x_j \alpha_j^\lambda \}$. Once the $\alpha_j^\lambda$ have been found, the thermal diffusion ratios can be computed as $\chi_i = 5 / 2 \sum_{j \in S} \{ \Lambda_{ij} \alpha_j^\lambda \}, \, \forall \, (i \in S)$, where the matrix $\Lambda$ is made up by the entries:
\begin{align}
 \Lambda_{ii} = & \frac{1}{\mathrm{k_B}} \sum_{\substack{j \in S \\ i \ne j}} \biggl\{ \frac{x_i x_j}{n \mathcal{D}_{ij}} \frac{m_j}{m_i + m_j} \left[ \frac{2}{5} - \frac{12}{25} C_{ij} \right] \biggr\}, \quad i \in S, \\
 \Lambda_{ij} = & \frac{1}{\mathrm{k_B}} \frac{x_i x_j}{n \mathcal{D}_{ij}} \frac{m_i}{m_i + m_j} \left[ \frac{12}{25} C_{ij} - \frac{2}{5} \right], \quad \begin{array}{c}
                                                                                                            i, j \in S \\
                                                                                                            (i \ne j),
                                                                                                           \end{array}
\end{align}
which in turn depend on the additional collision integral ratios $C_{ij} = \bar{Q}_{ij}^{(1,2)} / \bar{Q}_{ij}^{(1,1)}$. Note that the thermal diffusion ratios verify the consistency relation $\sum_{i \in S} \chi_i = 0$.


\section{Collision integrals for viscous transport properties} \label{app:collision_integrals}

Our goal is to ensure consistency between the transport phenomena modeled at the hydrodynamic scale of Navier-Stokes and the kinetic-scale DSMC simulations. To this end one must compute the relevant transport properties using collision integrals consistent with the set of cross sections and scattering laws used in DSMC.

For sake of simplicity, in this work we assume that all (pseudo-) species involved in the \emph{fast} processes of Table~\ref{tab:collisional_processes} scatter isotropically. We use the variable hard sphere (VHS) model of Bird~\cite{bird80a, bird94a} for N-N elastic scattering and $\mathrm{N_2} (k)$-$\mathrm{N_2} (l)$ intra-bin scattering. For the VHS model the differential cross section takes on the form:
\begin{equation}
 \sigma_{ij \, [\mathrm{VHS}]} \left( g, \chi \right) = \frac{d_{\mathrm{ref}, ij}^2}{4 \, \Gamma \left( 5/2 - \omega_{ij} \right)} \left( \frac{2 \, \mathrm{k_B} T_\mathrm{ref}}{\mu_{ij} \, g^2} \right)^{\omega_{ij} - 1/2} \label{eq:vhs_differential_cross_section}
\end{equation}
where $d_{\mathrm{ref}, ij}$, $\omega_{ij}$ and $T_\mathrm{ref}$ are species-pair-specific model parameters used to adjust the shape of the cross section. The species-dependent parameters used in our work are taken from Stephani et al.~\cite{stephani12a} and listed in Table~\ref{tab:bin_model_vhs_parameters}. Furthermore, in Eq.~(\ref{eq:vhs_differential_cross_section}) $\mathrm{k_B}$ is Boltzmann's constant, $\mu_{ij} = m_i \, m_j / ( m_i + m_j )$ is the reduced mass for the species pair $ij$ and $\Gamma \left( . \right)$ is the gamma function. Since the VHS model assumes isotropic scattering, its differential cross section is actually independent of the post-collision deflection angle $\chi$. The corresponding integrated cross section is obtained as $\sigma_{ij}^\mathrm{I} = 2 \pi \int_{0}^{\pi} \sigma_{ij} \left( g, \chi \right) \, \sin \chi \, \mathrm{d} \chi$:
\begin{equation}
 \sigma_{ij \, [\mathrm{VHS}]}^\mathrm{I} \left( g \right) = 4 \pi \, \sigma_{ij \, [\mathrm{VHS}]} \left( g, \chi \right) \label{eq:vhs_cross_section}
\end{equation}

The integrated cross section is required at the moment of computing the collision probability $[ \sigma_{ij}^\mathrm{I} (g) \cdot g ]_\mathrm{pair} / [\sigma \cdot g]_\mathrm{max}$ for a given collision pair in the No Time Counter scheme~\cite{bird89a} of DSMC. Note that in using a single set of parameters for all $\mathrm{N_2} (k)$-$\mathrm{N_2} (l)$ collision pairs, we have implicitly assumed that the cross sections for all molecule-molecule intra-bin collisions possess the same value, regardless of the pre-collision internal states $\mathrm{N_2}(k)$ and $\mathrm{N_2}(l)$. 

\begin{table}[htb]
 \centering
 \caption{VHS parameters used in elastic N-N and $\mathrm{N_2} \left( k \right)$-$\mathrm{N_2} \left( l \right)$ intra-bin collisions} \label{tab:bin_model_vhs_parameters}
 \begin{tabular}{l c c c}
  pairing $i$-$j$           & $d_\mathrm{ref}$ [\AA] & $\omega$ & $T_\mathrm{ref}$ $[\mathrm{K}]$ \\ \hline
  & & & \\[-1em]
  $\mathrm{N}$-$\mathrm{N}$ & $2.60$ & $0.70$ & $2880$ \\
  $\mathrm{N_2} ( k )$-$\mathrm{N_2} ( l )$ & $3.20$ & $0.68$ & $2880$
 \end{tabular}
\end{table}

For consistent VHS transport properties at the hydrodynamic scale, we start from  Eq.~(\ref{eq:vhs_differential_cross_section}) to compute the integrated transport cross sections of the form: $Q_{ij}^{(l)} \left( g \right) = 2 \pi \int_{0}^{\pi} \left( 1 - \cos^l \chi \right) \sigma_{ij} \left( g, \chi \right) \sin \chi \, \mathrm{d} \chi$. For a first-order approximation of the transport coefficients only $l=1$ and $l=2$, i.e. \emph{momentum} and \emph{viscosity} cross sections, are needed. For the VHS model they take on the simple forms~\cite{bird94a} $Q_{ij \mathrm{[VHS]}}^{l=1} = \sigma_{ij \mathrm{[VHS]}}^\mathrm{I} ( g )$ and $Q_{ij \mathrm{[VHS]}}^{l=2} = \frac{2}{3} \sigma_{ij \mathrm{[VHS]}}^\mathrm{I} ( g )$ respectively. Further integration over relative collision speed $g$, yields temperature-dependent collision integrals~\cite{giovangigli99a, ferziger72a}: 
\begin{equation}
 \bar{Q}_{ij}^{(l,s)} \left( T \right)= [2 \left( l + 1 \right)] / [\left( s + 1 \right)! (2 l + 1 - (-1)^l )] \int_{0}^{\infty} Q_{ij}^l ( g ) \, \exp \left( \frac{-\mu_{ij} g^2}{2 \mathrm{k_B} T} \right) \left[ \frac{\mu_{ij} g^2}{2 \mathrm{k_B} T} \right]^{s + 1} \frac{\mu_{ij} g}{\mathrm{k_B} T} \, \mathrm{d} g.
\end{equation}

Analytical expressions of the VHS model for all necessary combinations $l=1,2$ and $s=1,2,3$ can be written as:
\begin{align}
 \bar{Q}_{ij \, [\mathrm{VHS}]}^{(1,1)} & = \frac{1}{2} \left( \frac{5}{2} - \omega_{ij} \right) f_{ij} (T) \label{eq:q_11_vhs} \\
 \bar{Q}_{ij \, [\mathrm{VHS}]}^{(1,2)} & = \frac{1}{6} \left( \frac{7}{2} - \omega_{ij} \right) \left( \frac{5}{2} - \omega_{ij} \right) f_{ij} (T) \label{eq:q_12_vhs} \\
 \bar{Q}_{ij \, [\mathrm{VHS}]}^{(1,3)} & = \frac{1}{24} \left( \frac{9}{2} - \omega_{ij} \right) \left( \frac{7}{2} - \omega_{ij} \right) \times \nonumber \\
 & \qquad \ldots \, \times \left( \frac{5}{2} - \omega_{ij} \right) f_{ij} (T) \label{eq:q_13_vhs} \\
 \bar{Q}_{ij \, [\mathrm{VHS}]}^{(2,2)} & = \frac{1}{6} \left( \frac{7}{2} - \omega_{ij} \right) \left( \frac{5}{2} - \omega_{ij} \right) f_{ij} (T) \label{eq:q_22_vhs}
\end{align}
with the common factor $f_{ij} (T) = \pi d_{\mathrm{ref}, ij}^2 \left( T / T_\mathrm{ref} \right)^{1/2 - \omega_{ij}}$.

For the \emph{fast} $\mathrm{N_2} (k)$-$\mathrm{N}$ intra-bin collisions, we also assume isotropic scattering. However, instead of defining the cross sections in terms of VHS parameters, we determine them directly based on the coarse-grained cross section database of Ref.~\cite{torres20a}. An analytical expression for the integrated cross section $\sigma_{\mathrm{N_2} (k), \mathrm{N}}^\mathrm{I} \left( g \right) = \sigma_{k \rightarrow k}^\mathrm{E} \left( g \right)$ was proposed in that reference and can be written as:
\begin{equation}
 \begin{split}
  \sigma_{k \rightarrow k}^\mathrm{E} \left( g \right) & = \frac{A_{k \rightarrow k}^\mathrm{E} \, \mathrm{k_B}^{-b_{k \rightarrow k}^\mathrm{E}}}{2 \, \Gamma \left( 3/2 + b_{k \rightarrow k}^\mathrm{E} \right)} \times \\
  & \qquad \times \sqrt{ \frac{\pi \mu_{_{\mathrm{N_2} \, \mathrm{N}}}}{2}} \left( \frac{\mu_{_{\mathrm{N_2} \, \mathrm{N}}} \, g^2}{2} \right)^{b_{k \rightarrow k}^\mathrm{E} - 1/2}
 \end{split}
\end{equation}
where the notation $\sigma_{k \rightarrow k}^E$ is shorthand for the intra-bin scattering cross section of collision the pair $\mathrm{N_2} (k) + \mathrm{N}$ and $A_{k \rightarrow k}^\mathrm{E}$, $b_{k \rightarrow k}^\mathrm{E}$ are parameters derived in Ref.~\cite{torres20a} by post-processing the NASA Ames N3 database. Given the assumption of isotropic scattering, the corresponding \emph{momentum} and \emph{viscosity} cross sections turn out to be $Q_{k,\mathrm{N}}^{l=1} = \sigma_{k \rightarrow k}^\mathrm{E}$ and $Q_{k,\mathrm{N}}^{l=2} = \frac{2}{3} \sigma_{k \rightarrow k}^\mathrm{E}$ respectively, analogous to the VHS case. The resulting analytical collision integrals are:
\begin{align}
 \bar{Q}_{k, \mathrm{N}}^{(1,1)} & = \frac{1}{2} \left( \frac{3}{2} + b_{k\rightarrow k}^\mathrm{E} \right) h_{k, \mathrm{N}} (T) \label{eq:q_11_bins} \\
 \bar{Q}_{k, \mathrm{N}}^{(1,2)} & = \frac{1}{6} \left( \frac{5}{2} + b_{k\rightarrow k}^\mathrm{E} \right) \left( \frac{3}{2} + b_{k\rightarrow k}^E \right) h_{k, \mathrm{N}} (T) \label{eq:q_12_bins} \\
 \bar{Q}_{k, \mathrm{N}}^{(1,3)} & = \frac{1}{24} \left( \frac{7}{2} + b_{k\rightarrow k}^\mathrm{E} \right) \left( \frac{5}{2} + b_{k\rightarrow k}^E \right) \times \nonumber \\
 & \qquad \ldots \times \, \left( \frac{3}{2} + b_{k\rightarrow k}^E \right) h_{k, \mathrm{N}} (T) \label{eq:q_13_bins} \\
 \bar{Q}_{k, \mathrm{N}}^{(2,2)} & = \frac{1}{6} \left( \frac{5}{2} + b_{k\rightarrow k}^\mathrm{E} \right) \left( \frac{3}{2} + b_{k\rightarrow k}^\mathrm{E} \right) h_{k, \mathrm{N}} (T) \label{eq:q_22_bins}
\end{align}
with the common bin-specific factor $h_{k, \mathrm{N}} (T) = A_{k\rightarrow k}^\mathrm{E} \, T^{b_{k\rightarrow k}^\mathrm{E}} ( \pi \mu_{_\mathrm{N_2,N}} / (8 \, \mathrm{k_B} T) )^{1/2}$. 

\bibliography{rgd}
\bibliographystyle{unsrt}


\end{document}